\documentclass[conference,draftclsnofoot]{IEEEtran}
\IEEEoverridecommandlockouts
\usepackage{amsmath,amssymb,amsfonts}
\usepackage{algorithmic}
\usepackage{graphicx}
\usepackage{textcomp}
\usepackage{xcolor}
\def\BibTeX{{\rm B\kern-.05em{\sc i\kern-.025em b}\kern-.08em
    T\kern-.1667em\lower.7ex\hbox{E}\kern-.125emX}}

\usepackage[numbers]{natbib}
\PassOptionsToPackage{options}{natbib}

\usepackage[utf8]{inputenc} 
\usepackage[T1]{fontenc}    
\usepackage{hyperref}       
\usepackage{url}            
\usepackage{booktabs}       
\usepackage{amsfonts}       
\usepackage{nicefrac}       
\usepackage{microtype}      
\usepackage{xcolor}  
\usepackage{caption}

\usepackage{amsmath}
\usepackage{amsthm}
\usepackage{amssymb}
\usepackage[capitalise]{cleveref}
\crefname{loss}{Loss}{Losses}
\usepackage{xcolor}
\usepackage{subcaption}
\usepackage{graphicx}

\newtheorem{theorem}{Theorem}
\newtheorem{proposition}{Proposition}

\newtheorem{lemma}{Lemma}

\DeclareMathOperator*{\argmin}{arg\,min}
\newcommand{\red}[1]{\textcolor{red}{#1}}

\begin{document}

\title{A Theoretical Analysis of Recommendation Loss Functions under Negative Sampling\\
{\large \red{Preprint. To appear in the Proceedings of IJCNN 2025.}}
\thanks{This work was partially supported by projects FAIR (PE0000013) and SERICS (PE00000014) under the MUR National Recovery and Resilience Plan funded by the European Union - NextGenerationEU. Supported also by the project NEREO (Neural Reasoning over Open Data) project funded by the Italian Ministry of Education and Research (PRIN) Grant no. 2022AEF-HAZ.}
}


\author{
\IEEEauthorblockN{ 
Giulia Di Teodoro\IEEEauthorrefmark{2},
Federico Siciliano\IEEEauthorrefmark{3},
Nicola Tonellotto\IEEEauthorrefmark{2}, 
Fabrizio Silvestri\IEEEauthorrefmark{3}}
\IEEEauthorblockA{
\IEEEauthorrefmark{2}\textit{Information Engineering Department, University of Pisa}, Italy\\
\IEEEauthorrefmark{3}\textit{Department of Computer, Control and Management Engineering, Sapienza University of Rome}, Italy\\
\{siciliano, fsilvestri\}@diag.uniroma1.it, giulia.di.teodoro@ing.unipi.it, nicola.tonellotto@unipi.it
}
}

\maketitle

\begin{abstract}
Loss functions like Categorical Cross Entropy (CCE), Binary Cross Entropy (BCE), and Bayesian Personalized Ranking (BPR) are commonly used in training Recommender Systems (RSs) to differentiate positive items - those interacted with by users - and negative items. While prior works empirically showed that CCE outperforms BCE and BPR when using the full set of negative items, we provide a theoretical explanation for this by proving that CCE offers the tightest lower bound on ranking metrics like Normalized Discounted Cumulative Gain (\text{NDCG}) and Mean Reciprocal Rank (MRR), followed by BPR and BCE.
However, using the full set of negative items is computationally infeasible for large-scale RSs, prompting the use of negative sampling techniques. Under negative sampling, we reveal that BPR and CCE are equivalent when a single negative sample is drawn, and all three losses converge to the same global minimum. We further demonstrate that the sampled losses remain lower bounds for \text{NDCG} (MRR), albeit in a probabilistic sense. Our worst-case analysis shows that BCE offers the strongest bound on \text{NDCG} (MRR).
Experiments on five datasets and four models empirically support these theoretical findings.
Our code and supplementary material are available at \url{https://github.com/federicosiciliano/recsys_losses.git}.
\end{abstract}


\section{Introduction} \label{sec:introduction}

Recommender Systems (RSs) are designed to predict user preferences for items, aiming to provide personalized suggestions quickly and accurately.
A common approach in RSs is to use datasets with implicit feedback, where user interactions like clicks, purchases, or views are treated as positive or target items, while non-interacted items are considered negative.
This framework forms the basis for learning to rank items according to user preferences, with the evaluation of RS models relying heavily on ranking metrics such as Normalized Discounted Cumulative Gain (NDCG) ~\citep{ndcg_2002} and Mean Reciprocal Rank (MRR). These metrics measure how well a model orders items to reflect user preferences, rewarding models that give higher scores to positive items.

Various loss functions have been proposed to optimize ranking performance. Among the most commonly used are Categorical Cross Entropy (CCE), Binary Cross Entropy (BCE), and Bayesian Personalized Ranking (BPR) \citep{Steffen2009_BPR}. Each of these losses learns to distinguish a user’s positive interactions from a set of negative items. Prior work \cite{klenitskiy2023,Petrov2023_gSAS,xu_2024} has shown that, when the full set of negative items is used in the loss computation (referred to as \textit{full losses}), CCE achieves superior performance compared to BCE and BPR. This raises a crucial question: how do these losses relate to the ranking metrics like NDCG or MRR?

In this paper, we build upon existing findings \citep{Petrov2023_gSAS,xu_2024,Wu_2024_ssm,bruch2019analysis,pu2024,yang2024pls} by providing a formal analysis of the relationship between CCE, BCE, and BPR. First, we prove that these losses act as lower bounds on NDCG (MRR) when all negative items are considered, with the tightness of the bound being highest for CCE, followed by BPR and then BCE.

In practice, however, the use of full losses is computationally infeasible due to the massive size of item catalogs in real-world RSs. RSs often require to store a collection of vector embeddings corresponding to each item in their catalog, which are updated during training. These catalogs can grow to millions or even billions of items.
In such scenarios, the embedding matrix can become the largest component of the model in terms of number of parameters. Another challenge posed by large catalogs is the size of the output scores tensor. For instance, in models like BERT4Rec \citep{Fei2019_IKM}, this tensor contains a score for each item at every position across all sequences in a batch, making it computationally impractical to train the model with catalogs exceeding one million items.
To overcome these challenges, negative sampling is commonly employed. Instead of considering all possible negative items, the model randomly samples a smaller set of negative items for each user interaction. This reduces the computational cost while still allowing the model to learn effective item rankings. Given the widespread use of negative sampling in RSs, we shift our focus to the sampled losses derived from CCE, BCE, and BPR. Various negative sampling strategies have been proposed \citep{Ding2020_NEURIPS,lian2020personalized,liu2023bayesian, Zhang2013_SIGIR,Zhao2023_RecSys,Zhu2022_ACMW}.
We focus on the uniform random sampling that is used in many RSs like Caser~\citep{Tang2018}, SASRec~\citep{kang2018_ICDM}, and GRU4Rec~\citep{hidasi_2016_gru4rec}. 

Our work shows that when only one negative item is sampled, CCE and BPR are equivalent. We theoretically prove that they generally converge to the same global optimum as BCE when the item scores are bounded.
However, this theoretical insight is limited in practice, given the inherent challenges of optimizing Deep Neural Networks (DNNs). Reaching a global minimum in DNNs is highly unlikely due to the non-convex nature of the loss landscape, and gradient-based optimization methods typically guarantee convergence only to a stationary point \citep{Baldi_2012,Palagi_2019}.
For this reason, we extended our study to investigate the behaviour of the sampled loss functions, i.e. when a subset of negative samples are considered for each target item, with respect to the \text{NDCG} (MRR) when they are sampled loss.
First, we identify distinct lower bounds for BCE, CCE, and BPR.
We demonstrate that optimizing these losses is equivalent to maximizing NDCG (MRR), but only in probabilistic terms when sampling is involved. 
Although the probabilities of these losses being bounds of the ranking metrics are not directly comparable, we analyze worst-case scenarios by examining the smallest probabilities associated with each bound. Our findings show that the lower bound provided by CCE on \text{NDCG} is weaker than that of BPR, which in turn is weaker than BCE. 

We further conduct experiments to show that our theoretical analysis is consistent with empirical evidence. We use the presented losses to train both sequential RSs such as SASRec \citep{kang2018_ICDM} and GRU4Rec \citep{hidasi_2016_gru4rec}, and non-sequential RSs like Neural Collaborative Filtering (NCF) \citep{he2017NCF} and LightGCN \citep{He2020LightGCN}  on five different datasets (MovieLens-1M, Amazon Beauty, FourSquare NYC, Amazon Books and Yelp). 

The main contributions of this work are the following:
\begin{enumerate}
    \item CCE, BPR, and BCE act as lower bounds on \text{NDCG} (MRR) when all negative items are considered, with CCE being the tighter surrogate.
    \item CCE and BPR are equivalent for samples with one positive and one negative item per user.
    \item Using the three losses leads to obtaining the same global minimum when scores are bounded.
    \item The analysis of how losses bound ranking metrics is extended to consider sampled losses, proving that optimizing these functions is equivalent to maximizing NDCG (MRR) in probabilistic terms and that, in edge cases, the CCE bound on NDCG is weaker than that of BPR, which in turn is weaker than that of BCE.
    \item Our theoretical analysis is showed to be consistent with empirical evidence.
\end{enumerate}

To our knowledge, this is the first work to formally prove the relationships between BCE, BPR, and CCE losses in RSs and to address the following important research question: what is the effect of the optimization process on \text{NDCG} (and MRR) when using BCE, CCE, or BPR losses, in full or sampled form, and is it possible to state which loss is better than the others?

\section{Related work} \label{sec:related_work}


RSs capture dynamic user interests via item interactions. Related works are detailed in the supplementary material.

Negative sampling in pointwise/pairwise losses is a common approach to address this issue: models are trained using all positive interactions while sampling only a small subset of negative interactions. 
Consequently, models that do not incorporate sampling, like BERT4Rec, cannot be utilized. Additionally, recent studies \citep{Petrov2023_gSAS,xu_2024,betello2024reproducible} indicate that LLM-based RSs are less effective than previously believed, with BERT4Rec's performance gains over SASRec being due to the absence of negative sampling and full CCE loss, rather than architectural differences. \\
BPR \citep{Steffen2009_BPR} was an early effort using negative sampling to train RSs. BPR addresses model overconfidence, i.e. the tendency to predict scores almost one for all positive items, by sampling one negative item and optimizing their relative order. However, BPR aims to optimize the Area Under the Curve (AUC) metric and the Partial AUC, which isn't ideal for ranking tasks \citep{Rendle2022_book,shi2023theories}.\\
To improve BPR, methods like WARP \citep{Weston_2011}, LambdaRank \citep{burges_2010ranknet}, LambdaFM \citep{Yuan_2016},  adaptive item sampling \citep{Rendle_2014}, importance sampling \citep{lian2020personalized} and Augmented Negative Sampling \citep{Zhao2023_RecSys} target \textit{hard} negative samples, i.e. the most informative ones, with erroneously high scores. These methods, often relying on iterative techniques, aren't suitable for NN-based sequential RSs, which perform well with parallel computing on GPUs but poorly with iterative methods. Hence, sequential RSs use simple heuristics like uniform random sampling (e.g. SASRec \citep{kang2018_ICDM}) or no negative sampling (e.g. BERT4Rec \citep{Fei2019_IKM}).
Other sampling techniques are: popularity-based negative sampling \citep{Pellegrini_2022}, that showed to be beneficial for popularity-based evaluation metrics but not for full metrics; \textit{in-batch} sampling (e.g. GRU4Rec \citep{hidasi_2016_gru4rec}) that is comparable to popularity-based sampling; and a Bayesian optimal sampling rule to do unbiased negative sampling \citep{liu2023bayesian}. Given recent studies \citep{Dallman_2021,Krichene_2022} advising against sampled metrics, we focus on the simplest uniform negative sampling technique.\\
BCE approaches the ranking problem by treating it as a series of independent binary classification tasks. BCE evaluates each probability separately, meaning their total does not need to sum to 1. As recently showed \citep{Petrov2023_gSAS}, when BCE is combined with negative sampling, the model could be overconfident in predicting the probabilities of the top-ranked items, not being able to properly distinguish between them. More recently introduced losses \citep{Petrov2023_gSAS,xu_2024} modify sampled BCE to adjust positive score magnitudes. While they can reduce training time, they usually require additional hyperparameter tuning and lack proven theoretical superiority.\\
Unlike BCE, CCE addresses the ranking problem as a multi-class classification task, accounting for the probability distribution across all sampled items. The probabilities calculated by sampled CCE sum to 1 but are higher than that from full CCE due to a smaller denominator, making CCE prone to overconfidence \citep{wei_2022}. However, recent findings \citep{Wu_2024_ssm} indicate that CCE offers model-agnostic benefits in \textit{(i)} mitigating popularity bias when using in-batch sampling, \textit{(ii)} mining hard negative samples, and \textit{(iii)} optimizing ranking metrics.
Nevertheless, it also has a drawback in inaccurately estimating score magnitudes in RSs.
In terms of ranking metrics optimization, the study demonstrates that $-log DCG$ serves as a bound for CCE when all catalogue items are considered. 
Comparable findings are discussed in \citep{bruch2019analysis,yang2024pls,Wu_2024_ssm} and in \citep{pu2024}, where a link between squared loss and ranking metrics is demonstrated via a Taylor expansion of the DCG-consistent surrogate loss, specifically the full CCE.
Our theoretical examination, however, specifically delves into the correlation between loss and ranking metrics in the context of negative sampling, which adds complexity to the analysis.
As in \citep{Petrov2023_gSAS}, we do not consider contrastive learning losses because they serve as auxiliary contrastive objectives to improve sequence representations but still rely on primary losses that require negative sampling for large item catalogs. Thus, they address an orthogonal direction to our focus.
\section{Methodology} \label{sec:methodology}
    \subsection{Background}\label{sec:background}
        This section formally presents the recommendation problem and the notation used in the paper.
        
        We consider a set of users $\mathcal{U} = \{1,2,\dots,n\} \subset \mathbb{N}$ and a set of items $\mathcal{I} = \{1,2,\dots,m\} \subset \mathbb{N}$. Let's $\mathcal{D} = \{\mathcal{I}_1,\mathcal{I}_2,\dots,\mathcal{I}_n\}$ be the set of interactions, where $\mathcal{I}_u = \{i_1,i_2,\dots,i_{m_u}\} \subset \mathcal{I}$ represents the set of interactions for user $u$.
        
        The core task in recommendation is to predict the interactions $\mathcal{I}_u$ for a given user $u$. When considering the temporal aspect of interactions, we can split $\mathcal{I}_u$ into two subsets: $\mathcal{I}_u^x$ and $\mathcal{I}_u^y$, where $\mathcal{I}_u^x$ are the interactions that occurred before those in $\mathcal{I}_u^y$. The recommendation task then becomes predicting the remaining unseen interactions $\mathcal{I}_u^y = \{i \in \mathcal{I}_u \mid i \notin \mathcal{I}_u^x\}$ given the observed interactions $\mathcal{I}_u^x \subset \mathcal{I}_u$ for user $u$. 
        Further simplification can be achieved by focusing solely on predicting the most recent item $i_u^+$ that user $u$ has interacted with: given all interactions except the last one $\mathcal{I}_u^x = \mathcal{I}_u \setminus \{i_u^+\}$, the task is to predict this last interaction $i_u^+ \in \mathcal{I}_u$.
        
        RSs typically use user and item embeddings in a latent space. The user embedding table is represented as $\mathcal{E}_u: \mathcal{U} \rightarrow \mathbb{R}^d$ and the item embedding table as $\mathcal{E}_i: \mathcal{I} \rightarrow \mathbb{R}^d$, mapping users and items to $d$-dimensional vectors. The interaction score between user $u$ and item $i$ is the dot product of their embeddings: $s_{u,i} = \vec{h_u} \cdot \vec{h_i}$. The user embedding $\vec{h_u}$ is computed from past interactions $\mathcal{I}_u^x$ using a model $\mathcal{M}$, which ideally learns embeddings that capture user preferences for accurate predictions.
        
        Maximizing the score $s_{u,i_u^+}$ for the positive item $i_u^+$ alone is insufficient, as a trivial solution could set all user and positive item embeddings to the same vector. To prevent this, negative items, which are items a user has not interacted with ($\mathcal{I}_u^- = \{i \in \mathcal{I} \;|\, i \notin \mathcal{I}_u\}$), are introduced. The model aims to maximize the score for the positive item and minimize the scores for negative items, guiding the learning of meaningful representations.
        
        \subsubsection{Losses}
            \looseness -1 \textit{Binary Cross-Entropy (BCE)}, typically applied to binary classification, can be adapted to recommendation by considering the positive item label as $1$ and negative items as $0$. The loss function is formulated as:
           \looseness -1 \begin{align*}
                \mathcal{L}_{BCE} &= \sum_{u=1}^U \ell_{BCE}^u \\
                \ell_{BCE}^u &= -\log\;\sigma\left(s_{u,i_u^+}\right) 
                - \sum_{i\in\mathcal{I}_u^-} \log\left(1-\sigma\left(s_{u,i}\right)\right)
                \end{align*}

            which, considering a single user, can be rewritten as:
            \begin{align}
                 \ell_{BCE} &= log\left(1+e^{-s_+}\right) + \sum_{i\in\mathcal{I}_u^-} log\left(1+e^{s_i}\right)
            \end{align}

            \textit{Categorical Cross-Entropy (CCE)} is commonly employed for multiclass classification:
            
            \begin{align*}
                \mathcal{L}_{CCE}&= \sum_{u=1}^{U} \ell_{CCE}^u \\
                \ell_{CCE}^u &=-log\left(\frac{e^{s_{u,i_u^+}}}{e^{s_{u,i_u^+}}-\sum_{i\in\mathcal{I}_u^-}e^{s_{u,i}}}\right)
            \end{align*}

            which, considering a single user, can be rewritten as:
            \begin{align}
                 \ell_{CCE}&= log\left(1+\sum_{i\in\mathcal{I}_u^-}e^{\left(s_i-s_+\right)}\right)
            \end{align}
            \textit{Bayesian personalized ranking (BPR)} is a pairwise ranking loss derived from the maximum posterior estimator. We present it here without its $L2$ weight normalization.
            
            \begin{align*}
                \mathcal{L}_{BPR} &= \sum_{u=1}^{U} \ell_{BPR}^u & \ell_{BPR}^u =-\sum_{i\in\mathcal{I}_u^-}log\; \sigma\left(s_{u,i_u^+}-s_{u,i}\right)
            \end{align*}
            which, considering a single user, can be rewritten as:
            \begin{align}
                 \ell_{BPR} &= \sum_{i\in\mathcal{I}_u^-}log\left(1+e^{\left(s_i-s_+\right)}\right)
            \end{align}

            Since the losses can be separated by user, we simplify the notation by omitting the user index $u$ whenever unambiguous. This replaces $s_{u,i_u^+}$ with $s_+$ and $s_{u,i}$ with $s_i$.

        \subsubsection{Metrics}
            Since the recommendation task involves ranking items, the crucial measure is the rank $r_+ = |\{i: s_i \geq s_+, i \in \mathcal{I^-}\}|+1$ of the positive item. This value represents the number of items (including itself) with scores greater than or equal to the positive item's score $s_+$.
            
            This value is part of various metrics. One of the most common metrics in recommendation is the Normalized Discounted Cumulative Gain (\text{NDCG}), which incorporates the graded relevance of items at different positions in the ranking:
            \begin{equation}
                NDCG(r_+)=\frac{1}{log_2(1+r_+)}
            \end{equation}

            Another ranking metric valuable for evaluating RSs is Mean Reciprocal Rank (MRR). MRR considers the reciprocal rank of the first relevant item in the recommendation list:
            \begin{equation}
                MRR(r_+)=\frac{1}{r_+}
            \end{equation}
            
        \subsubsection{Negative sampling}
            While the ideal scenario involves computing the true positive item rank $r_+$ and the true loss value, the sheer number of negative items $|I^-|$ makes it infeasible to compute all scores $s_i$ . Instead, a subset $\mathcal{I}^{-,K} \subseteq \mathcal{I}^-$, with $|\mathcal{I}^{-,K}| =K$, is usually sampled for computational cost.

            Negative sampling, though, introduces its own limitations. Because we only sample a subset of negative items, directly calculating the true rank is not possible. However, we can estimate the rank using an approximation $\hat r_+$. Let's define $\Gamma^K$ as the set of $K$ sampled negative items with scores greater than or equal to the positive item score $s_+$: $\Gamma^K = \{i: s_i \geq s_+, i \in \mathcal{I}^{-,K}\}$.
            The cardinality of this set provides a lower bound for the positive item rank $r_+$: $|\Gamma^K| + 1 = \leq r_+$.
            Another set worth considering is the set of all sampled negative items with non-negative scores: $\Gamma^{K}_0 = \{i: s_i \geq 0,i \in \mathcal{I}^{-,K}\}$.
            This set includes \textit{hard negative} items (clearly irrelevant) and \textit{marginally relevant} items that the model doesn't outright reject, suggesting they may have some relevance to the user.
    
    \subsection{Theoretical results} \label{sec:theoretical_results}
     Due to space constraints, not all proofs of the presented theorems, lemmas, and propositions are included here. These can be found in the supplementary material linked on GitHub.
        \subsubsection{Full loss ranking capability} Here we establish a theoretical link between full losses and standard ranking metrics.
        \begin{theorem} \label{thr:full_loss_ranking}
        Let $\ell_{CCE}$, $\ell_{BPR}$, and $\ell_{BCE}$ denote the full forms of the losses. Then the following inequalities hold:
        \begin{align*}
            -\log \text{NDCG}(r_+) \leq \ell_{CCE} \leq \ell_{BPR},
        \end{align*}
        and, if $s_+ \geq 0$, we further have:
        \begin{align*}
            \ell_{BPR} \leq \ell_{BCE}.
        \end{align*}
        \end{theorem}
         \begin{proof}
        Similar to \citep{bruch2019analysis,yang2024pls,Wu_2024_ssm,pu2024}, it is easy to see that:
        \[
        -\log \text{NDCG}(r_+) = -\log \left(\frac{1}{\log_2(1 + r_+)}\right) \leq \log(r_+),
        \]
        where \( r_+ = \sum_{i \in \mathcal{I}_u^-} \delta(s_i \geq s_+) \) and \( \delta(\cdot) \) is the Heaviside step function. Using the inequality \( x \leq 1 + x \) for \( x \geq 0 \), we get:
        $\log(r_+) \leq \log\left(1 + \sum_{i \in \mathcal{I}_u^-} \delta(s_i \geq s_+)\right)$.\\
        Since \( \delta(s_i \geq s_+) \leq e^{s_i - s_+} \), it follows that:
        \[
        \log\left(1 + \sum_{i \in \mathcal{I}_u^-} \delta(s_i \geq s_+)\right) \leq \log\left(1 + \sum_{i \in \mathcal{I}_u^-} e^{s_i - s_+}\right),
        \]
        which gives $-\log \text{NDCG}(r_+) \leq \ell_{CCE}$.
                
        To prove that \( \ell_{CCE} \leq \ell_{BPR} \), note that the inequality:
        \[
        \prod_{i \in \mathcal{I}_u^-} \left(1 + e^{s_i - s_+}\right) \geq 1 + \sum_{i \in \mathcal{I}_u^-} e^{s_i - s_+}
        \]
        holds because the exponential function is positive. Applying the logarithm yields:
        \[
        \ell_{BPR} = \log\left(\prod_{i \in \mathcal{I}_u^-} \left(1 + e^{s_i - s_+}\right)\right) \geq \ell_{CCE}.
        \]
        Thus, we conclude that:
        \[
        -\log \text{NDCG}(r_+) \leq \ell_{CCE} \leq \ell_{BPR}.
        \]
        Since \( s_+ \geq 0 \) and the exponential function is monotonically increasing, 
        $\log(1 + e^{s_i}) \geq \log(1 + e^{s_i - s_+}) \quad \forall i \in \mathcal{I}_u^-$. 
        Summing over \( \mathcal{I}_u^- \):  $\sum_{i \in \mathcal{I}_u^-} \log(1 + e^{s_i}) \geq \sum_{i \in \mathcal{I}_u^-} \log(1 + e^{s_i - s_+})$. \\ 
        Since \( \log(1 + e^{-s_+}) \geq 0 \), we have:
        
        \[
        \ell_{BCE} \geq \sum_{i \in \mathcal{I}_u^-} \log(1 + e^{s_i - s_+}) = \ell_{BPR}.
        \]

        \end{proof}

        The previous theorem establishes that all three losses serve as surrogates for \text{NDCG}, under the condition that all negative samples are used, i.e., when the losses are in their full forms. While prior works \citep{bruch2019analysis,yang2024pls,Wu_2024_ssm,pu2024} have already explored the connection between the full $\ell_{CCE}$ and ranking metrics, we are the first to highlight that the bound provided by $\ell_{CCE}$ is tighter than those given by $\ell_{BPR}$ and $\ell_{BCE}$. This implies that $\ell_{CCE}$ aligns better with ranking-based evaluation metrics and can potentially lead to faster convergence during training.  
        
        In practice, however, losses are often computed using a sampled subset of negative samples for computational efficiency. While the relative relationships between the losses are maintained with sampling, the exact bounds with respect to \text{NDCG} no longer hold. Despite this, the sampled losses remain probabilistically bounded for ranking metrics, and we will further explore this behaviour.
        
        \subsubsection{Global minimum} \label{sec:global_minimum}
        This section explores the theoretical properties of loss functions commonly used in RSs with negative sampling. 

        It can be easily seen that $\ell_{BPR}$ is equivalent to $\ell_{CCE}$ when sampling one negative item.

        \begin{proposition} \label{proposition_BPR_CCE}
            $\ell_{BPR}$ = $\ell_{CCE}$ if one negative item $K=1$ is sampled for each user.
        \end{proposition}

        Next, we present a proposition that formally states the equivalence of global minima for the three loss functions. The proposition demonstrates that when one negative item is sampled and item scores are bounded, BPR, BCE, and CCE will all achieve the same global minimum.
        
        \begin{proposition} \label{proposition_BPR_min}
            If $s_+,s_i \in [-S,S]$ with $S \in \mathbb{R^+}$ then:
            \begin{align*}
                &\argmin_{s_+} \ell_{BCE} = \argmin_{s_+} \ell_{BPR} = \argmin_{s_+} \ell_{CCE} = S\\
                &\argmin_{s_i} \ell_{BCE} = \argmin_{s_i} \ell_{BPR} = \argmin_{s_i} \ell_{CCE} = -S
            \end{align*}
        \end{proposition}

        This result has significant implications for training RSs. Firstly, it suggests consistency across BPR, BCE, and CCE loss functions in this specific scenario. This simplifies the choice of loss function in practical applications, as all three lead to the same optimal solution, reducing hyperparameter tuning complexity.

        However, it's important to acknowledge limitations in applying this finding to DNNs used in RSs. RSs inherently incorporate inductive biases for generalization and the choice of the best model based on the validation set, avoids overfitting making such extreme values $S$ and $-S$ unattainable.
        Moreover, DNNs typically have non-convex error surfaces with numerous stationary points, including local minima with identical objective values \citep{Palagi_2019}. Achieving a global minimum in DNNs with nonlinear activation functions is difficult, and the gradient descent methods typically only guarantee convergence to a stationary point \citep{Baldi_2012,Palagi_2019}. Considering these limitations, in the following sections we explore how these loss functions interact with ranking metrics under negative sampling.

        \subsubsection{Bounds} \label{sec:bounds}
        Our goal is to achieve a deeper understanding of losses' behaviour in the context of item recommendation. We establish lower bounds for each loss function in the negative sampling regime. These bounds will be instrumental in deriving probabilistic bounds for \text{NDCG}.

        \begin{lemma} \label{thm:bounds}
            Given $\Gamma^K$ and $\Gamma^K_0$, we have
            \begin{align*}
                & \ell_{BPR} \geq  |\Gamma^K| \log 2
                &\ell_{BCE} \geq |\Gamma^K_0|\log2 \\
                &\ell_{CCE} \geq  \log(|\Gamma^K|).
            \end{align*}
        \end{lemma}

        \begin{proof}
            According to the definition of BPR, it follows that
            \begin{align*}
                \ell_{BPR} &= \sum_{i\in \mathcal{I^{-,K}}}\log \left(1+ e^{(s_{i} - s_+)}\right) \\
                & = \sum_{i\in \Gamma^K} \log \left(1+ e^{(s_{i} - s_+)} \right) + \sum_{i \notin \Gamma^K} \log \left(  1+ e^{(s_{i} - s_+)} \right) \\
                & \geq \sum_{i\in \Gamma^K} \log \left( 1+ e^{(s_{i} - s_+)} \right) \\
                & \geq |\Gamma^K|\log 2
            \end{align*}

            The proof for the bounds of $\ell_{BCE}$ and $\ell_{CCE}$ can be obtained in a similar manner, and their details are provided in the supplementary material.
        \end{proof}

        \subsubsection{Loss ranking capability} \label{sec:losses_ranking_capabilities}
            The established bounds on the losses depend on the cardinality of sets: \(\Gamma^K\) and \(\Gamma^K_0\), which are influenced by the negative sampling method. Here, we introduce a way of computing these probabilities.

            \begin{lemma} \label{lemma_prob} 
                Let consider the sets $\Gamma^K$ and $\Gamma^K_0$. If the $K$ negative items are \textit{uniformly} sampled without replacement from the set $I^-$ of all negative items, then $|\Gamma^K|$ and $|\Gamma^K_0|$ are described by two hypergeometrics:
                \begin{align*}
                    &|\Gamma^K |\sim Hypergeometric\left(|I^-|,|\Gamma|,K\right)\\
                    &|\Gamma^K_0| \sim Hypergeometric\left(|I^-|,|\Gamma_0|,K\right)
                \end{align*}
            \end{lemma}
            
            The following theorem establishes the minimum probability that the three losses are an upper bound on the negative logarithm of \text{NDCG}. This tells us than minimising a loss leads to maximising the \text{NDCG}.
            
            \begin{theorem} \label{thr:loss_ndcg}
                When uniformly sampling $K$ negative items, we have:
                \begin{align} \label{prob_relation_BCE}
                &\mathbb{P}(-\log \text{NDCG}(r_+) \leq \ell_{BCE}) \nonumber\\
                &\geq 1- CDF_{|\Gamma^K_0|}\left(\log_2 \left(\log_2 (1+r_+)\right)\right)\\
                \label{prob_relation_BPR}
                &\mathbb{P}(-\log \text{NDCG}(r_+) \leq \ell_{BPR}) \nonumber\\
                &\geq 1- CDF_{|\Gamma^K|}\left(\log_2 \left(\log_2 (1+r_+)\right)\right)\\
                \label{prob_relation_CCE}
                &\mathbb{P}(-\log \text{NDCG}(r_+) \leq \ell_{CCE}) \nonumber\\
                &\geq 1- CDF_{|\Gamma^K|}\left(\log_2 (1+r_+)\right)
                \end{align}

            \end{theorem}

        \begin{proof} 
            According to \cref{thm:bounds} we know that
            $\ell_{BPR} \geq  |\Gamma^K| \log 2 $
            Therefore, $\mathbb{P}(-\log \text{NDCG}(r_+) \leq \ell_{BCE})$ can be rewritten as:
            \begin{small}
            \begin{align}  
             & P(-\log \text{NDCG}(r_+) \leq \ell_{BPR})  \nonumber\\
             &\geq P\left(-\log \left(\frac{1}{\log_2\left(1+r_+\right)}\right) 
             \leq |\Gamma^K|\log2 \right) \nonumber\\
             &=P\left(\log_2\left(1+r^+\right) \leq 2^{|\Gamma^K|}\right) \nonumber\\
             &= P\left(|\Gamma^K| \geq \log_2 \left(\log_2 (1+r_+)\right)\right)  \label{eq:prob_BPR}
            \end{align}
            \end{small}
            According to \cref{lemma_prob}, $|\Gamma^K|$ is an hypergeometric variable with population size $|\mathcal{I^-}|$, $|\Gamma|$ number of successes and $K$ number of draws. Thus, by defining $\rho = \log_2(\log_2(1+r_+))$, we can rewrite \cref{eq:prob_BPR} using the cumulative distribution function of an hypergeometric:
            \begin{small}
            \begin{align*}
                P\left(|\Gamma^K| \geq \rho \right) = 1- CDF_{|\Gamma^K|}\left(\rho \right)
            \end{align*}
            \end{small}
            that is 
            \begin{small}
            \begin{align*}
                P\left(- log (NDCG(r_+))\leq \ell_{BPR}\right) \geq 1 - \sum_{i=0}^{\lfloor\rho\rfloor} \frac{\binom{|\Gamma|}{i} \binom{|\mathcal{I^-}|-|\Gamma|}{K-i}}{\binom{|\mathcal{I^-}|}{K}}
            \end{align*}
            \end{small}
            The proof for $\ell_{BCE}$ and $\ell_{CCE}$ can be obtained in a similar manner, and their details are provided in the supplementary material.
            \end{proof}

        \subsubsection{Comparing losses} \label{sec:losses_comparative_analysis}
            This section compares the probabilistic bounds established for each loss function in their ability to upper bound $-\log(\text{NDCG})$. We focus on the worst-case scenario, where the probability of achieving the bound is the smallest possible.

            \begin{theorem} \label{thr:loss_comparison}
                When uniformly sampling $K$ negative items, in the worst-case scenario, and $s_+ \geq 0$:
                \begin{align*} 
                \mathbb{P}(-\log \text{NDCG}(r_+) \leq \ell_{BCE}) \geq \\
                \mathbb{P}(-\log \text{NDCG}(r_+) \leq \ell_{BPR})  \geq \\
                \mathbb{P}(-\log \text{NDCG}(r_+) \leq \ell_{CCE})
                \end{align*}
            \end{theorem}

            \cref{thr:loss_comparison} shows that the bound for BPR is inherently weaker than the bound for BCE because the corresponding term in BPR is smaller. This suggests that BPR is less likely to achieve a good bound on \text{NDCG} compared to BCE. In addition, for the same predicted rank $r_+ $ and number of negative samples $K$, the bound probability for BPR is higher than for CCE.

            As training progresses, the true rank $r_+$ of the positive item generally decreases, meaning that it ranks higher in relation to other items. This reduction in $r_+$ 	can decrease the size of $|\Gamma^K|$, which in turn influences the probability bound for the ranking model. This interplay makes it challenging to determine a definitive trend.
            Additionally, during training, models trained using BPR and CCE may begin to predict different values of $r_+$ due to their distinct optimization objectives. This difference in predicted ranks can alter the relationship between the probabilities derived from the bounds, further complicating the trend analysis.
            

            In contrast, the BCE bound appears more favorable. If the data does not induce embedding collapse — e.g., in the absence of strong popularity bias \cite{zhao2023embedding} — item embeddings are expected to remain well distributed, keeping the set of negative items with non-negative scores $|\Gamma_0|$ relatively stable during training. In this case, as the true rank $r_+$ decreases, BCE offers a tighter bound to ranking metrics. However, since real-world data may force embeddings to concentrate in specific regions, the interplay between $|\Gamma_0|$ and $r_+$ makes it difficult to draw a general conclusion on the relative effectiveness of BCE.

        \subsection{Mean Reciprocal Rank} \label{sec:reciprocal_rank}
            The results for MRR are analogous to those for \text{NDCG} and reported integrally in the supplementary material.
            
\section{Experiments} \label{sec:experiments}
    In this section, we outline the experimental design employed to empirically validate our theoretical findings regarding recommendation loss functions under negative sampling.

    \subsection{Experimental setup}\label{sec:experimental_results}
    We follow a next-item prediction approach, aligning with established works \citep{hidasi_2016_gru4rec, kang2018_ICDM}. This signifies that the positive instance under evaluation is the next item to be predicted, which corresponds to the last item in a user's test sequence.

    \textbf{Datasets.} To ensure the generalizability of our results, we leverage three well-regarded real-world benchmark datasets for RSs:

    \begin{itemize}
        \item \textit{Amazon Beauty \citep{10.1145/2766462.2767755}:} This dataset, on the smaller side, encompasses product reviews from 1,274 users for 1,076 items, totalling 7,113 interactions.
        \item \textit{Foursquare NYC \citep{6844862}:} A widely adopted dataset containing check-in records within New York City, encompassing 227,42 entries.
        \item \textit{MovieLens 1M \citep{10.1145/2827872}:} This dataset features 1 million movie ratings from various users.
    \end{itemize}

    The details of the larger datasets, Yelp and Amazon Books, and the related results of experiments conducted on these datasets are given in the supplementary material.

    \textbf{Baselines.} For a thorough evaluation, we incorporate two well-established, state-of-the-art RSs:
    (\textit{i}) \textbf{GRU4Rec \citep{hidasi_2016_gru4rec}} that leverages a Gated Recurrent Unit architecture for recommendation;
    (\textit{ii}) \textbf{SASRec \citep{kang2018_ICDM}} that employs a self-attention mechanism for recommendation tasks.

    Results for non-sequential RSs, NCF and LightGCN, are presented in the supplementary material.

    \textbf{Losses.} We evaluate the performance of the three introduced loss functions: BCE, BPR, and CCE.

    \textbf{Negative Sampling.} We consider different number of uniformly sampled negative items per sample. In particular, the considered values are within the set $\{1, 2, 5, 20, 50, 100\}$. 

    \textbf{Hyperparameter Settings}
    For consistency across all experiments, we utilize the following hyperparameters:
    Embedding Size: 64;
    Input Sequence Length: 200;
    Batch Size: 128;
    Optimizer: Adam;
    Learning Rate: 0.001;
    Maximum Training Epochs: 600.\\
    These hyperparameter selections were established based on prevailing practices within the RSs domain, as evidenced in prior works like \citep{hidasi_2016_gru4rec,kang2018_ICDM}.
    
    \subsection{Empirical results} \label{sec:empirical_results}

    Due to space constraints, we present the full set of results only for the ML-1M dataset and losses comparison for Foursquare dataset. The results for the other datasets can be found in the supplementary material. Similar considerations to those drawn in this section also apply to the other datasets.

    \subsubsection{Number of negatives comparison} \label{sec:number_neg_comparison}
    \begin{figure*}[!ht]
    \centering
        \begin{subfigure}{0.32\textwidth}
            \centering
            \includegraphics[width=\textwidth]{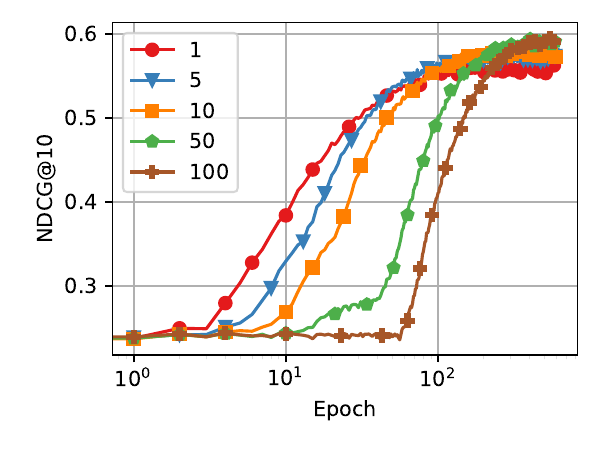}
            \caption{BCE}
        \end{subfigure}
        \begin{subfigure}{0.32\textwidth}
            \centering
            \includegraphics[width=\textwidth]{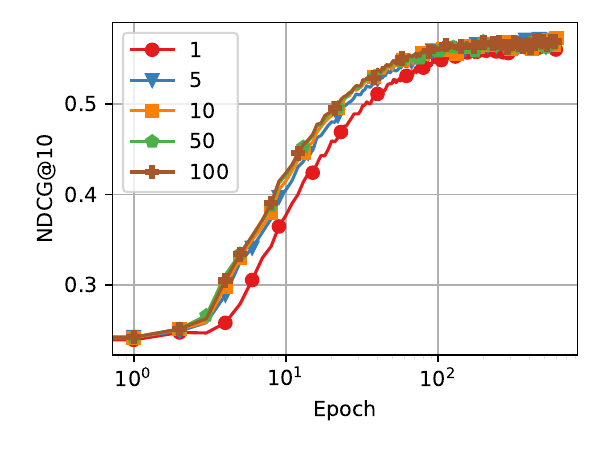}
            \caption{BPR}
        \end{subfigure}
        \begin{subfigure}{0.32\textwidth}
            \centering
            \includegraphics[width=\textwidth]{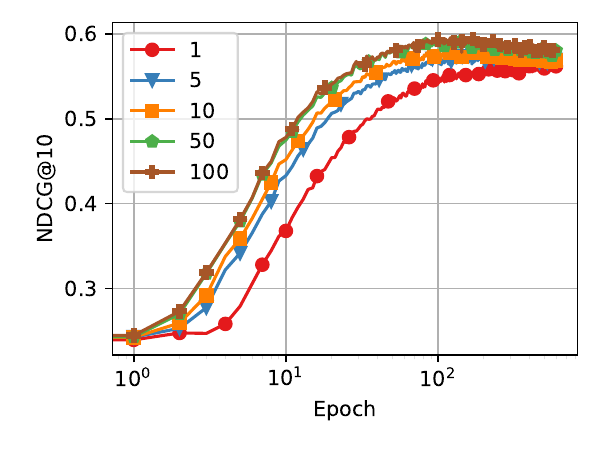}
            \caption{CCE}
        \end{subfigure}
        \caption{GRU4Rec NDCG@10 during training changing number of negative items and loss on ML-1M dataset.}
        \label{fig:neg_gru4rec}
    \end{figure*}


        \begin{figure*}[!ht]
        \centering
        \begin{subfigure}{0.32\textwidth}
            \centering
            \includegraphics[width=\textwidth]{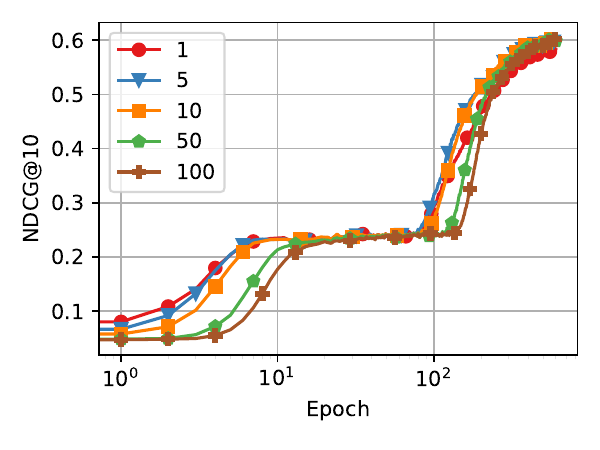}
            \caption{BCE}
        \end{subfigure}
        \begin{subfigure}{0.32\textwidth}
            \centering
            \includegraphics[width=\textwidth]{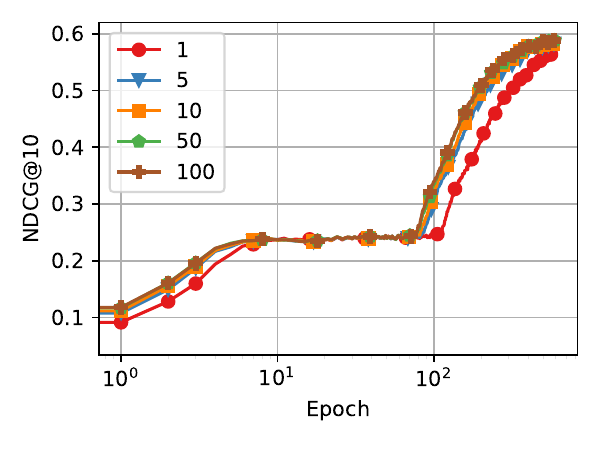}
            \caption{BPR}
        \end{subfigure}
        \begin{subfigure}{0.32\textwidth}
            \centering
            \includegraphics[width=\textwidth]{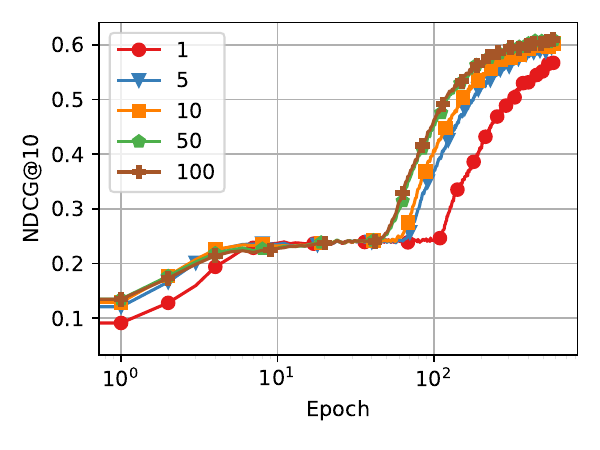}
            \caption{CCE}
        \end{subfigure}
        \caption{SASRec NDCG@10 during training changing number of negative items and loss on ML-1M dataset.}
        \label{fig:neg_sasrec}
    \end{figure*}

       \begin{figure}[ht!]
       \centering
        \begin{subfigure}[b]{0.24\textwidth}
            \centering
            \includegraphics[width=\textwidth]{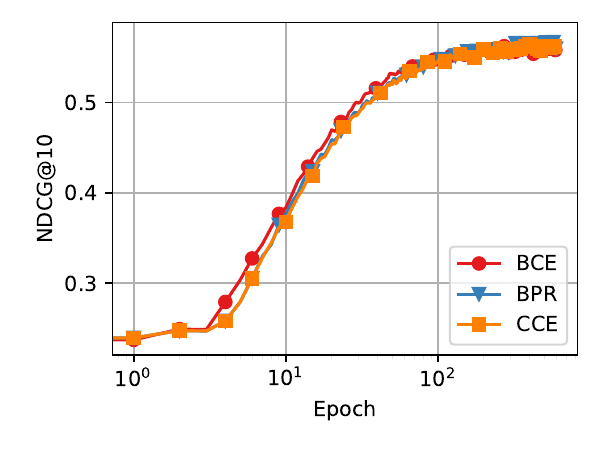}
            \caption{GRU4Rec}
        \end{subfigure}
        \hfill
        \begin{subfigure}[b]{0.24\textwidth}
            \centering
            \includegraphics[width=\textwidth]{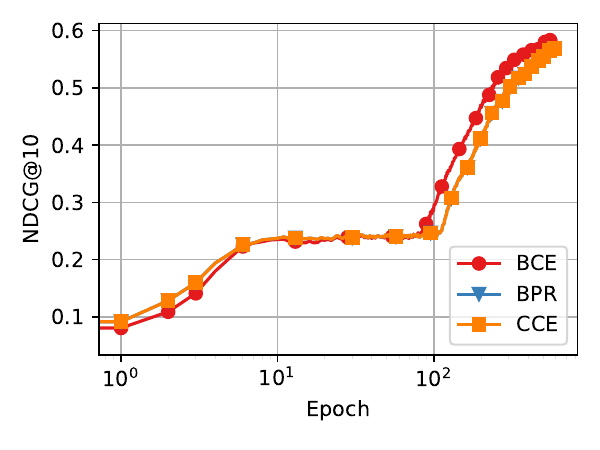}
            \caption{SASRec}
        \end{subfigure}
        
        \caption{SASRec and GRU4Rec NDCG@10 during training, using 1 negative item and changing loss on ML-1M dataset.}
        \label{fig:loss_1neg}
    \end{figure}

    \begin{figure}[ht!]
    \centering
        \begin{subfigure}[b]{0.24\textwidth}
            \centering
            \includegraphics[width=\textwidth]{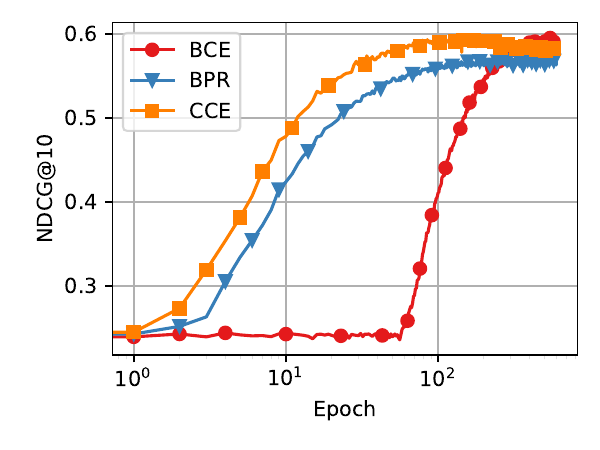}
            \caption{GRU4Rec}
        \end{subfigure}
        \hfill
        \begin{subfigure}[b]{0.24\textwidth}
            \centering
            \includegraphics[width=\textwidth]{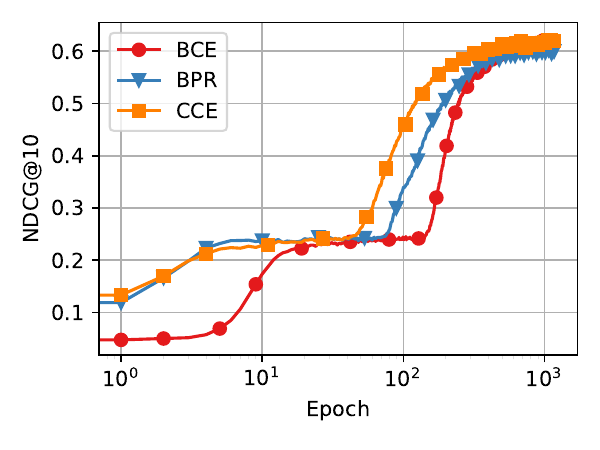}
            \caption{SASRec}
        \end{subfigure}
        
        \caption{SASRec and GRU4Rec NDCG@10 during training, using 100 negative items and changing loss on ML-1M dataset.}
        \label{fig:loss_100neg}
    \end{figure}

    \begin{figure}[ht!]
    \centering
    \begin{subfigure}[b]{0.24\textwidth} 
        \centering
        \includegraphics[width=\textwidth]{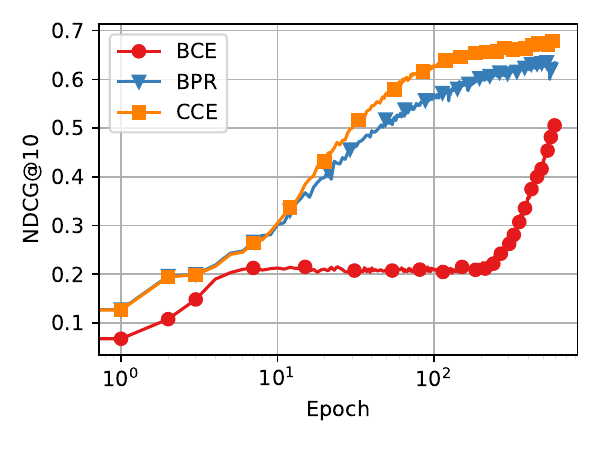}
        \caption{GRU4Rec}
    \end{subfigure}
    \hfill 
    \begin{subfigure}[b]{0.24\textwidth} 
        \centering
        \includegraphics[width=\textwidth]{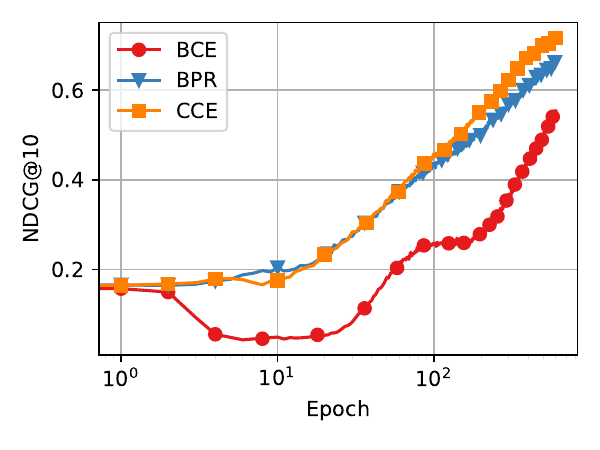}
        \caption{SASRec}
    \end{subfigure}

    \caption{SASRec and GRU4Rec NDCG@10 during training changing loss, using 100 negative items on Foursquare dataset.}
    \label{fig:fs_loss_100neg}
\end{figure}

    \cref{fig:neg_gru4rec,fig:neg_sasrec} visualize the results for GRU4Rec and SASRec. The results show NDCG@10 as a function of training epochs and the number of negative items, for the three losses. 

    For BCE, we can observe that in the early training epochs, using fewer negatives leads to better results. In particular, NDCG@10 exhibits an initial plateau that extends the more negatives there are; so much that when $100$ are used, GRU4Rec does not show improvements in the metric until almost epoch $100$. This could be due to the fact that the model is focusing too much on the negatives and too little on the positive item. In the initial training stages, the model is still learning to distinguish between relevant and irrelevant items and the embedding representations in the latent space are poor. This is even more challenging when more negative items and only one positive item are considered.

    The final part of the training shows a different situation. In fact, even though the increase in the metric is postponed for many negatives, the rise is faster, so much that the final metric benefits from the use of more negatives: the best final metrics are obtained when $100$ negatives are used. This is likely because, once the model is trained, it is more difficult to do meaningful sampling using only one negative, while it is easier to draw items with scores higher than the positive one if more are used. 
    
    BPR behaves similarly at the end of training, but shows that using more negatives is even better initially. Our hypothesis is that in BPR the positive item is used for the loss calculation paired with each negative, so it is effectively weighted the same number of times as the negatives.
    
    Finally, CCE shows results similar to BPR, with the use of more negatives improving both the final and initial phase of training. This is the only model where it is possible to notice a slight difference between the two models, GRU4Rec and SASRec, specifically in the early training epochs, where using $5$ or $10$ negatives seems to be better than using either $1$ or $100$, even if the difference is small.
    
    We can conclude that, in general, the use of more negatives is almost always advantageous, especially for the end performance of the models. However, since the use of a single negative is a special case of interest, the next section will compare the losses with one negative and $100$ negatives, the two limiting cases of the analysis we have carried out.

    \subsubsection{Losses Comparison} \label{sec:losses_comparison}
    The experimental results on ML-1M and Foursquare provide valuable insights into the behavior of different loss functions when training SASRec and GRU4Rec. For ML-1M \cref{fig:loss_1neg,fig:loss_100neg} depict NDCG@10 as a function of training epochs, considering both one and 100 negative items. When one negative sample is used, BPR and CCE exhibit identical performance for SASRec, consistent with theoretical findings. On GRU4Rec, however, slight discrepancies between the two are likely due to machine-level approximations, as both losses remain implemented in their original formulations. In this scenario, BCE demonstrates better performance than BPR and CCE during the final training epochs for SASRec, whereas for GRU4Rec, BCE, BPR, and CCE exhibit smaller performance differences.

    With 100 negative samples, the results shift. For ML-1M, CCE generally outperforms BPR on both models, with a few exceptions where BPR briefly surpasses CCE on SASRec. BCE, while initially lagging behind CCE and BPR in early epochs, shows a steep and consistent improvement throughout training. Notably, BCE surpasses the other two losses during the final epochs for ML-1M, showing that this behavior is probably due to its theoretical advantage in aligning better with the ranking metric in the worst-case scenario. On Foursquare, with 100 negatives, as shown in \cref{fig:fs_loss_100neg}, BCE again ranks behind CCE and BPR at the beginning of training, but its strong upward trend suggests the possibility of overcoming them with prolonged training. However, since the number of epochs required for this passing is uncertain, CCE emerges as a more reliable loss due to its better overall performance during training.

\section{Conclusions} \label{sec:conclusion}



In this paper, we provided a comprehensive theoretical analysis of BCE, CCE, and BPR losses in the context of RSs. We have shown that, when using full losses, CCE provides the tightest bound to \text{NDCG} and MRR metrics, followed by BPR and then BCE. When these losses are used in sampled form, they continue to be bonds for these metrics only probabilistically. We studied the impact of uniform negative sampling when different numbers of negative items are considered. Our findings revealed that BPR and CCE share the same functional form when one negative item is considered. We first proved that the three losses share the same global minimum. Secondly, we proved that in worst-case scenario, CCE and BPR bounds on ranking metrics are generally weaker than those provided by BCE. Experimental results across various datasets validated our theoretical insights. They indicate that BCE could outperform the other two losses in the later training stages, likely due to its better alignment with the ranking metric under these conditions. Despite this, the overall performance during training suggests that CCE demonstrates more consistent behavior across datasets and models, making it a compelling choice for practical applications where the training dynamics play a significant role. This work advances our understanding of how different loss functions impact RSs performance, offering valuable guidance in selecting the appropriate loss function for reccomendation tasks.
Future research could investigate the impact of various negative sampling strategies on loss performance, as the current theorems are valid only with uniform sampling.
An open problem is extending our analysis to the multi-user scenario where an item can be positive for one user but negative for another, where the loss optimization becomes more complex and difficult to study.

\textbf{Limitations.}
This paper compares sampled loss functions under the worst-case scenario, where the probability that these losses serve as a bound for ranking metrics is the smallest possible. Additionally, our analysis assumes that the score for the target item is non-negative, a condition likely to occur during later stages of model training but probably not in the early phases of training. 

\bibliographystyle{plain}
\bibliography{biblio}

\begin{thebibliography}{10}

\bibitem{Baldi_2012}
P.~Baldi and Z.~Lu.
\newblock Complex-valued autoencoders.
\newblock {\em Neural Networks}, 33:136--147, 2012.

\bibitem{betello2024reproducible}
Filippo Betello, Antonio Purificato, Federico Siciliano, Giovanni Trappolini, Andrea Bacciu, Nicola Tonellotto, and Fabrizio Silvestri.
\newblock A reproducible analysis of sequential recommender systems.
\newblock {\em IEEE Access}, 2024.

\bibitem{bruch2019analysis}
S.~Bruch, X.~Wang, M.~Bendersky, and M.~Najork.
\newblock An analysis of the softmax cross entropy loss for learning-to-rank with binary relevance.
\newblock In {\em Proceedings of the 2019 ACM SIGIR international conference on theory of information retrieval}, pages 75--78, 2019.

\bibitem{burges_2010ranknet}
C.~J.~C. Burges.
\newblock From {RankNet} to {LambdaRank} to {LambdaMART}: An overview.
\newblock Technical report, Microsoft Research, 2010.

\bibitem{Dallman_2021}
A.~Dallmann, D.~Zoller, and A.~Hotho.
\newblock A case study on sampling strategies for evaluating neural sequential item recommendation models.
\newblock In {\em Proceedings of the 15th ACM Conference on Recommender Systems}, RecSys '21, page 505–514, New York, NY, USA, 2021. Association for Computing Machinery.

\bibitem{desouza_2021}
G.~de~Souza Pereira~Moreira, S.~Rabhi, J.~M. Lee, R.~Ak, and E.~Oldridge.
\newblock Transformers4rec: Bridging the gap between nlp and sequential / session-based recommendation.
\newblock In {\em Proceedings of the 15th ACM Conference on Recommender Systems}, RecSys '21, page 143–153, New York, NY, USA, 2021. Association for Computing Machinery.

\bibitem{Ding2020_NEURIPS}
J.~Ding, Y.~Quan, Q.~Yao, Y.~Li, and D.~Jin.
\newblock Simplify and robustify negative sampling for implicit collaborative filtering.
\newblock In {\em Proceedings of the 34th International Conference on Neural Information Processing Systems}, NIPS '20, Red Hook, NY, USA, 2020. Curran Associates Inc.

\bibitem{10.1145/2827872}
F.~Maxwell Harper and Joseph~A. Konstan.
\newblock The movielens datasets: History and context.
\newblock {\em ACM Trans. Interact. Intell. Syst.}, 5(4), dec 2015.

\bibitem{He2020LightGCN}
X.~He, K.~Deng, X.~Wang, Y.~Li, Y.~Zhang, and M.~Wang.
\newblock Lightgcn: Simplifying and powering graph convolution network for recommendation.
\newblock In {\em Proceedings of the 43rd International ACM SIGIR Conference on Research and Development in Information Retrieval}, SIGIR '20, page 639–648, New York, NY, USA, 2020. Association for Computing Machinery.

\bibitem{he2017NCF}
X.~He, L.~Liao, H.~Zhang, L.~Nie, X.~Hu, and T.~Chua.
\newblock Neural collaborative filtering.
\newblock In {\em Proceedings of the 26th international conference on world wide web}, pages 173--182, 2017.

\bibitem{hidasi_2016_gru4rec}
B.~Hidasi, A.~Karatzoglou, L.~Baltrunas, and D.~Tikk.
\newblock Session-based recommendations with recurrent neural networks.
\newblock In {\em International Conference on Learning Representations (ICLR)}, 2016.

\bibitem{hu2008cf}
Y.~Hu, Y.~Koren, and C.~Volinsky.
\newblock Collaborative filtering for implicit feedback datasets.
\newblock In {\em 2008 Eighth IEEE international conference on data mining}, pages 263--272. Ieee, 2008.

\bibitem{ndcg_2002}
K.~J\"{a}rvelin and J.~Kek\"{a}l\"{a}inen.
\newblock Cumulated gain-based evaluation of ir techniques.
\newblock {\em ACM Trans. Inf. Syst.}, 20(4):422–446, oct 2002.

\bibitem{kang2018_ICDM}
W.~Kang and J.~McAuley.
\newblock Self-attentive sequential recommendation.
\newblock In {\em 2018 IEEE International Conference on Data Mining (ICDM)}, pages 197--206, Los Alamitos, CA, USA, nov 2018. IEEE Computer Society.

\bibitem{klenitskiy2023}
A.~Klenitskiy and A.~Vasilev.
\newblock Turning dross into gold loss: is bert4rec really better than sasrec?
\newblock In {\em Proceedings of the 17th ACM Conference on Recommender Systems}, pages 1120--1125, 2023.

\bibitem{koren2009}
Y.~Koren, R.~Bell, and C.~Volinsky.
\newblock Matrix factorization techniques for recommender systems.
\newblock {\em Computer}, 42(8):30--37, 2009.

\bibitem{Krichene_2022}
W.~Krichene and S.~Rendle.
\newblock On sampled metrics for item recommendation.
\newblock {\em Commun. ACM}, 65(7):75–83, jun 2022.

\bibitem{lee1999}
D.~D Lee and H~S. Seung.
\newblock Learning the parts of objects by non-negative matrix factorization.
\newblock {\em nature}, 401(6755):788--791, 1999.

\bibitem{Li_2017}
J.~Li, P.~Ren, Z.~Chen, Z.~Ren, T.~Lian, and J.~Ma.
\newblock Neural attentive session-based recommendation.
\newblock In {\em Proceedings of the 2017 ACM on Conference on Information and Knowledge Management}, CIKM '17, page 1419–1428, New York, NY, USA, 2017. Association for Computing Machinery.

\bibitem{lian2020personalized}
D.~Lian, Q.~Liu, and E.~Chen.
\newblock Personalized ranking with importance sampling.
\newblock In {\em Proceedings of The Web Conference 2020}, pages 1093--1103, 2020.

\bibitem{liu2023bayesian}
B.~Liu and B.~Wang.
\newblock Bayesian negative sampling for recommendation.
\newblock In {\em 2023 IEEE 39th International Conference on Data Engineering (ICDE)}, pages 749--761. IEEE, 2023.

\bibitem{10.1145/2766462.2767755}
J.~McAuley, C.~Targett, Q.~Shi, and A.~van~den Hengel.
\newblock Image-based recommendations on styles and substitutes.
\newblock In {\em Proceedings of the 38th International ACM SIGIR Conference on Research and Development in Information Retrieval}, SIGIR '15, page 43–52, New York, NY, USA, 2015. Association for Computing Machinery.

\bibitem{Palagi_2019}
L.~Palagi.
\newblock Global optimization issues in deep network regression: an overview.
\newblock {\em J. Glob. Optim.}, 73(2):239--277, February 2019.

\bibitem{Pellegrini_2022}
R.~Pellegrini, W.~Zhao, and I.~Murray.
\newblock Don’t recommend the obvious: estimate probability ratios.
\newblock In {\em Proceedings of the 16th ACM Conference on Recommender Systems}, RecSys '22, page 188–197, New York, NY, USA, 2022. Association for Computing Machinery.

\bibitem{Petrov2023_gSAS}
A.~V. Petrov and C.~Macdonald.
\newblock gsasrec: Reducing overconfidence in sequential recommendation trained with negative sampling.
\newblock In {\em Proceedings of the 17th ACM Conference on Recommender Systems}, RecSys '23, page 116–128, New York, NY, USA, 2023. Association for Computing Machinery.

\bibitem{pu2024}
Y.~Pu, X.~Chen, X.~Huang, J.~Chen, D.~Lian, and E.~Chen.
\newblock Learning-efficient yet generalizable collaborative filtering for item recommendation.
\newblock In {\em Forty-first International Conference on Machine Learning (ICML)}, 2024.

\bibitem{Rajput_2023}
S.~Rajput, N.~Mehta, A.~Singh, R.~Hulikal~Keshavan, T.~Vu, L.~Heldt, L.~Hong, Y.~Tay, V.~Tran, J.~Samost, M.~Kula, E.~Chi, and M.~Sathiamoorthy.
\newblock Recommender systems with generative retrieval.
\newblock In A.~Oh, T.~Naumann, A.~Globerson, K.~Saenko, M.~Hardt, and S.~Levine, editors, {\em Advances in Neural Information Processing Systems}, volume~36, pages 10299--10315. Curran Associates, Inc., 2023.

\bibitem{Rendle2022_book}
S.~Rendle.
\newblock {\em Item Recommendation from Implicit Feedback}, pages 143--171.
\newblock Springer US, New York, NY, 2022.

\bibitem{Rendle_2014}
S.~Rendle and C.~Freudenthaler.
\newblock Improving pairwise learning for item recommendation from implicit feedback.
\newblock In {\em Proceedings of the 7th ACM International Conference on Web Search and Data Mining}, WSDM '14, page 273–282, New York, NY, USA, 2014. Association for Computing Machinery.

\bibitem{Steffen2009_BPR}
S.~Rendle, C.~Freudenthaler, Z.~Gantner, and L.~Schmidt-Thieme.
\newblock Bpr: Bayesian personalized ranking from implicit feedback.
\newblock In {\em Proceedings of the Twenty-Fifth Conference on Uncertainty in Artificial Intelligence}, UAI '09, page 452–461, Arlington, Virginia, USA, 2009. AUAI Press.

\bibitem{shi2023theories}
W.~Shi, J.~Chen, F.~Feng, J.~Zhang, J.~Wu, C.~Gao, and X.~He.
\newblock On the theories behind hard negative sampling for recommendation.
\newblock In {\em Proceedings of the ACM Web Conference 2023}, pages 812--822, 2023.

\bibitem{Fei2019_IKM}
F.~Sun, J.~Liu, J.~Wu, C.~Pei, X.~Lin, W.~Ou, and P.~Jiang.
\newblock Bert4rec: Sequential recommendation with bidirectional encoder representations from transformer.
\newblock In {\em Proceedings of the 28th ACM International Conference on Information and Knowledge Management}, CIKM '19, page 1441–1450, New York, NY, USA, 2019. Association for Computing Machinery.

\bibitem{Tang2018}
J.~Tang and K.~Wang.
\newblock Personalized top-n sequential recommendation via convolutional sequence embedding.
\newblock In {\em Proceedings of the Eleventh ACM International Conference on Web Search and Data Mining}, WSDM '18, page 565–573, New York, NY, USA, 2018. Association for Computing Machinery.

\bibitem{Vaswani2017_NIPS}
A.~Vaswani, N.~Shazeer, N.~Parmar, J.~Uszkoreit, L.~Jones, A.~N Gomez, L.~Kaiser, and I.~Polosukhin.
\newblock Attention is all you need.
\newblock In I.~Guyon, U.~Von Luxburg, S.~Bengio, H.~Wallach, R.~Fergus, S.~Vishwanathan, and R.~Garnett, editors, {\em Advances in Neural Information Processing Systems}, volume~30. Curran Associates, Inc., 2017.

\bibitem{wei_2022}
H.~Wei, R.~Xie, H.~Cheng, L.~Feng, B.~An, and Y.~Li.
\newblock Mitigating neural network overconfidence with logit normalization.
\newblock In {\em International conference on machine learning}, pages 23631--23644. PMLR, 2022.

\bibitem{Weston_2011}
J.~Weston, S.~Bengio, and N.~Usunier.
\newblock Wsabie: scaling up to large vocabulary image annotation.
\newblock In {\em Proceedings of the Twenty-Second International Joint Conference on Artificial Intelligence - Volume Volume Three}, IJCAI'11, page 2764–2770. AAAI Press, 2011.

\bibitem{Wu_2024_ssm}
J.~Wu, X.~Wang, X.~Gao, J.~Chen, H.~Fu, and T.~Qiu.
\newblock On the effectiveness of sampled softmax loss for item recommendation.
\newblock {\em ACM Trans. Inf. Syst.}, 42(4), mar 2024.

\bibitem{xu_2024}
C.~Xu, Z.~Zhu, J.~Wang, J.~Wang, and W.~Zhang.
\newblock Understanding the role of cross-entropy loss in fairly evaluating large language model-based recommendation, 2024.

\bibitem{6844862}
D.~Yang, D.~Zhang, V.~W. Zheng, and Z.~Yu.
\newblock Modeling user activity preference by leveraging user spatial temporal characteristics in lbsns.
\newblock {\em IEEE Transactions on Systems, Man, and Cybernetics: Systems}, 45(1):129--142, 2015.

\bibitem{yang2024pls}
W.~Yang, J.~Chen, X.~Xin, S.~Zhou, B.~Hu, Y.~Feng, C.~Chen, and C.~Wang.
\newblock Psl: Rethinking and improving softmax loss from pairwise perspective for recommendation.
\newblock In {\em The Thirty-eighth Annual Conference on Neural Information Processing Systems}, 2024.

\bibitem{Yuan_2016}
F.~Yuan, G.~Guo, J.~M. Jose, L.~Chen, H.~Yu, and W.~Zhang.
\newblock Lambdafm: Learning optimal ranking with factorization machines using lambda surrogates.
\newblock In {\em Proceedings of the 25th ACM International on Conference on Information and Knowledge Management}, CIKM '16, page 227–236, New York, NY, USA, 2016. Association for Computing Machinery.

\bibitem{Zhang2013_SIGIR}
W.~Zhang, T.~Chen, J.~Wang, and Y.~Yu.
\newblock Optimizing top-n collaborative filtering via dynamic negative item sampling.
\newblock In {\em Proceedings of the 36th International ACM SIGIR Conference on Research and Development in Information Retrieval}, SIGIR '13, page 785–788, New York, NY, USA, 2013. Association for Computing Machinery.

\bibitem{zhao2023embedding}
Xiangyu Zhao, Maolin Wang, Xinjian Zhao, Jiansheng Li, Shucheng Zhou, Dawei Yin, Qing Li, Jiliang Tang, and Ruocheng Guo.
\newblock Embedding in recommender systems: A survey.
\newblock {\em arXiv preprint arXiv:2310.18608}, 2023.

\bibitem{Zhao2023_RecSys}
Y.~Zhao, R.~Chen, R.~Lai, Q.~Han, H.~Song, and L.~Chen.
\newblock Augmented negative sampling for collaborative filtering.
\newblock In {\em Proceedings of the 17th ACM Conference on Recommender Systems}, RecSys '23, page 256–266, New York, NY, USA, 2023. Association for Computing Machinery.

\bibitem{Zhou_2020}
K.~Zhou, H.~Wang, W.~X. Zhao, Y.~Zhu, S.~Wang, F.~Zhang, Z.~Wang, and J.~Wen.
\newblock S3-rec: Self-supervised learning for sequential recommendation with mutual information maximization.
\newblock In {\em Proceedings of the 29th ACM International Conference on Information \& Knowledge Management}, CIKM '20, page 1893–1902, New York, NY, USA, 2020. Association for Computing Machinery.

\bibitem{Zhu2022_ACMW}
Q.~Zhu, H.~Zhang, Q.~He, and Z.~Dou.
\newblock A gain-tuning dynamic negative sampler for recommendation.
\newblock In {\em Proceedings of the ACM Web Conference 2022}, WWW '22, page 277–285, New York, NY, USA, 2022. Association for Computing Machinery.

\end{thebibliography}

\onecolumn
\appendix
\section*{Supplementary material}

\section{Related work (cont.)} \label{sec:related_work_cont}
RSs aim to capture the compatibility between users and items by leveraging historical user interactions, such as purchases, likes and clicks. Feedback from users can be explicit, like ratings, or implicit, like clicks. Interpreting implicit feedback and understanding how to model unobserved interactions present challenges. To handle this, point-wise \citep{hu2008cf} and pair-wise \cite{Steffen2009_BPR} methods have been proposed. Collaborative filtering (CF) \citep{hu2008cf} methods, including user-based and item-based techniques, have long been popular for their straightforward implementation and ability to effectively capture patterns in user-item interactions. More sophisticated matrix factorization methods, such as Singular Value Decomposition (SVD) \citep{koren2009} and Non-negative Matrix Factorization (NMF) \citep{lee1999}, have advanced CF by generating compact, latent representations that allow for more precise modeling of user and item preferences.\\
Beyond CF, neural network-based recommender systems, particularly those leveraging deep learning, have gained prominence for their ability to capture complex, non-linear relationships within data. 
Sequential RSs focuses on understanding dynamic user interests through sequences of item interactions.
Early approaches such as GRU4Rec \citep{hidasi_2016_gru4rec} and Neural Attentive Session-based Recommendation (NARM)\citep{Li_2017} used Recurrent Neural Networks (RNNs) for sequence modeling. 
Concurrently, Neural Collaborative Filtering (NCF) \citep{he2017NCF} and Caser \citep{Tang2018} were introduced.
More recently, the Transformer architecture \citep{Vaswani2017_NIPS} has gained popularity in RSs due to its parallel processing capabilities and superior performance. For instance, SASRec \citep{kang2018_ICDM} utilizes a unidirectional self-attention mechanism, while BERT4Rec \citep{Fei2019_IKM} and Transformers4Rec \citep{desouza_2021} employ bidirectional self-attention for sequential recommendation tasks. $S^{3}$-Rec \citep{Zhou_2020} advances beyond masking techniques by pre-training on four self-supervised tasks to enhance data representation. Graph-based approaches, such as LightGCN \citep{He2020LightGCN}, have also been effective in modeling higher-order relationships in recommendation data, using graph convolutions to capture structural dependencies between users and items.
Lately TIGER \citep{Rajput_2023} has been proposed that directly predicts the Semantic ID of the next item using Generative Retrieval, bypassing the need for Approximate Nearest Neighbor search in a Maximum Inner Product Search (MIPS) space. \\
It is important to note that next-token prediction for Large Language Model (LLM)-based RSs pre-training and fine-tuning employs a cross-entropy loss with a Full Softmax over the entire corpus. However, due to the vast size of item catalogues in many real-world applications, using this loss becomes computationally prohibitive.

\section{Global minimum} \label{sec:global_minimum_appendix}
\begin{proposition} \label{proposition_BPR_CCE2}
            $\ell_{BPR}$ = $\ell_{CCE}$ if $K=1$
         \end{proposition}
        \begin{proof}
            If $K=1$, then we can lose the summation on both losses, getting:
            \begin{align*}
                &\ell_{BPR} = log\left(1+e^{\left(s_i-s_+\right)}\right)
                &\ell_{CCE} = log\left(1+e^{\left(s_i-s_+\right)}\right)
            \end{align*}
        \end{proof}

Proposition 2 in the main paper appears in the appendix as \cref{proposition_BPR_min2,proposition_BCE_min,proposition_CCE_min}.
\begin{proposition} \label{proposition_BPR_min2}
            Let us consider the $\ell_{BPR}$ and let's bound the scores $-S\leq s_+,s_i \leq S$. We have that:
            \begin{align*}
                &\argmin_{s_+} \ell_{BPR} = S
                &\argmin_{s_i} \ell_{BPR} = -S
            \end{align*}
        \end{proposition}
        \begin{proof}
            We have
            \begin{align*}
            \argmin_{s_+} \ell_{BPR} = \argmin_{s_+} \sum_{i\in\mathcal{I}^{-,K}}log\left(1+e^{\left(s_i-s_+\right)}\right) = \argmin_{s_+}-s_+ = \max_{s_+}s_+ = S
            \end{align*}
            \begin{align*}
            \argmin_{s_i} \ell_{BPR} = \argmin_{s_i} \sum_{i\in\mathcal{I}^{-,K}}log\left(1+e^{\left(s_i-s_+\right)}\right) = \argmin_{s_i}s_i = \min_{s_i}s_i = -S
            \end{align*}
        \end{proof}

\begin{proposition} \label{proposition_BCE_min}
            Let us consider the $\ell_{BCE}$ and let's bound the scores $-S\leq s_+,s_i \leq S$. We have that:
            \begin{align*}
                &\argmin_{s_+} \ell_{BCE} = S
                &\argmin_{s_i} \ell_{BCE} = -S
            \end{align*}
        \end{proposition}
        \begin{proof}
            We have
            \begin{align*}
            \argmin_{s_+} \ell_{BCE} = \argmin_{s_+} log\left(1+e^{-s_+}\right) + \sum_{i\in\mathcal{I}^{-,K}} log\left(1+e^{s_i}\right) = \argmin_{s_+}-s_+ = \max_{s_+}s_+ = S
            \end{align*}
            \begin{align*}
            \argmin_{s_i} \ell_{BCE} = \argmin_{s_i} log\left(1+e^{-s_+}\right) = \argmin_{s_i}s_i = \min_{s_i}s_i = -S
            \end{align*}
        \end{proof}

\begin{proposition} \label{proposition_CCE_min}
            Let us consider the $\ell_{CCE}$ and let's bound the scores $-S\leq s_+,s_i \leq S$. We have that:
            \begin{align*}
                &\argmin_{s_+} \ell_{CCE} = S
                &\argmin_{s_i} \ell_{CCE} = -S
            \end{align*}
        \end{proposition}
        \begin{proof}
            We have
            \begin{align*}
            \argmin_{s_+} \ell_{CCE} = \argmin_{s_+} log\left(1+\sum_{i\in\mathcal{I}^{-,K}}e^{\left(s_i-s_+\right)}\right) = \argmin_{s_+}-s_+ = \max_{s_+}s_+ = S
            \end{align*}
            \begin{align*}
            \argmin_{s_i} \ell_{CCE} = \argmin_{s_i} log\left(1+\sum_{i\in\mathcal{I}^{-,K}}e^{\left(s_i-s_+\right)}\right) = \argmin_{s_i}s_i = \min_{s_i}s_i = -S
            \end{align*}
        \end{proof}

\section{Loss bounds} \label{sec:loss_bounds_appendix}
Lemma 1 in the main paper appears in the appendix as \cref{lemma_BPR,lemma_BCE,lemma_CCE}.
\begin{lemma} \label{lemma_BPR}
            Given $\Gamma^K$, for BPR loss we have
            \begin{equation*}
                \ell_{BPR} \geq  |\Gamma^K| \log 2.
            \end{equation*}
        \end{lemma}
        
        \begin{proof}
            According to the definition of BPR, it follows that
            \begin{align*}
                \ell_{BPR} &= \sum_{i\in \mathcal{I^{-,K}}}\log \left(1+ e^{(s_{i} - s_+)}\right) \\
                & = \sum_{i\in \Gamma^K} \log \left(1+ e^{(s_{i} - s_+)} \right) + \sum_{i \notin \Gamma^K} \log \left(  1+ e^{(s_{i} - s_+)} \right) \\
                & \geq \sum_{i\in \Gamma^K} \log \left( 1+ e^{(s_{i} - s_+)} \right) \\
                & \geq |\Gamma^K|\log 2
            \end{align*}
        \end{proof}
        
        \begin{lemma} \label{lemma_BCE}
            Given $\Gamma^K_0$, for BCE loss we have:
            \begin{equation*}
                \ell_{BCE} \geq |\Gamma^K_0|\log2.
            \end{equation*}
        \end{lemma} 
        
        \begin{proof}
            According to the definition of BCE, it follows that
            \begin{align*} 
                \ell_{BCE} & = log\left(1+e^{-s_+}\right) + \sum_{i\in\mathcal{I}^{-,K}} log\left(1+e^{s_i}\right) \\
                &=  \log \left(1 + e^{-s_+}\right) +\sum_{i \in \Gamma^K_0} \log \left(1 + e^{s_{i}}\right) +\sum_{i \notin \Gamma^K_0}  \log \left(1 + e^{s_{i}}\right) \\
                &\geq \sum_{i \in \Gamma^K_0} \log \left(1 + e^{s_{i}}\right) \\
                & \geq |\Gamma^K_0|\log2
            \end{align*}
        \end{proof}

        \begin{lemma} \label{lemma_CCE}
            Given $\Gamma^K$, for CCE loss we have
            \begin{equation*}
                \ell_{CCE} \geq  \log(|\Gamma^K|).
            \end{equation*}
        \end{lemma}
        
        \begin{proof}
            According to the definition of CCE, it follows that
            \begin{align*}
                \ell_{CCE} & = log\left(1+\sum_{i\in\mathcal{I}^{-,K}}e^{\left(s_i-s_+\right)}\right)\\
                & = \log\left(1+\sum_{i \in \Gamma^K} e^{(s_i-s_+)}+\sum_{i \notin \Gamma^K} e^{(s_i-s_+)}\right)\\
                & \geq \log\left(\sum_{i \in \Gamma^K} e^{(s_i-s_+)}\right)\\
                & \geq \log\left(|\Gamma^K|\right)
            \end{align*}
        \end{proof}

\section{Ranking capability} \label{sec:ranking_capability_appendix}
Lemma 2 in the main paper appears in the appendix as \cref{lemma_prob2}.
\begin{lemma} \label{lemma_prob2} 
                Let consider the sets $\Gamma^K$ and $\Gamma^K_0$. The $K$ negative items are \textit{uniformly} sampled from the set $I^-$ of all negative items. Let $\Gamma =\{i \in I^-: s_i > s_+\}$ be the set of all negative items meeting the condition of the set $\Gamma^K$, with the cardinality $|\Gamma| = r_+-1$. Let $\Gamma_0 =\{i \in I^-: s_i > s_+\}$ be the set of all items meeting the condition of the set $\Gamma^K_0$, with the cardinality $|\Gamma_0|$.
                If the negative items are drawn without replacement, the two discrete random variables are described by two hypergeometrics:
                $$|\Gamma^K |\sim Hypergeometric\left(|I^-|,|\Gamma|,K\right) \;\;\; |\Gamma^K_0| \sim Hypergeometric\left(|I^-|,|\Gamma_0|,K\right)$$
            \end{lemma}
            
            \begin{proof}
                The hypergeometric distribution is a discrete probability distribution that describes the probability of k successes in n draws, without replacement, from a finite population of size N that contains exactly K objects with that feature, wherein each draw is either a success or a failure. In this context, the population consists of negative items, i.e. is the set $\mathcal{I^-}$. In both cases, items are divided into two sets: those meeting the condition of \(\Gamma^K\) or \(\Gamma^K_0\) and their respective complements. This division matches the hypergeometric distribution, where the population is split into two groups. We consider the probability of uniformly drawing a subset of $K$ negative items that meets the specific condition of the set \(\Gamma\) or \(\Gamma_0\) from the finite population, given by $\frac{|\Gamma|}{|\mathcal{I^-}|}$ or $\frac{|\Gamma_0|}{|\mathcal{I^-}|}$, respectively. 
            \end{proof}


            Theorem 2 in the main paper appears in the appendix as \cref{theorem_BPR,theorem_BCE,theorem_CCE}
            \begin{theorem} \label{theorem_BPR}
                Let consider the BPR loss when, for each user, we uniformly sample $K$ negative items for training. Let $\Gamma =\{i \in I^-: s_i \geq s_+\}$ be the set of all negative items that have a score greater or equal to the score of the positive item that is ranked $r_+$. Then we have
                \begin{equation}
                    \label{bound_BPR}
                    - log (NDCG(r_+))\leq \ell_{BPR}
                \end{equation}
                with probability at least 
                \begin{equation} \label{prob_BPR}
                    1 - \sum_{i=0}^{\log_2(\log_2(1+r_+))} \frac{\binom{|\Gamma|}{i} \binom{|\mathcal{I^-}|-|\Gamma|}{K-i}}{\binom{|\mathcal{I^-}|}{K}} 
                \end{equation}
            \end{theorem}
            
            \begin{proof} 
            According to \cref{lemma_BPR} we know that
            $$\ell_{BPR} \geq  |\Gamma^K| \log 2 $$
            Therefore, \cref{bound_BPR} can be rewritten as:
            \begin{align}  
             P(-\log \text{NDCG}(r_+) \leq \ell_{BPR}) &\geq P\left(-\log \left(\frac{1}{\log_2\left(1+r_+\right)}\right) \leq |\Gamma^K|\log2 \right)  \\ \notag
             =P\left(\log_2\left(1+r^+\right) \leq 2^{|\Gamma^K|}\right) & = P\left(|\Gamma^K| \geq \log_2 \left(\log_2 (1+r_+)\right)\right)\\ \label{eq:prob_BPR_}
            \end{align}
            According to Lemma 2 of the main paper, $|\Gamma^K|$ is an hypergeometric variable with population size $|\mathcal{I^-}|$, $|\Gamma^K|$ number of successes and $K$ number of draws. Thus, we can rewrite \cref{eq:prob_BPR_} using the cumulative distribution function of an hypergeometric: 
            \begin{align*}
                P\left(|\Gamma^K| \geq \log_2 \left(\log_2 (1+r_+)\right)\right) = 1- CDF_{|\Gamma^K|}\left(\log_2 \left(\log_2 (1+r_+)\right)\right)
            \end{align*}
            that is 
            \begin{align*}
                P\left(- log (NDCG(r_+))\leq \ell_{BPR}\right) \geq 1 - \sum_{i=0}^{\log_2(\log_2(1+r_+))} \frac{\binom{|\Gamma|}{i} \binom{|\mathcal{I^-}|-|\Gamma|}{K-i}}{\binom{|\mathcal{I^-}|}{K}}
            \end{align*}
            \end{proof}
            
            \begin{theorem} \label{theorem_BCE}
                Let consider the BCE loss when, for each user, we uniformly sample $K$ negative items for training. Let $\Gamma_0 =\{i \in I^-: s_i \geq 0\}$ be the set of all negative items that have a non-negative score. Then we have
                \begin{equation} \label{bound_BCE}
                    - log (NDCG(r_+))\leq \ell_{BCE}
                \end{equation}
                with probability at least 
                \begin{equation} \label{prob_BCE}
                    1 - \sum_{i=0}^{\log_2(\log_2(1+r_+))} \frac{\binom{|\Gamma_0|}{i} \binom{|\mathcal{I^-}|-|\Gamma^0|}{K-i}}{\binom{|\mathcal{I^-}|}{K}} 
                \end{equation}
            \end{theorem}
            
            \begin{proof} 
            According to \cref{lemma_BCE} we know that
            $$\ell_{BCE} \geq  |\Gamma^K_0| \log 2 $$
            Thus, the demonstration is equivalent to that of \cref{theorem_BCE}, using the set $\Gamma_0^K$ instead of $\Gamma^K$.
            \end{proof}

            \begin{theorem} \label{theorem_CCE}
                Let consider the CCE loss when, for each user, we uniformly sample $K$ negative items for training. Let $\Gamma =\{i \in I^-: s_i \geq s_+\}$ be the set of all negative items that have a score greater or equal to the score of the positive item that is ranked $r_+$. Then we have
                \begin{equation} \label{bound_CCE}
                    - log (NDCG(r_+))\leq \ell_{CCE}
                \end{equation}
                with probability at least 
                \begin{equation} \label{prob_CCE}
                    1 - \sum_{i=0}^{\log_2(1+r_+)} \frac{\binom{|\Gamma|}{i} \binom{|\mathcal{I^-}|-|\Gamma|}{K-i}}{\binom{|\mathcal{I^-}|}{K}} 
                \end{equation}
            \end{theorem}
            
            \begin{proof} 
            According to \cref{lemma_CCE} we know that
            $$\ell_{CCE} \geq   \log (|\Gamma^K|) $$
            Therefore, \cref{bound_CCE} can be rewritten as follows:
            \begin{align}  
             P(-\log \text{NDCG}(r_+) \leq \ell_{CCE}) &\geq P\left(-\log \left(\frac{1}{\log_2\left(1+r_+\right)}\right) \leq \log(|\Gamma^K|) \right)  \\ \notag
             = P\left(|\Gamma^K| \geq \log_2 (1+r_+)\right)\\ \label{eq:prob_CCE}
            \end{align}
            According to Lemma 2 of the main paper, the set $\Gamma^K$ is an hypergeometric variable with population size $|\mathcal{I^-}|$, $|\Gamma^K|$ number of successes and $K$ number of draws. Thus, we can rewrite \cref{eq:prob_BPR_} using the cumulative distribution function of an hypergeometric: 
            \begin{align*}
                P\left(|\Gamma^K| \geq \log_2 (1+r_+)\right) = 1- CDF_{|\Gamma^K|}\left(\log_2 (1+r_+)\right)
            \end{align*}
            that is 
            \begin{align*}
                P\left(- log (NDCG(r_+))\leq \ell_{CCE}\right) \geq \left(1 - \sum_{i=0}^{\log_2(1+r_+)} \frac{\binom{|\Gamma|}{i} \binom{|\mathcal{I^-}|-|\Gamma|}{K-i}}{\binom{|\mathcal{I^-}|}{K}}\right) 
            \end{align*}
            \end{proof}

        \section{Losses comparative analysis} \label{sec:losses_comparative_appendix}
        Theorem 3 in the main paper appears in the appendix as \cref{thr:loss_comparison2}.
            \begin{theorem} \label{thr:loss_comparison2}
                \looseness -1 When uniformly sampling $K$ negative items, in the worst-case scenario and $s_+ \geq 0$:
                \begin{align*}
                \mathbb{P}(-\log \text{NDCG}(r_+) \leq \ell_{BCE}) \geq
                \mathbb{P}(-\log \text{NDCG}(r_+) \leq \ell_{BPR}) \geq 
                \mathbb{P}(-\log \text{NDCG}(r_+) \leq \ell_{CCE})
                \end{align*}
            \end{theorem}

            We assume the inequalities for BCE, BPR, and CCE hold with equality, representing the worst-case scenario for the bound's probability. We also consider a non-negative score for the positive item, i.e. $s_+ \geq 0$, implying \(\Gamma \subseteq \Gamma_0\), meaning $|\Gamma| \leq |\Gamma_0|$. This is a plausible assumption as training progresses, as the loss incentivises increasing $s_+$.
            
            When comparing BPR and BCE in the worst-case scenario, we can show that BPR has a weaker bound on NDCG with a lower probability than BCE: \( P(-\log \text{NDCG}(r_+) \leq \ell_{BPR}) \leq P(-\log \text{NDCG}(r_+) \leq \ell_{BCE}) \). This holds true because both BPR and BCE bounds rely on hypergeometric distributions with the same population size and number of negative samples. The only difference is the number of successes in the population: \(|\Gamma|\) for BPR and \( |\Gamma_0|\) for BCE, with \(|\Gamma| \leq |\Gamma_0|\). This makes the probability of drawing a "success" in each sample for BCE greater than that of BPR. Thus, it follows that $ CDF_{|\Gamma^K|}\left(\log_2\left(\log_2 (1+r_+)\right)\right) \geq CDF_{|\Gamma_0^K|}\left(\log_2\left(\log_2 (1+r_+)\right)\right) $, which implies $ 1 - CDF_{|\Gamma^K|}\left(\log_2\left(\log_2 (1+r_+)\right)\right) \leq 1 - CDF_{|\Gamma_0^K|}\left(\log_2\left(\log_2 (1+r_+)\right)\right) $. This translates to a higher probability of achieving a lower bound on NDCG with BCE compared to BPR.

\begin{figure}[!ht]

                \begin{subfigure}{\textwidth}
                    \centering
                    \includegraphics[width=\textwidth]{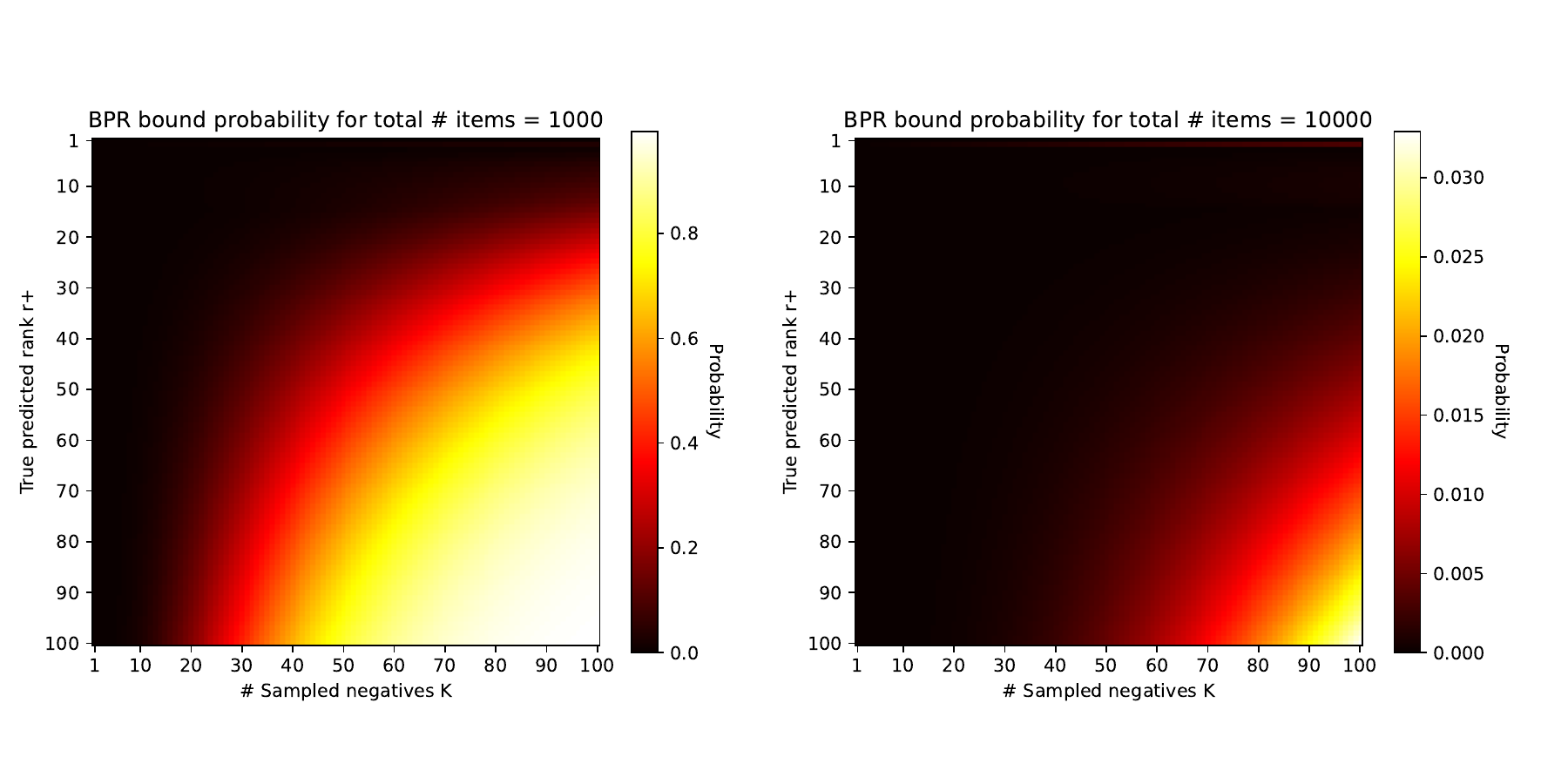}
                    \caption{BPR bound probabilities}
                \end{subfigure}
                \begin{subfigure}{\textwidth}
                    \centering
                    \includegraphics[width=\textwidth]{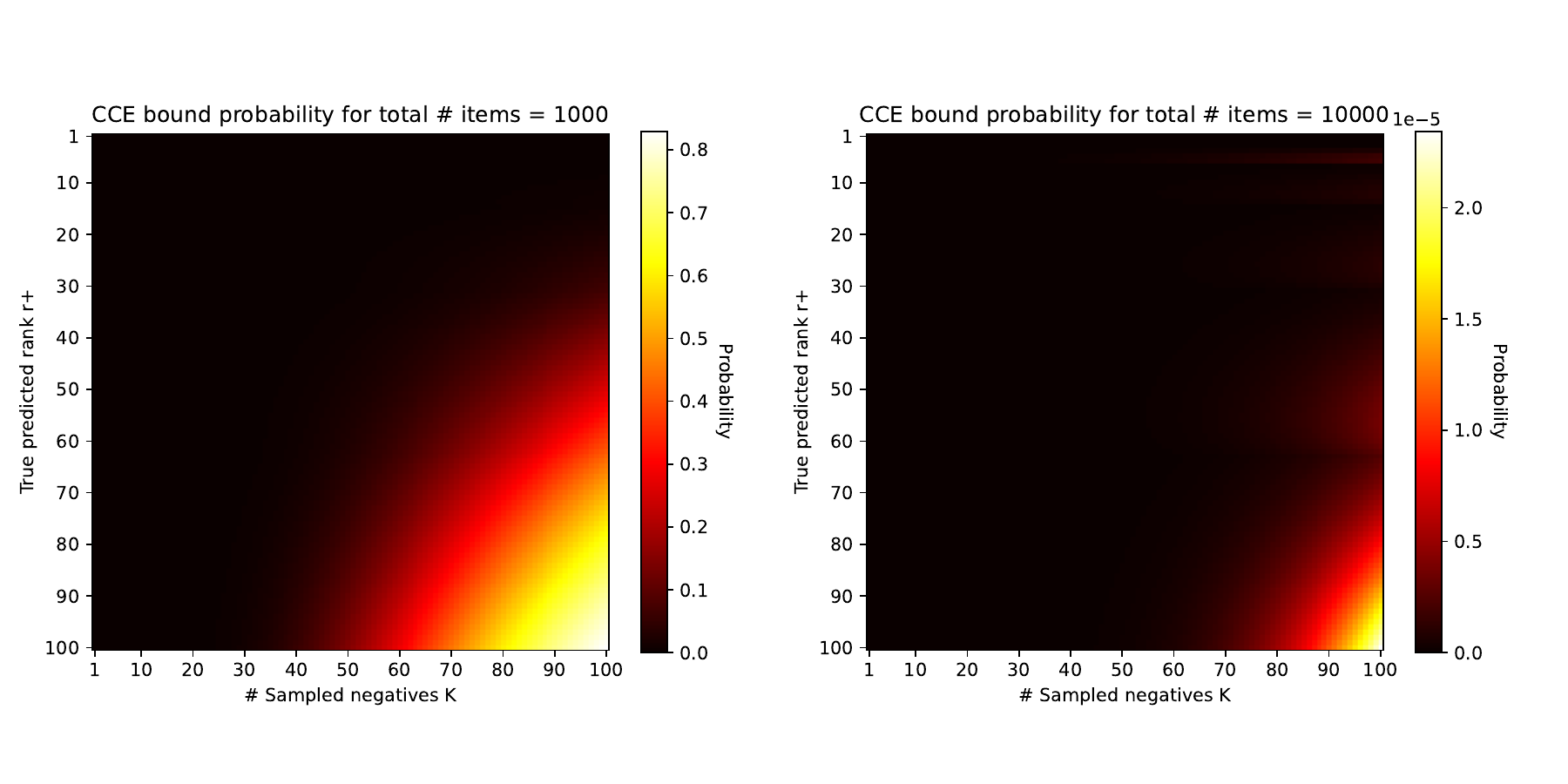}
                    \caption{CCE bound probabilities}
                \end{subfigure}
                \caption{BPR and CCE bound probabilities as a function of the number of sampled negatives $K$ and the predicted rank $r_+$.}
                \label{fig:bpr_cce_bound_prob}
            \end{figure}

            Comparing BPR and CCE involves different points for the hypergeometric distribution's CDF, while its parameters remain constant. It can be seen that point used for BPR's bound is always greater than or equal to the point used for CCE's bound: $1 - CDF_{|\Gamma^K|}\left(\log_2 (1+r_+)\right) \leq 1 - CDF_{|\Gamma^K|}\left(\log_2\left(\log_2 (1+r_+)\right)\right)$, since $\left(\log_2 (1+r_+)\right) \geq \left(\log_2\left(\log_2 (1+r_+)\right)\right)$. This implies that CCE could suffer of a weaker bound than BPR and consequently also than BCE. This can be easily seen in \cref{fig:bpr_cce_bound_prob} that shows Eqs. (7) and (8) of the main paper in the worst-case scenario as the number of negative items sampled \( K \) and the predicted rank \( r_+ \) vary. For both BPR and CCE, an increase in \( K \) generally increases the probability that the model will accurately rank a positive item, demonstrating the impact of larger negative sampling on the effectiveness of ranking. 
            The graphs clearly show that, given equal \( r_+ \) and \( K \), the relative probabilities for BPR are always higher or equal to those for CCE, consistent with the theoretical results.

\section{Reciprocal Rank}\label{sec:rr_appendix}
           The following theorems state that the $\ell_{BPR}$, $\ell_{BCE}$, and $\ell_{gBCE}$ losses are soft-proxy to the RR, i.e. minimizing these losses is equivalent to maximizing a lower bound of the RR.
            
            \begin{theorem} \label{theorem_RR}
                Let consider the BPR loss when, for each user, we uniformly sample $K$ negative items for training. Let $\Gamma =\{i \in I^-: s_i \geq s_+\}$ be the set of all negative items that have a score greater or equal to the score of the positive item that is ranked $r_+$. Then we have
                \begin{equation} \label{bound_RR}
                    - log (RR(r_+))\leq \ell_{BPR}
                \end{equation}
                with probability at least 
                \begin{equation} \label{prob_RR}
                    1 - \sum_{i=0}^{\log_2(r_+)} \frac{\binom{|\Gamma|}{i} \binom{|\mathcal{I^-}|-|\Gamma|}{K-i}}{\binom{|\mathcal{I^-}|}{K}} 
                \end{equation}
            \end{theorem}
            
            \begin{proof} 
            According to \cref{lemma_BPR} we know that
            $$\ell_{BPR} \geq  |\Gamma^K| \log 2 $$
            Therefore, \cref{bound_RR} can be rewritten as follows:
            \begin{align}  
             P(-\log \text{RR}(r_+) \leq \ell_{BPR}) &\geq P\left(-\log \left(\frac{1}{r_+}\right) \leq |\Gamma^K|\log2 \right)  \\ \notag
             =P\left(r^+ \leq 2^{|\Gamma^K|}\right) & = P\left(|\Gamma^K| \geq \log_2 (r_+)\right)\\ \label{eq:prob_BPR_RR}
            \end{align}
            According to Lemma 2 of the main paper, the $|\Gamma^K|$ is an hypergeometric variable with population size $|\mathcal{I^-}|$, $|\Gamma^K|$ number of successes and $K$ number of draws. Thus, we can rewrite \cref{eq:prob_BPR_RR} using the cumulative distribution function of an hypergeometric: 
            \begin{align*}
                P\left(|\Gamma^K| \geq \log_2 (r_+)\right) = 1- CDF_{|\Gamma^K|}\left(\log_2 (r_+)\right)
            \end{align*}
            that is 
            \begin{align*}
                P\left(- log (RR(r_+))\leq \ell_{BPR}\right) \geq \left(1 - \sum_{i=0}^{\log_2(r_+)} \frac{\binom{|\Gamma|}{i} \binom{|\mathcal{I^-}|-|\Gamma|}{K-i}}{\binom{|\mathcal{I^-}|}{K}}\right) 
            \end{align*}
            \end{proof}
            
            \begin{theorem} \label{theorem_BCE_RR}
                Let consider the BCE loss when, for each user, we uniformly sample $K$ negative items for training. Let $\Gamma_0 =\{i \in I^-: s_i \geq 0\}$ be the set of all negative items that have a non-negative score. Then we have
                \begin{equation} \label{bound_BCE_RR}
                    - log (RR(r_+))\leq \ell_{BCE}
                \end{equation}
                with probability at least 
                \begin{equation} \label{prob_BCE_RR}
                    1 - \sum_{i=0}^{\log_2(r_+)} \frac{\binom{|\Gamma_0|}{i} \binom{|\mathcal{I^-}|-|\Gamma^0|}{K-i}}{\binom{|\mathcal{I^-}|}{K}} 
                \end{equation}
            \end{theorem}
            
            \begin{proof} 
            According to \cref{lemma_BCE} we know that
            $$\ell_{BCE} \geq  |\Gamma^K_0| \log 2 $$
            Thus, the demonstration is equivalent to that of \cref{theorem_BCE}, using the set $\Gamma_0^K$ instead of $\Gamma^K$.
            \end{proof}
            
            \begin{theorem} \label{theorem_CCE_RR}
                Let consider the CCE loss when, for each user, we uniformly sample $K$ negative items for training. Let $\Gamma =\{i \in I^-: s_i \geq s_+\}$ be the set of all negative items that have a score greater or equal to the score of the positive item that is ranked $r_+$. Then we have
                \begin{equation} \label{bound_CCE_RR}
                    - log (RR(r_+))\leq \ell_{CCE}
                \end{equation}
                with probability at least 
                \begin{equation} \label{prob_CCE2}
                    1 - \sum_{i=0}^{r_+} \frac{\binom{|\Gamma|}{i} \binom{|\mathcal{I^-}|-|\Gamma|}{K-i}}{\binom{|\mathcal{I^-}|}{K}} 
                \end{equation}
            \end{theorem}
            
            \begin{proof} 
            According to \cref{lemma_CCE} we know that
            $$\ell_{BPR} \geq   \log (|\Gamma^K|) $$
            Therefore, \cref{bound_CCE} can be rewritten as follows:
            \begin{align}  
             P(-\log \text{RR}(r_+) \leq \ell_{CCE}) &\geq P\left(-\log \left(\frac{1}{r_+}\right) \leq \log(|\Gamma^K|) \right)  \\ \notag
             = P\left(|\Gamma^K| \geq r_+\right)\\ \label{eq:prob_CCE2}
            \end{align}
            According to Lemma 2 of the main paper, the set $\Gamma^K$ is an hypergeometric variable with population size $|\mathcal{I^-}|$, $|\Gamma^K|$ number of successes and $K$ number of draws. Thus, we can rewrite \cref{eq:prob_BPR} using the cumulative distribution function of an hypergeometric: 
            \begin{align*}
                P\left(|\Gamma^K| \geq r_+\right) = 1- CDF_{|\Gamma^K|}\left(r_+\right)
            \end{align*}
            that is 
            \begin{align*}
                P\left(- log (RR(r_+))\leq \ell_{CCE}\right) \geq \left(1 - \sum_{i=0}^{r_+} \frac{\binom{|\Gamma|}{i} \binom{|\mathcal{I^-}|-|\Gamma|}{K-i}}{\binom{|\mathcal{I^-}|}{K}}\right) 
            \end{align*}
            \end{proof}

\section{Results}\label{sec:results_appendix}

\textbf{Hardware}
All experiments were performed on a workstation equipped with an Intel Core i9-10940X (14-core CPU running at 3.3GHz) and 256GB of RAM, and a single Nvidia RTX A6000 with 48GB of VRAM.

\textbf{Additional datasets.} To ensure the generalizability of our results, we incorporated two larger real-world benchmark datasets for RSs:

    \begin{itemize}
        \item \textit{Amazon Books \citep{10.1145/2766462.2767755}:} This dataset contains user reviews and ratings specifically for books available on Amazon. It includes detailed information on user interactions, allowing for a nuanced analysis of preferences and behaviors across a wide selection of book titles. It encompasses reviews from 52,643 users on 91,599 items, totalling 2,984,108 interactions.

        \item \textit{Yelp (\url{http://www.yelp.com/dataset_challenge}):} This dataset consists of user reviews, ratings, and business information from Yelp, covering a diverse range of categories such as restaurants, nightlife, and shopping. It captures user interactions within the local business landscape, providing insights into user preferences in various service sectors.  It encompasses reviews from 31,668 users on 38,048 items, totalling 1,561,406 interactions.

    \end{itemize}

    \textbf{Additional baselines.} To ensure a comprehensive evaluation, we include two prominent, state-of-the-art non-sequential RSs:
    \begin{itemize}
        \item (\textit{i}) \textbf{NCF \citep{he2017NCF}} that utilizes a neural collaborative filtering approach to model user-item interactions;
        \item (\textit{ii}) \textbf{LightGCN \citep{He2020LightGCN}} that employs a simplified graph convolutional network designed specifically for recommendation tasks.
    \end{itemize}

\begin{table}[!ht]
\centering
\caption{NDCG@10 for all the configuration considered, training the models for 600 epochs. Best results for each model are highlighted in \textbf{bold}. $\dagger$ indicates a statistically significant result at level 0.01 using a Mann–Whitney–Wilcoxon test.}
\begin{tabular}{l|l||lll|lll}
\toprule
        &     & \multicolumn{3}{c|}{GRU4Rec}                                                 & \multicolumn{3}{c}{SASRec}                                                  \\
\midrule
Dataset & Negative items   & \multicolumn{1}{c}{BCE} & \multicolumn{1}{c}{BPR} & \multicolumn{1}{c|}{CCE} & \multicolumn{1}{c}{BCE} & \multicolumn{1}{c}{BPR} & \multicolumn{1}{c}{CCE} \\
\midrule\midrule
Beauty  & 1   & \textbf{0.6848}                  & 0.6836                  & 0.6788                  & 0.6304                  & 0.6723                  & \textbf{0.6758}$^\dagger$                  \\
Beauty  & 5   & 0.6806                  & \textbf{0.6843}$^\dagger$                  & 0.6784                  & 0.5719                  & 0.6753                  & \textbf{0.6783}                  \\
Beauty  & 10  & 0.6822                  & \textbf{0.6865}$^\dagger$                  & 0.6802                  & \textbf{0.6776}                  & 0.6753                  & 0.6734                  \\
Beauty  & 50  & 0.6834                  & \textbf{0.6848}                  & 0.6819                  & 0.5968                  & 0.6702                  & \textbf{0.6781}$^\dagger$                  \\
Beauty  & 100 & 0.6808                  & \textbf{0.6848}                  & 0.6835                  & 0.5070                  & 0.6745                  & \textbf{0.6819}$^\dagger$                  \\
\midrule
FS-NYC  & 1   & \textbf{0.6369}                  & \textbf{0.6369}                  & 0.6174                  & \textbf{0.6562}$^\dagger$                  & 0.6009                  & 0.6107                  \\
FS-NYC  & 5   & 0.6051                  & 0.6331                  & \textbf{0.6484}$^\dagger$                  & \textbf{0.6713}                  & 0.6242                  & 0.6676                  \\
FS-NYC  & 10  & 0.5887                  & 0.6249                  & \textbf{0.6476}$^\dagger$                  & 0.6600                  & 0.6288                  & \textbf{0.6782}$^\dagger$                  \\
FS-NYC  & 50  & 0.5713                  & 0.6321                  & \textbf{0.6513}$^\dagger$                  & 0.5973                  & 0.6502                  & \textbf{0.6994}$^\dagger$                  \\
FS-NYC  & 100 & 0.4969                  & 0.6332                  & \textbf{0.6751}$^\dagger$                  & 0.5470                  & 0.6519                  & \textbf{0.7150}$^\dagger$                  \\
\midrule
ML-1M   & 1   & \textbf{0.5627}                  & \textbf{0.5627}                  & 0.5611                  & \textbf{0.5704}                  & 0.5640                  & 0.5640                  \\
ML-1M   & 5   & 0.5661                  & \textbf{0.5692}                  & 0.5710                  & \textbf{0.5961}$^\dagger$                  & 0.5807                  & 0.5955                  \\
ML-1M   & 10  & \textbf{0.5808}$^\dagger$                  & 0.5669                  & 0.5694                  & \textbf{0.5992}$^\dagger$                  & 0.5790                  & 0.5966                  \\
ML-1M   & 50  & \textbf{0.5877}                  & 0.5628                  & 0.5834                  & 0.6010                   & 0.5807                  & \textbf{0.6060}                  \\
ML-1M   & 100 & \textbf{0.5909}                  & 0.5604                  & 0.5907                  & 0.5988                  & 0.5822                  & \textbf{0.6053}                  \\
\bottomrule
\end{tabular}
\end{table}

    \begin{figure}[!ht]
        \begin{subfigure}{0.33\textwidth}
            \centering
            \includegraphics[width=\textwidth]{img/epochs_negatives_NDCG10_SequentialBCEWithLogitsLoss_GRU_SAS_ml-1m_0.pdf}
            \caption{GRU4Rec}
        \end{subfigure}
        \begin{subfigure}{0.33\textwidth}
            \centering
            \includegraphics[width=\textwidth]{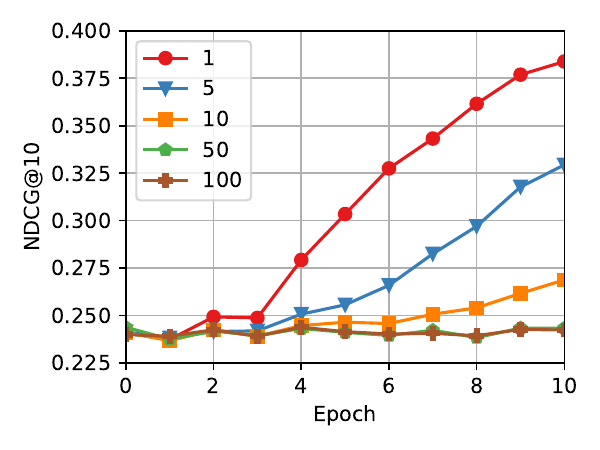}
            \caption{GRU4Rec - Start}
        \end{subfigure}
        \begin{subfigure}{0.33\textwidth}
            \centering
            \includegraphics[width=\textwidth]{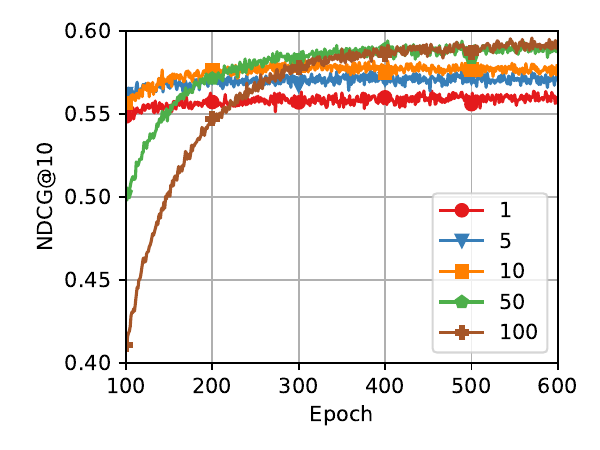}
            \caption{GRU4Rec - End}
        \end{subfigure}
        \begin{subfigure}{0.33\textwidth}
            \centering
            \includegraphics[width=\textwidth]{img/epochs_negatives_NDCG10_SequentialBCEWithLogitsLoss_GRU_SAS_ml-1m_1.pdf}
            \caption{SASRec}
        \end{subfigure}
        \begin{subfigure}{0.33\textwidth}
            \centering
            \includegraphics[width=\textwidth]{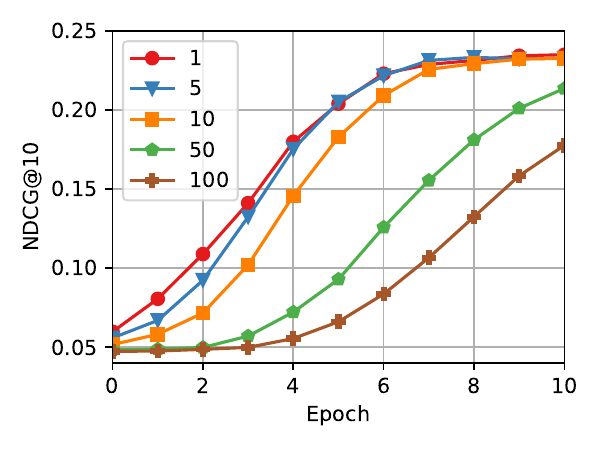}
            \caption{SASRec - Start}
        \end{subfigure}
        \begin{subfigure}{0.33\textwidth}
            \centering
            \includegraphics[width=\textwidth]{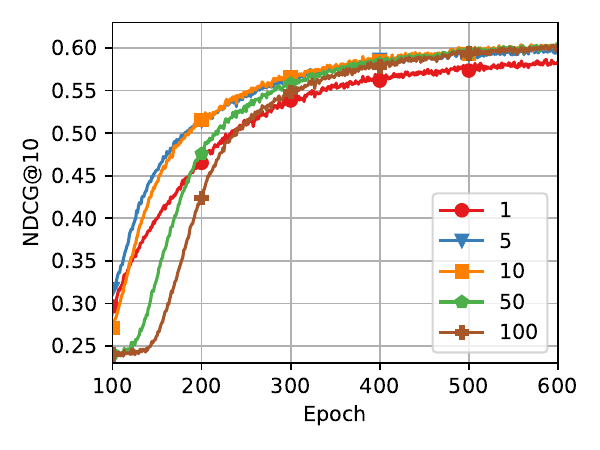}
            \caption{SASRec - End}
        \end{subfigure}
        \caption{SASRec and GRU4Rec NDCG@10 during training changing number of negative items, using BCE loss on ML-1M dataset.}
        \label{fig:neg_bce}
    \end{figure}

\begin{figure}[!ht]
    \begin{subfigure}{0.33\textwidth}
        \centering
        \includegraphics[width=\textwidth]{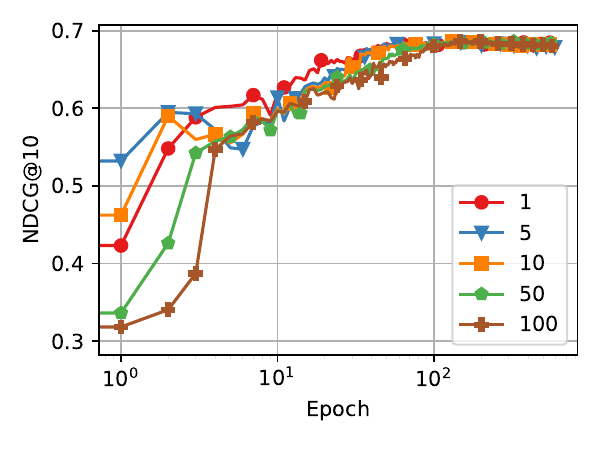}
        \caption{GRU4Rec}
    \end{subfigure}
    \begin{subfigure}{0.33\textwidth}
        \centering
        \includegraphics[width=\textwidth]{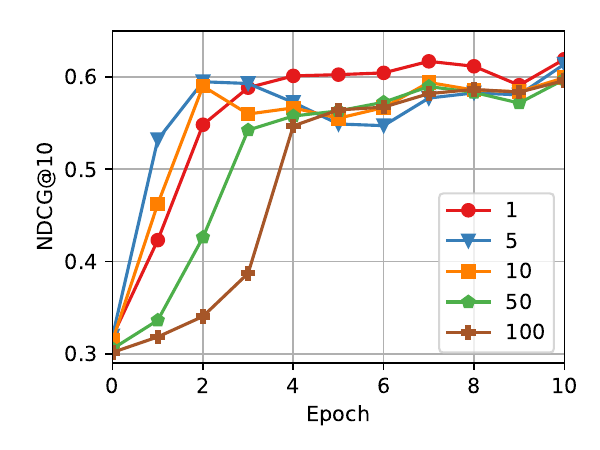}
        \caption{GRU4Rec - Start}
    \end{subfigure}
    \begin{subfigure}{0.33\textwidth}
        \centering
        \includegraphics[width=\textwidth]{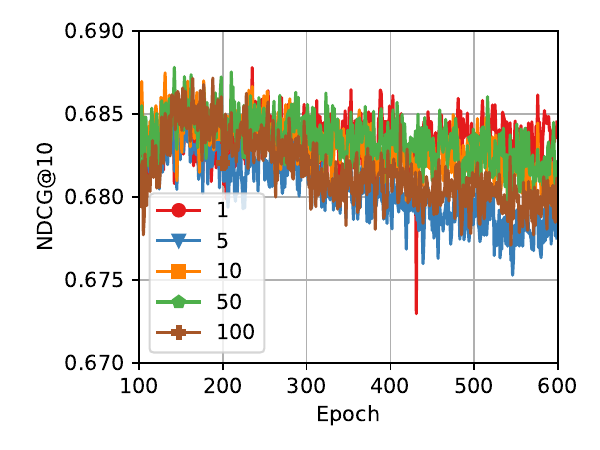}
        \caption{GRU4Rec - End}
    \end{subfigure}
    \begin{subfigure}{0.33\textwidth}
        \centering
        \includegraphics[width=\textwidth]{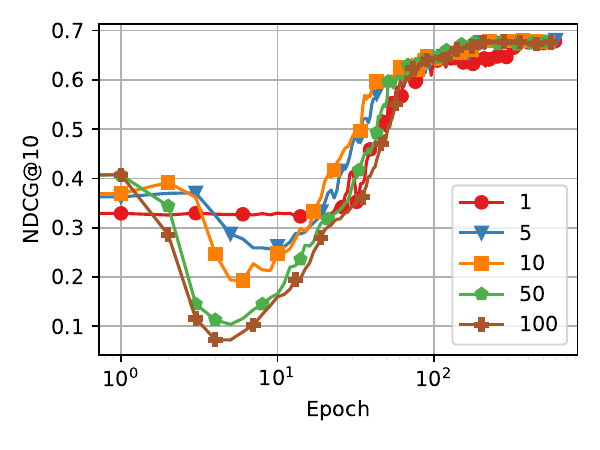}
        \caption{SASRec}
    \end{subfigure}
    \begin{subfigure}{0.33\textwidth}
        \centering
        \includegraphics[width=\textwidth]{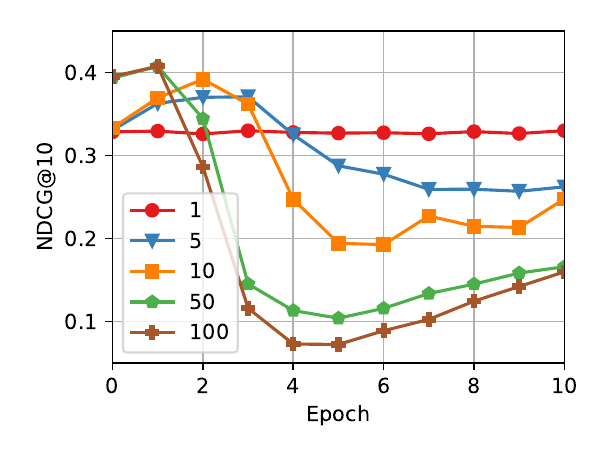}
        \caption{SASRec - Start}
    \end{subfigure}
    \begin{subfigure}{0.33\textwidth}
        \centering
        \includegraphics[width=\textwidth]{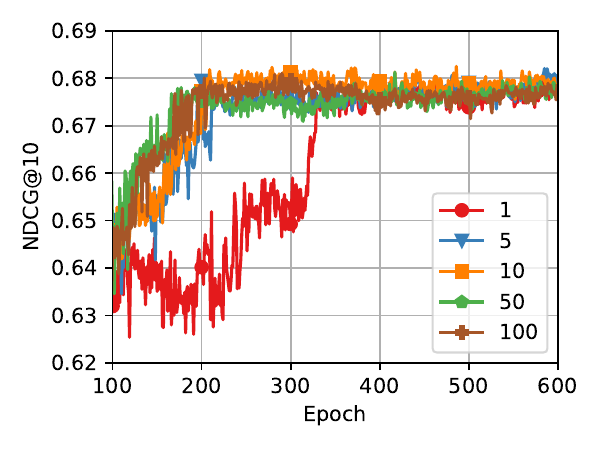}
        \caption{SASRec - End}
    \end{subfigure}
    \caption{SASRec and GRU4Rec NDCG@10 during training changing number of negative items, using BCE loss on Amazon Beauty dataset.}
\end{figure}

\begin{figure}[!ht]
    \begin{subfigure}{0.33\textwidth}
        \centering
        \includegraphics[width=\textwidth]{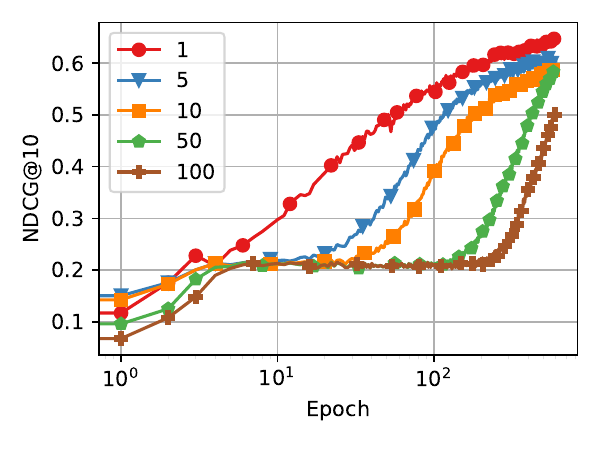}
        \caption{GRU4Rec}
    \end{subfigure}
    \begin{subfigure}{0.33\textwidth}
        \centering
        \includegraphics[width=\textwidth]{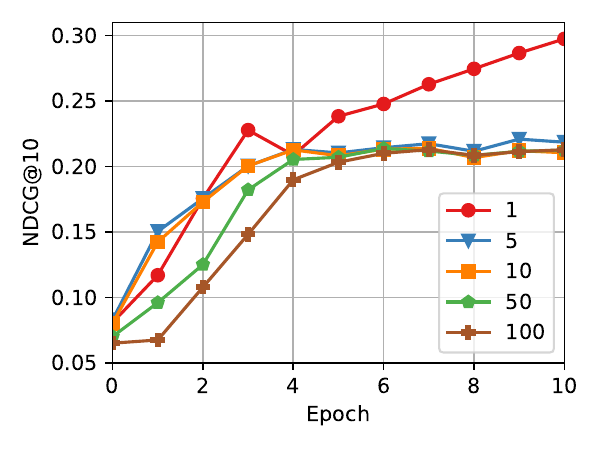}
        \caption{GRU4Rec - Start}
    \end{subfigure}
    \begin{subfigure}{0.33\textwidth}
        \centering
        \includegraphics[width=\textwidth]{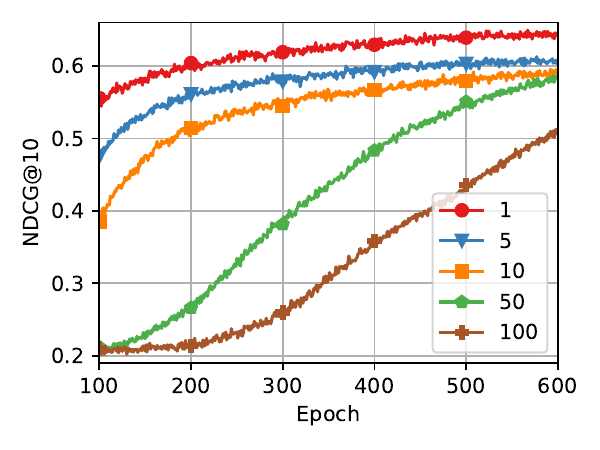}
        \caption{GRU4Rec - End}
    \end{subfigure}
    \begin{subfigure}{0.33\textwidth}
        \centering
        \includegraphics[width=\textwidth]{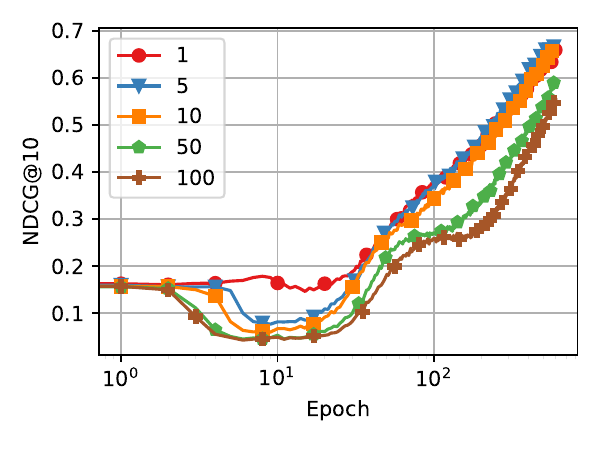}
        \caption{SASRec}
    \end{subfigure}
    \begin{subfigure}{0.33\textwidth}
        \centering
        \includegraphics[width=\textwidth]{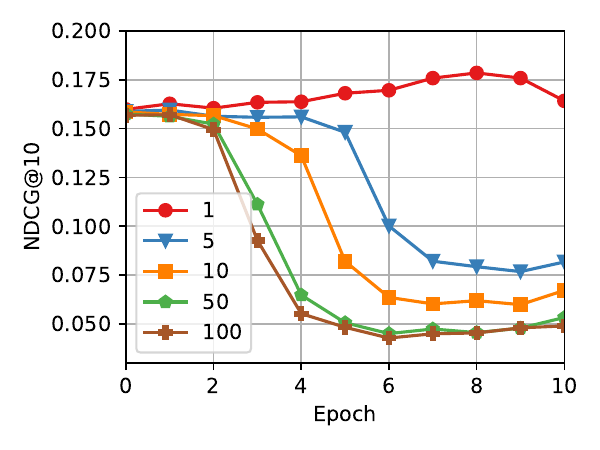}
        \caption{SASRec - Start}
    \end{subfigure}
    \begin{subfigure}{0.33\textwidth}
        \centering
        \includegraphics[width=\textwidth]{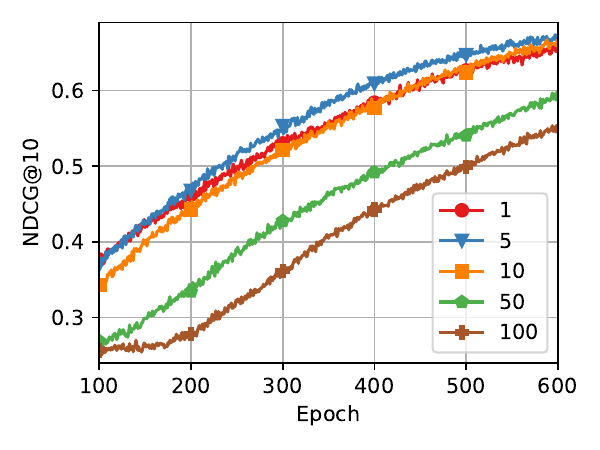}
        \caption{SASRec - End}
    \end{subfigure}
    \caption{SASRec and GRU4Rec NDCG@10 during training changing number of negative items, using BCE loss on Foursquare-NYC dataset.}
\end{figure}

    \begin{figure}[!ht]
        \begin{subfigure}{0.33\textwidth}
            \centering
            \includegraphics[width=\textwidth]{img/epochs_negatives_NDCG10_SequentialBPR_GRU_SAS_ml-1m_0.pdf}
            \caption{GRU4Rec}
        \end{subfigure}
        \begin{subfigure}{0.33\textwidth}
            \centering
            \includegraphics[width=\textwidth]{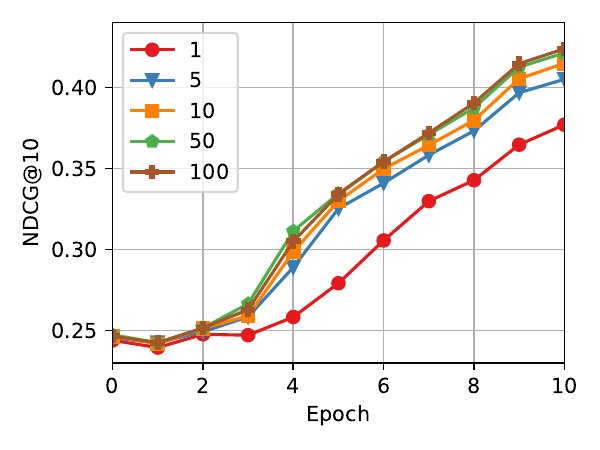}
            \caption{GRU4Rec - Start}
        \end{subfigure}
        \begin{subfigure}{0.33\textwidth}
            \centering
            \includegraphics[width=\textwidth]{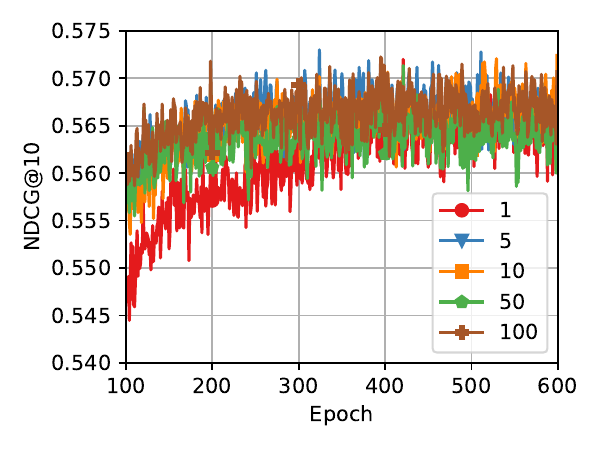}
            \caption{GRU4Rec - End}
        \end{subfigure}
        \begin{subfigure}{0.33\textwidth}
            \centering
            \includegraphics[width=\textwidth]{img/epochs_negatives_NDCG10_SequentialBPR_GRU_SAS_ml-1m_1.pdf}
            \caption{SASRec}
        \end{subfigure}
        \begin{subfigure}{0.33\textwidth}
            \centering
            \includegraphics[width=\textwidth]{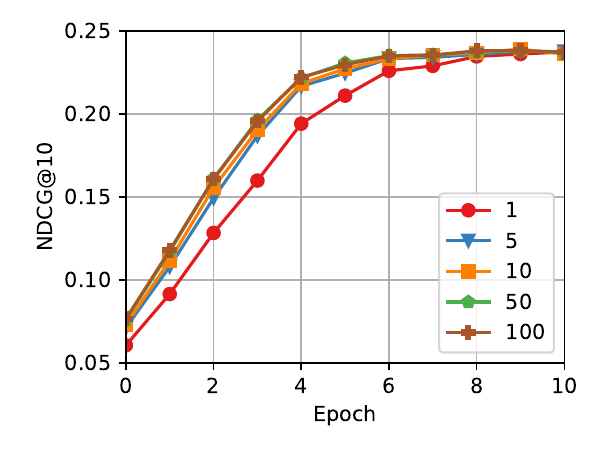}
            \caption{SASRec - Start}
        \end{subfigure}
        \begin{subfigure}{0.33\textwidth}
            \centering
            \includegraphics[width=\textwidth]{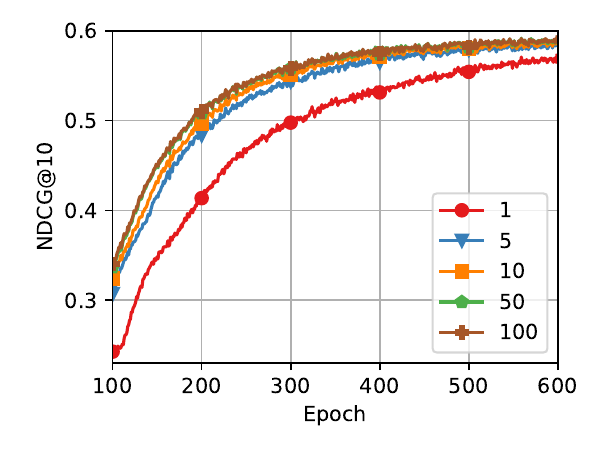}
            \caption{SASRec - End}
        \end{subfigure}
        \caption{SASRec and GRU4Rec NDCG@10 during training changing number of negative items, using BPR loss on ML-1M dataset.}
        \label{fig:neg_bpr}
    \end{figure}

\begin{figure}[!ht]
    \begin{subfigure}{0.33\textwidth}
        \centering
        \includegraphics[width=\textwidth]{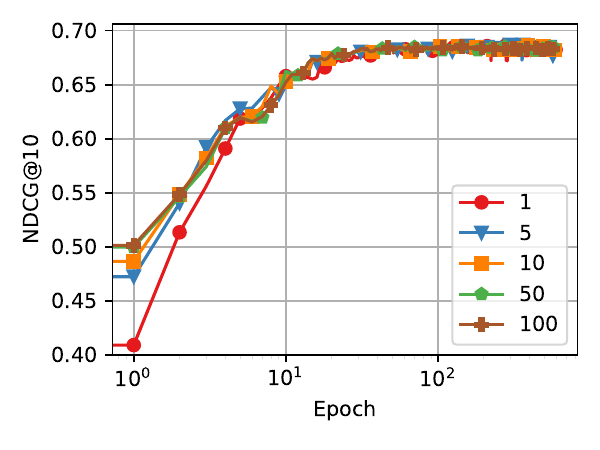}
        \caption{GRU4Rec}
    \end{subfigure}
    \begin{subfigure}{0.33\textwidth}
        \centering
        \includegraphics[width=\textwidth]{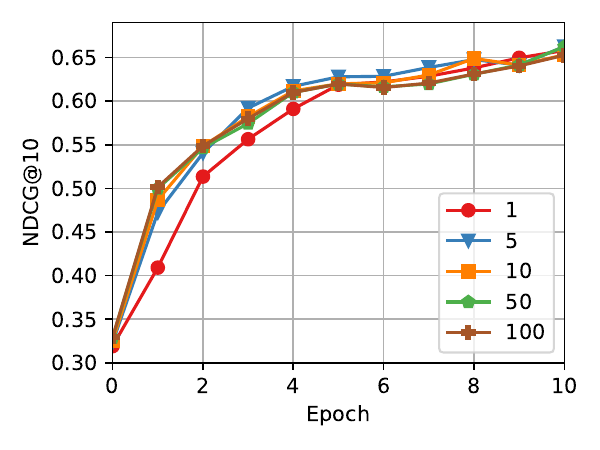}
        \caption{GRU4Rec - Start}
    \end{subfigure}
    \begin{subfigure}{0.33\textwidth}
        \centering
        \includegraphics[width=\textwidth]{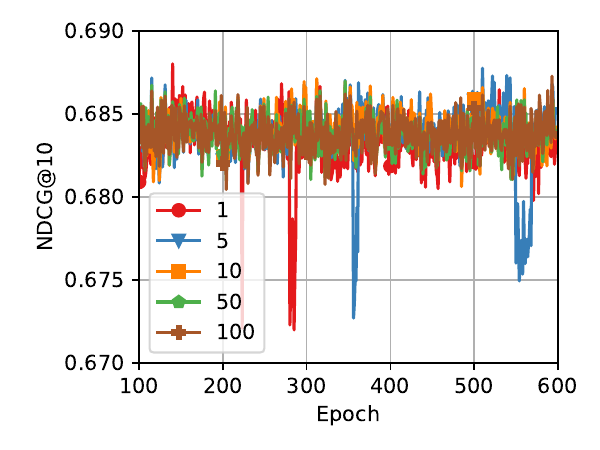}
        \caption{GRU4Rec - End}
    \end{subfigure}
    \begin{subfigure}{0.33\textwidth}
        \centering
        \includegraphics[width=\textwidth]{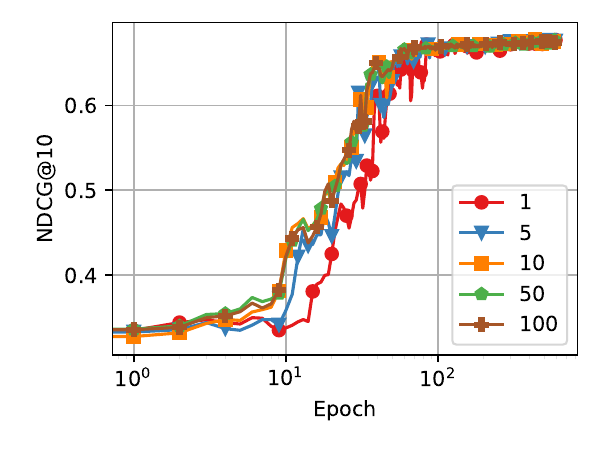}
        \caption{SASRec}
    \end{subfigure}
    \begin{subfigure}{0.33\textwidth}
        \centering
        \includegraphics[width=\textwidth]{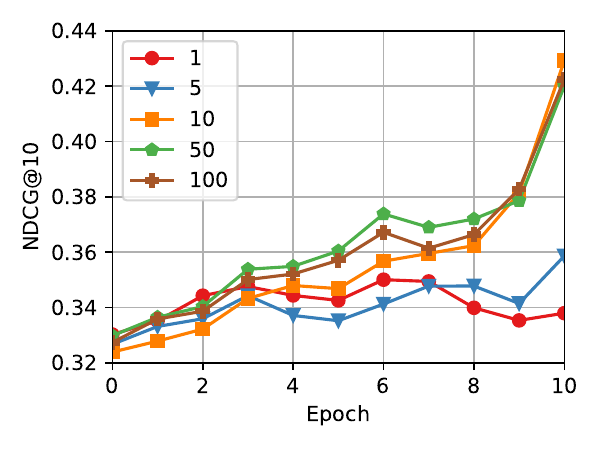}
        \caption{SASRec - Start}
    \end{subfigure}
    \begin{subfigure}{0.33\textwidth}
        \centering
        \includegraphics[width=\textwidth]{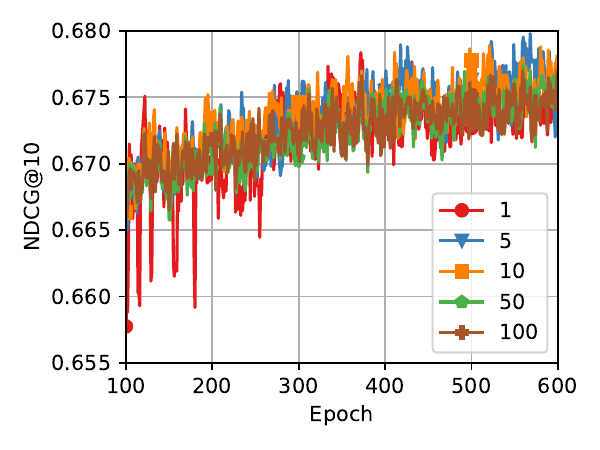}
        \caption{SASRec - End}
    \end{subfigure}
    \caption{SASRec and GRU4Rec NDCG@10 during training changing number of negative items, using BPR loss on Amazon Beauty dataset.}
\end{figure}

\begin{figure}[!ht]
    \begin{subfigure}{0.33\textwidth}
        \centering
        \includegraphics[width=\textwidth]{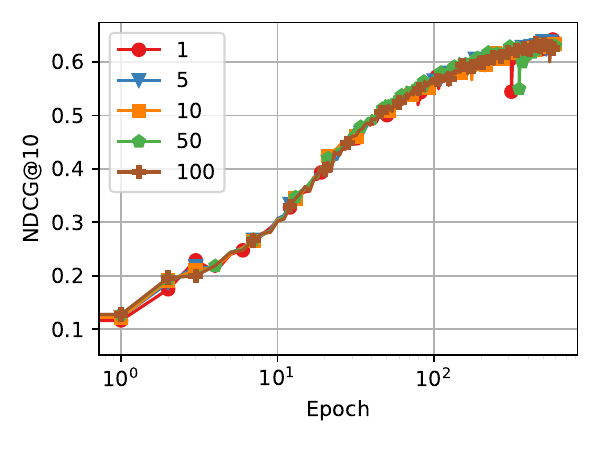}
        \caption{GRU4Rec}
    \end{subfigure}
    \begin{subfigure}{0.33\textwidth}
        \centering
        \includegraphics[width=\textwidth]{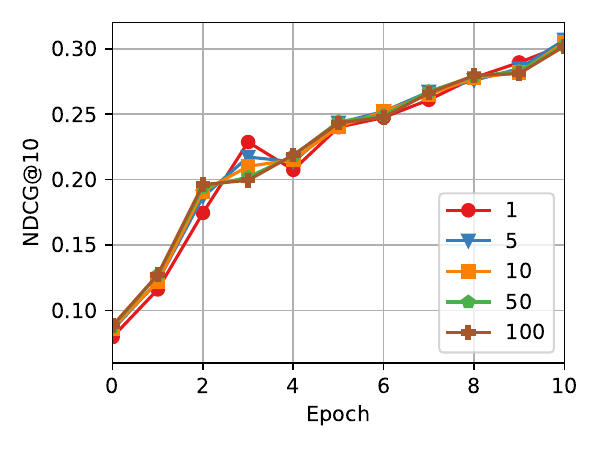}
        \caption{GRU4Rec - Start}
    \end{subfigure}
    \begin{subfigure}{0.33\textwidth}
        \centering
        \includegraphics[width=\textwidth]{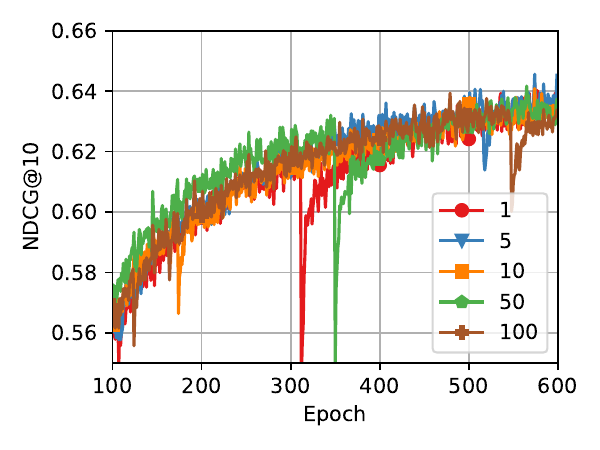}
        \caption{GRU4Rec - End}
    \end{subfigure}
    \begin{subfigure}{0.33\textwidth}
        \centering
        \includegraphics[width=\textwidth]{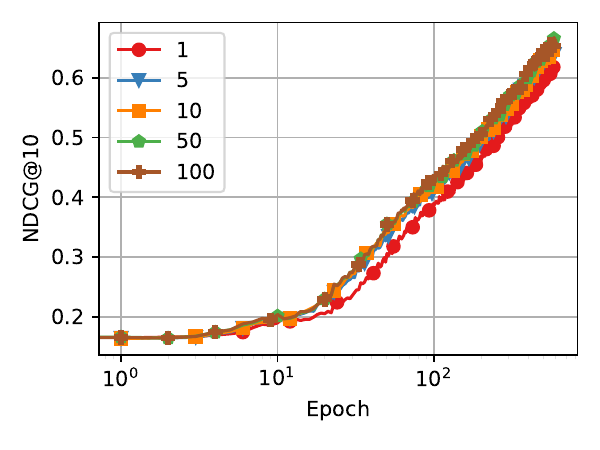}
        \caption{SASRec}
    \end{subfigure}
    \begin{subfigure}{0.33\textwidth}
        \centering
        \includegraphics[width=\textwidth]{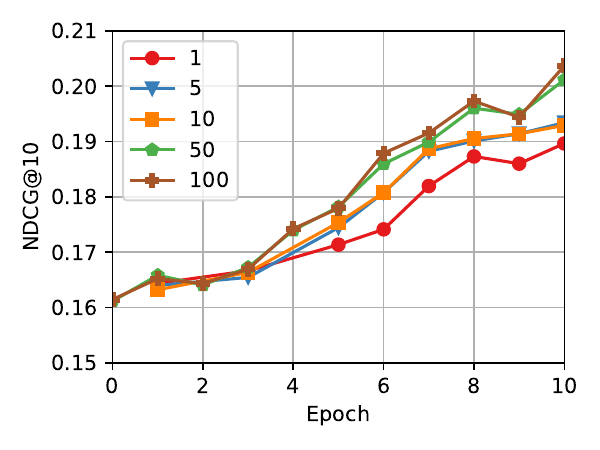}
        \caption{SASRec - Start}
    \end{subfigure}
    \begin{subfigure}{0.33\textwidth}
        \centering
        \includegraphics[width=\textwidth]{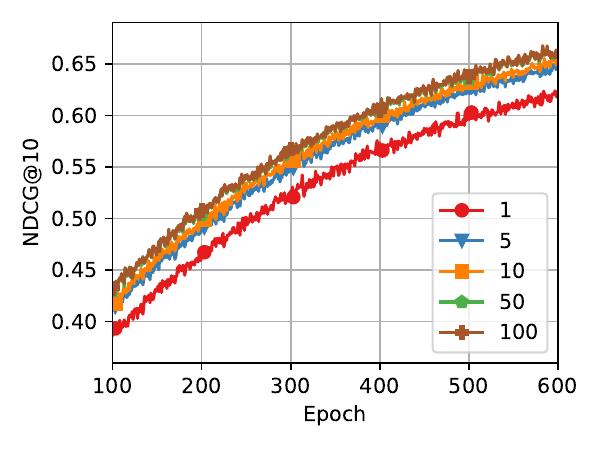}
        \caption{SASRec - End}
    \end{subfigure}
    \caption{SASRec and GRU4Rec NDCG@10 during training changing number of negative items, using BPR loss on Foursquare-NYC dataset.}
\end{figure}

    \begin{figure}[!ht]
        \begin{subfigure}{0.33\textwidth}
            \centering
            \includegraphics[width=\textwidth]{img/epochs_negatives_NDCG10_SequentialCrossEntropyLoss_GRU_SAS_ml-1m_0.pdf}
            \caption{GRU4Rec}
        \end{subfigure}
        \begin{subfigure}{0.33\textwidth}
            \centering
            \includegraphics[width=\textwidth]{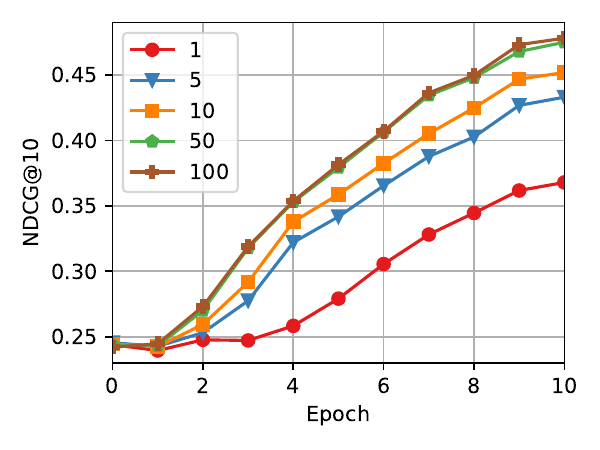}
            \caption{GRU4Rec - Start}
        \end{subfigure}
        \begin{subfigure}{0.33\textwidth}
            \centering
            \includegraphics[width=\textwidth]{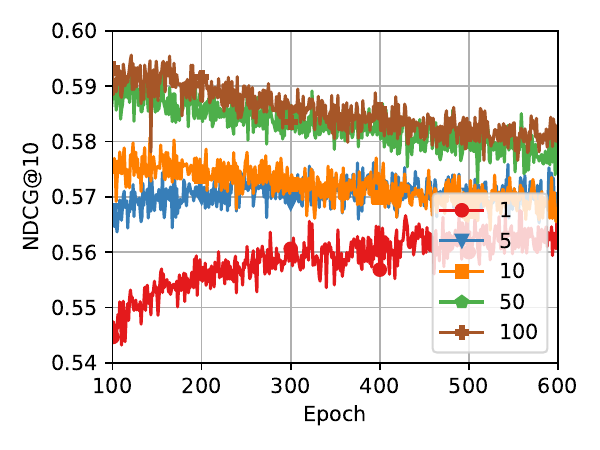}
            \caption{GRU4Rec - End}
        \end{subfigure}
        \begin{subfigure}{0.33\textwidth}
            \centering
            \includegraphics[width=\textwidth]{img/epochs_negatives_NDCG10_SequentialCrossEntropyLoss_GRU_SAS_ml-1m_1.pdf}
            \caption{SASRec}
        \end{subfigure}
        \begin{subfigure}{0.33\textwidth}
            \centering
            \includegraphics[width=\textwidth]{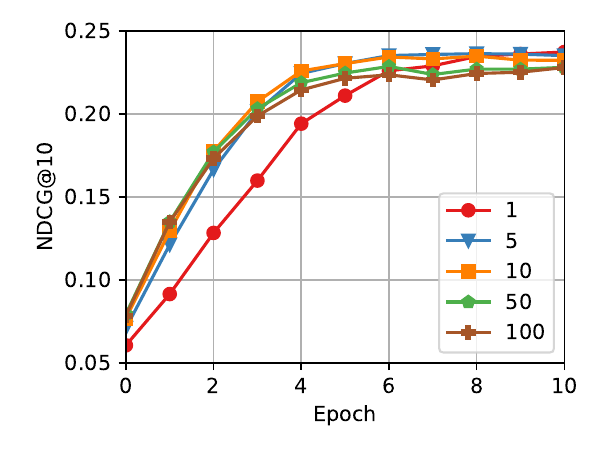}
            \caption{SASRec - Start}
        \end{subfigure}
        \begin{subfigure}{0.33\textwidth}
            \centering
            \includegraphics[width=\textwidth]{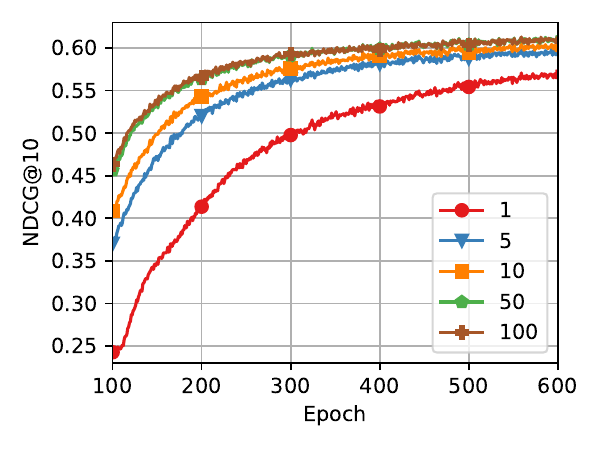}
            \caption{SASRec - End}
        \end{subfigure}
        \caption{SASRec and GRU4Rec NDCG@10 during training changing number of negative items, using CCE loss on ML-1M dataset.}
        \label{fig:neg_cce}
    \end{figure}
    
\begin{figure}[!ht]
    \begin{subfigure}{0.33\textwidth}
        \centering
        \includegraphics[width=\textwidth]{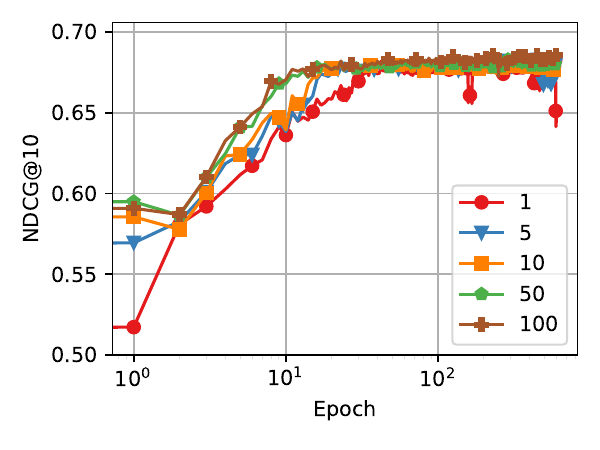}
        \caption{GRU4Rec}
    \end{subfigure}
    \begin{subfigure}{0.33\textwidth}
        \centering
        \includegraphics[width=\textwidth]{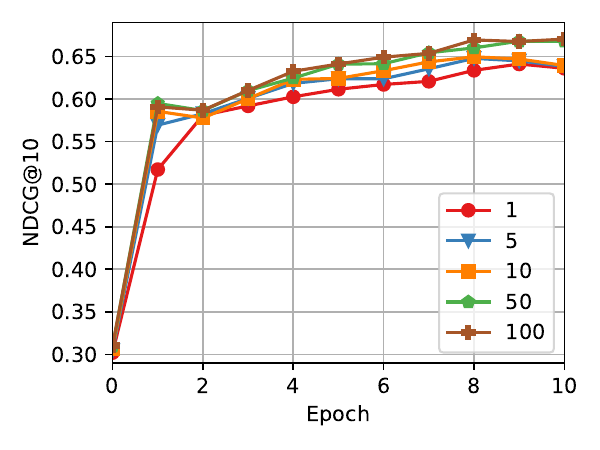}
        \caption{GRU4Rec - Start}
    \end{subfigure}
    \begin{subfigure}{0.33\textwidth}
        \centering
        \includegraphics[width=\textwidth]{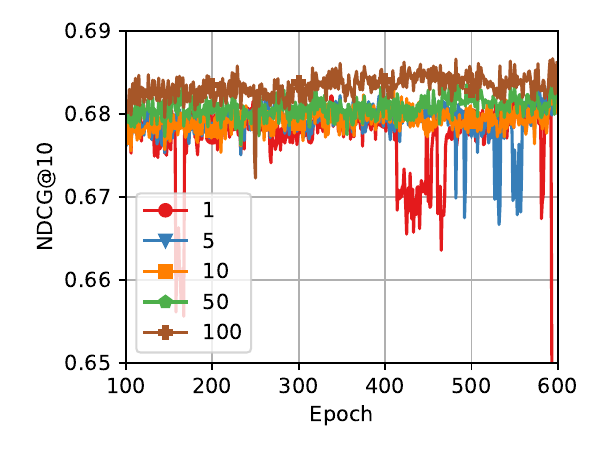}
        \caption{GRU4Rec - End}
    \end{subfigure}
    \begin{subfigure}{0.33\textwidth}
        \centering
        \includegraphics[width=\textwidth]{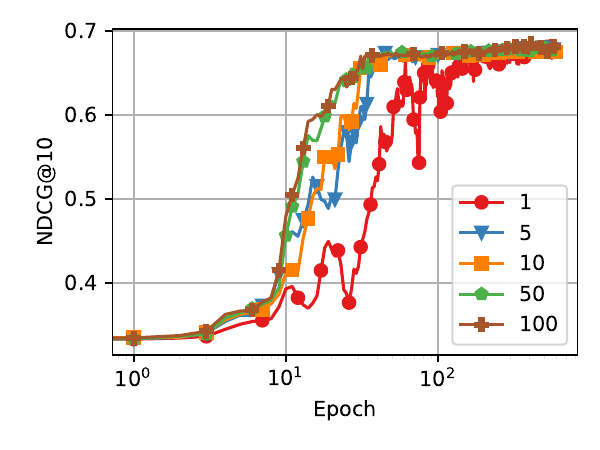}
        \caption{SASRec}
    \end{subfigure}
    \begin{subfigure}{0.33\textwidth}
        \centering
        \includegraphics[width=\textwidth]{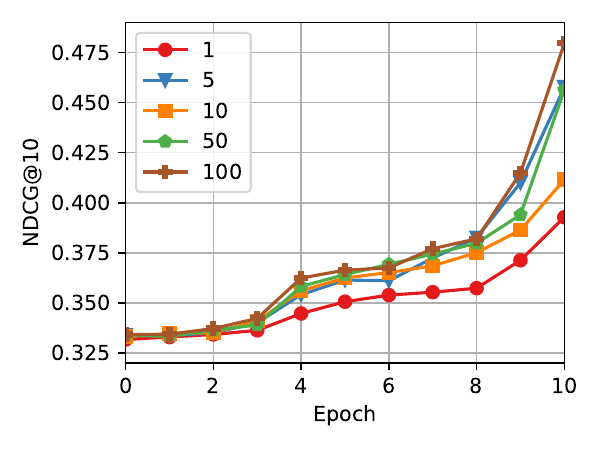}
        \caption{SASRec - Start}
    \end{subfigure}
    \begin{subfigure}{0.33\textwidth}
        \centering
        \includegraphics[width=\textwidth]{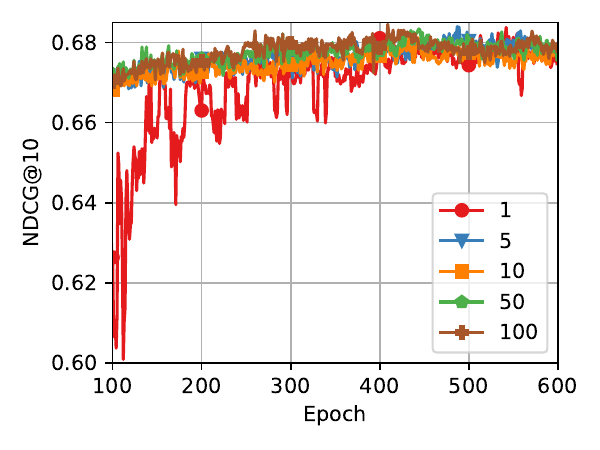}
        \caption{SASRec - End}
    \end{subfigure}
    \caption{SASRec and GRU4Rec NDCG@10 during training changing number of negative items, using CCE loss on Amazon Beauty dataset.}
\end{figure}

\begin{figure}[!ht]
    \begin{subfigure}{0.33\textwidth}
        \centering
        \includegraphics[width=\textwidth]{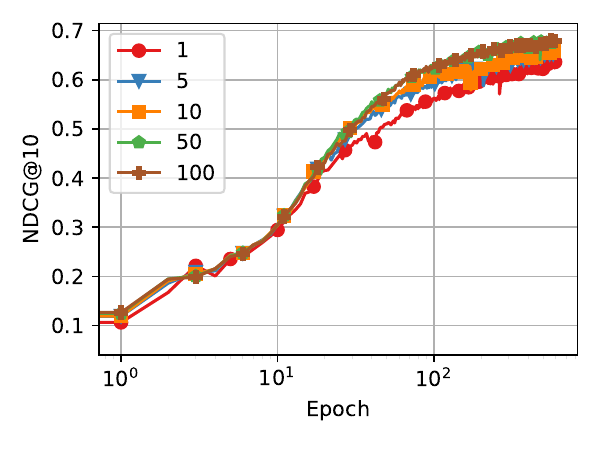}
        \caption{GRU4Rec}
    \end{subfigure}
    \begin{subfigure}{0.33\textwidth}
        \centering
        \includegraphics[width=\textwidth]{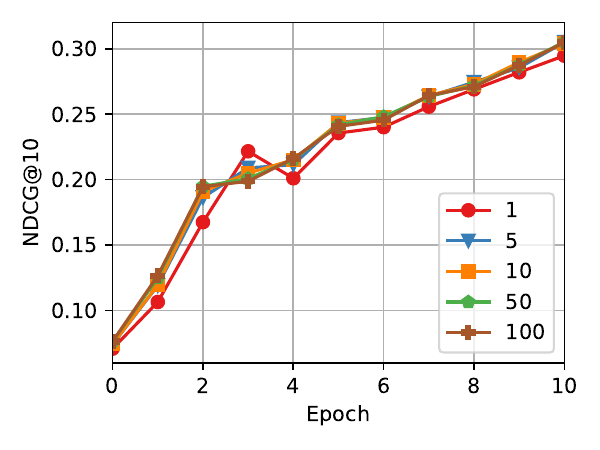}
        \caption{GRU4Rec - Start}
    \end{subfigure}
    \begin{subfigure}{0.33\textwidth}
        \centering
        \includegraphics[width=\textwidth]{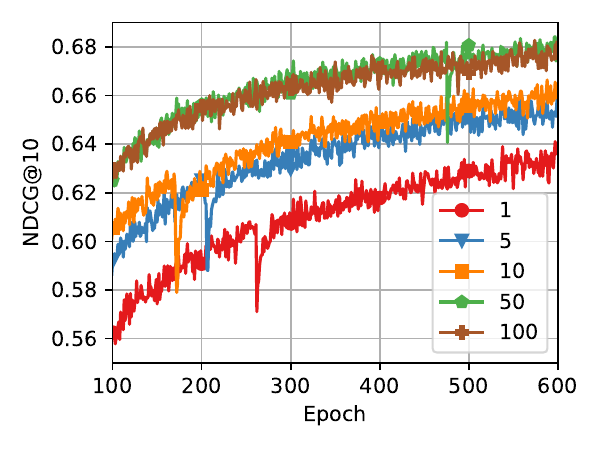}
        \caption{GRU4Rec - End}
    \end{subfigure}
    \begin{subfigure}{0.33\textwidth}
        \centering
        \includegraphics[width=\textwidth]{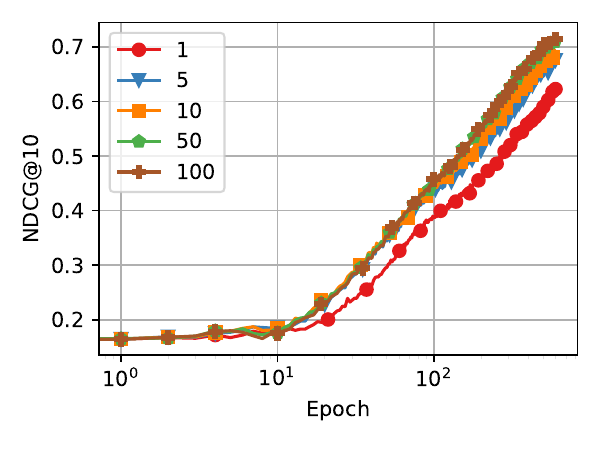}
        \caption{SASRec}
    \end{subfigure}
    \begin{subfigure}{0.33\textwidth}
        \centering
        \includegraphics[width=\textwidth]{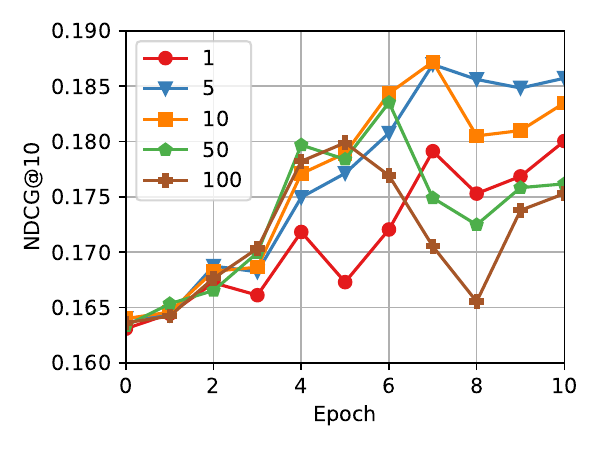}
        \caption{SASRec - Start}
    \end{subfigure}
    \begin{subfigure}{0.33\textwidth}
        \centering
        \includegraphics[width=\textwidth]{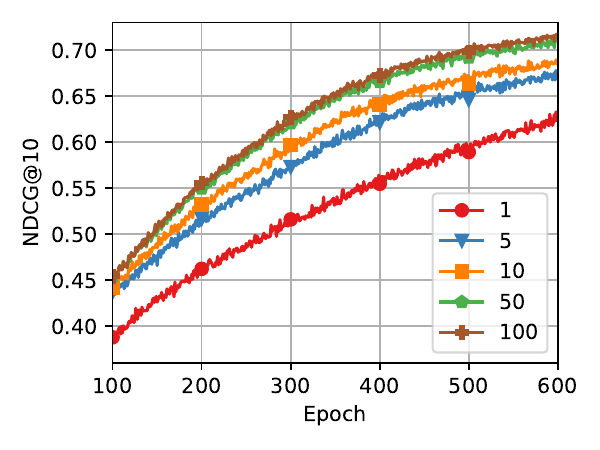}
        \caption{SASRec - End}
    \end{subfigure}
    \caption{SASRec and GRU4Rec NDCG@10 during training changing number of negative items, using CCE loss on Foursquare-NYC dataset.}
\end{figure}

    \begin{figure}[!ht]
        \begin{subfigure}{0.33\textwidth}
            \centering
            \includegraphics[width=\textwidth]{img/epochs_losses_NDCG10_1_GRU_SAS_ml-1m_0.pdf}
            \caption{GRU4Rec}
        \end{subfigure}
        \begin{subfigure}{0.33\textwidth}
            \centering
            \includegraphics[width=\textwidth]{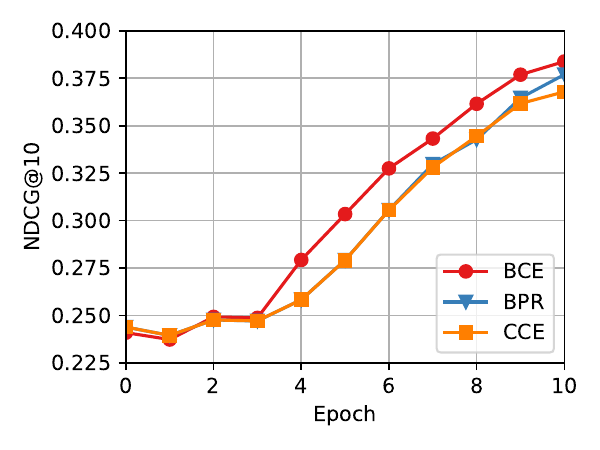}
            \caption{GRU4Rec - Start}
        \end{subfigure}
        \begin{subfigure}{0.33\textwidth}
            \centering
            \includegraphics[width=\textwidth]{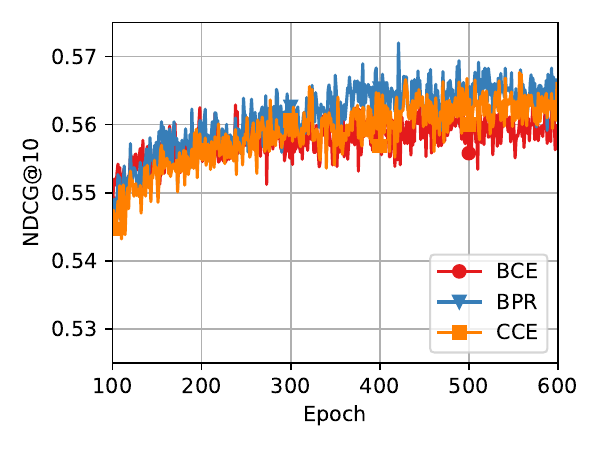}
            \caption{GRU4Rec - End}
        \end{subfigure}
        \begin{subfigure}{0.33\textwidth}
            \centering
            \includegraphics[width=\textwidth]{img/epochs_losses_NDCG10_1_GRU_SAS_ml-1m_1.pdf}
            \caption{SASRec}
        \end{subfigure}
        \begin{subfigure}{0.33\textwidth}
            \centering
            \includegraphics[width=\textwidth]{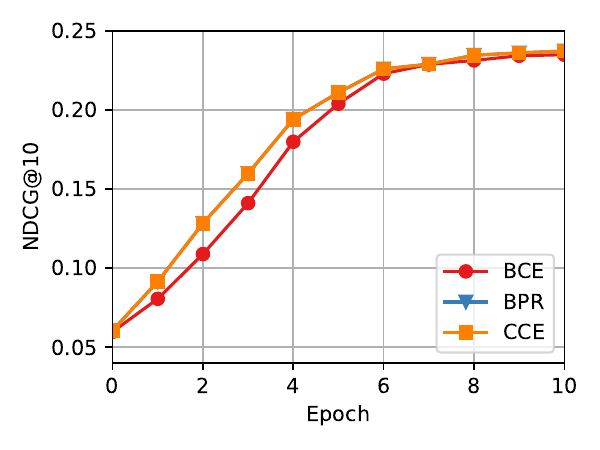}
            \caption{SASRec - Start}
        \end{subfigure}
        \begin{subfigure}{0.33\textwidth}
            \centering
            \includegraphics[width=\textwidth]{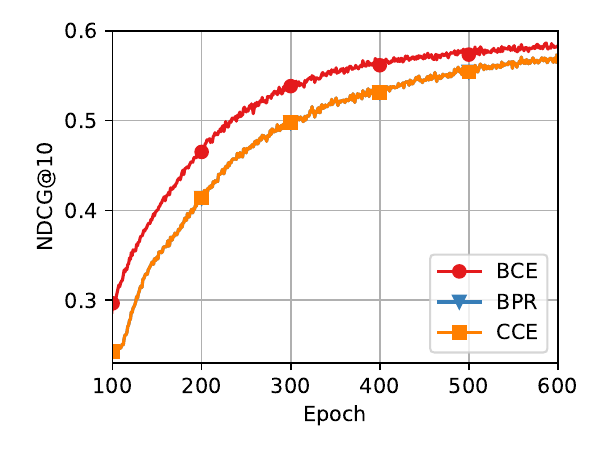}
            \caption{SASRec - End}
        \end{subfigure}
        \caption{SASRec and GRU4Rec NDCG@10 during training changing loss, using 1 negative item on ML-1M dataset.}
    \end{figure}
\begin{figure}[!ht]
    \begin{subfigure}{0.33\textwidth}
        \centering
        \includegraphics[width=\textwidth]{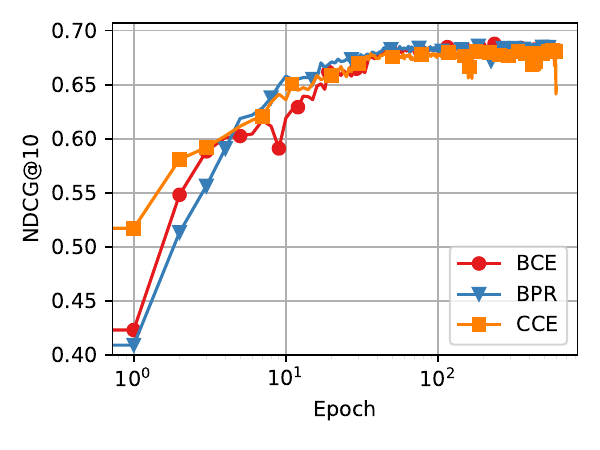}
        \caption{GRU4Rec}
    \end{subfigure}
    \begin{subfigure}{0.33\textwidth}
        \centering
        \includegraphics[width=\textwidth]{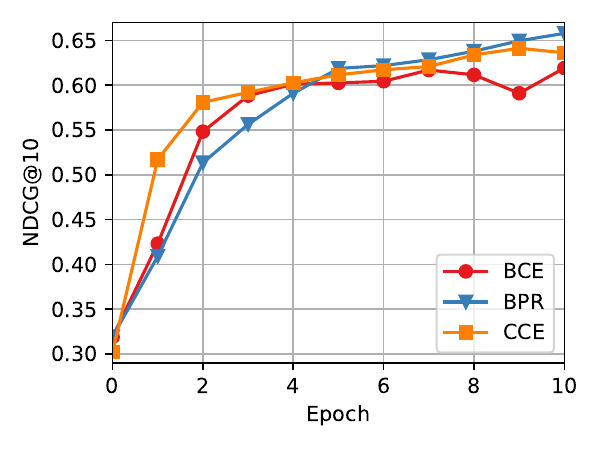}
        \caption{GRU4Rec - Start}
    \end{subfigure}
    \begin{subfigure}{0.33\textwidth}
        \centering
        \includegraphics[width=\textwidth]{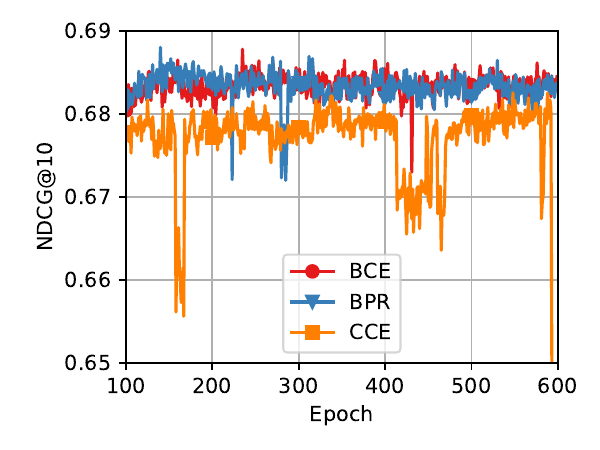}
        \caption{GRU4Rec - End}
    \end{subfigure}
    \begin{subfigure}{0.33\textwidth}
        \centering
        \includegraphics[width=\textwidth]{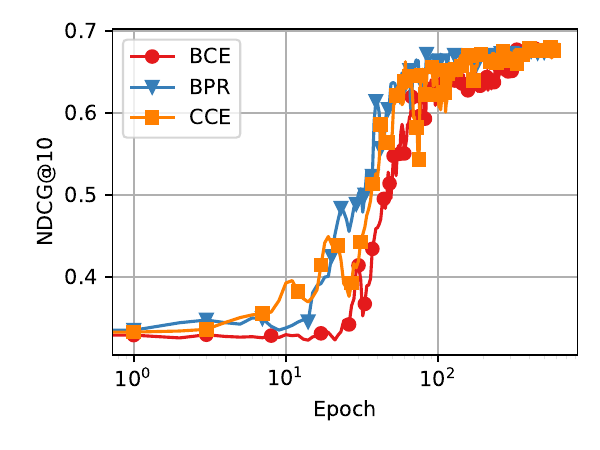}
        \caption{SASRec}
    \end{subfigure}
    \begin{subfigure}{0.33\textwidth}
        \centering
        \includegraphics[width=\textwidth]{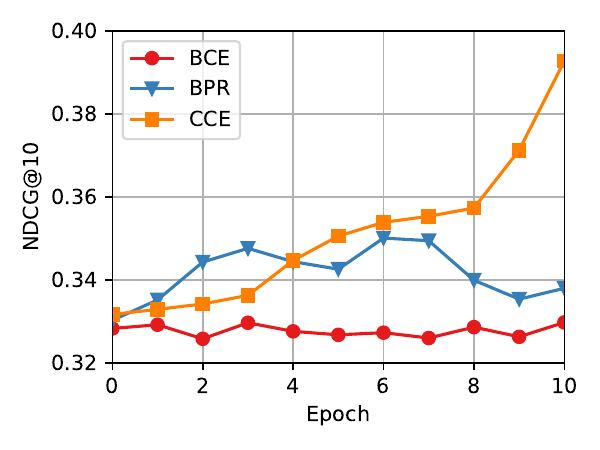}
        \caption{SASRec - Start}
    \end{subfigure}
    \begin{subfigure}{0.33\textwidth}
        \centering
        \includegraphics[width=\textwidth]{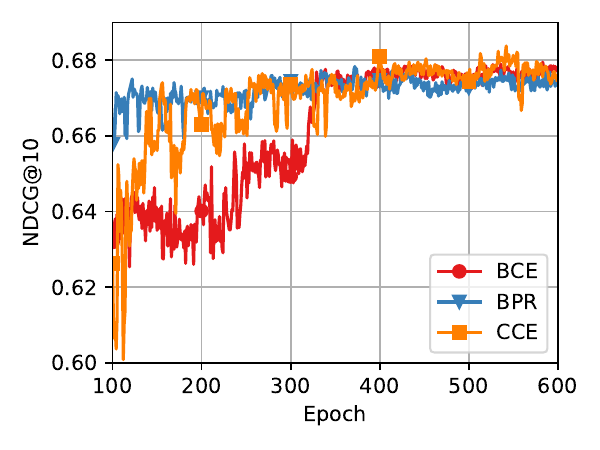}
        \caption{SASRec - End}
    \end{subfigure}
    \caption{SASRec and GRU4Rec NDCG@10 during training changing loss, using 1 negative item on Amazon Beauty dataset.}
\end{figure}

\begin{figure}[!ht]
    \begin{subfigure}{0.33\textwidth}
        \centering
        \includegraphics[width=\textwidth]{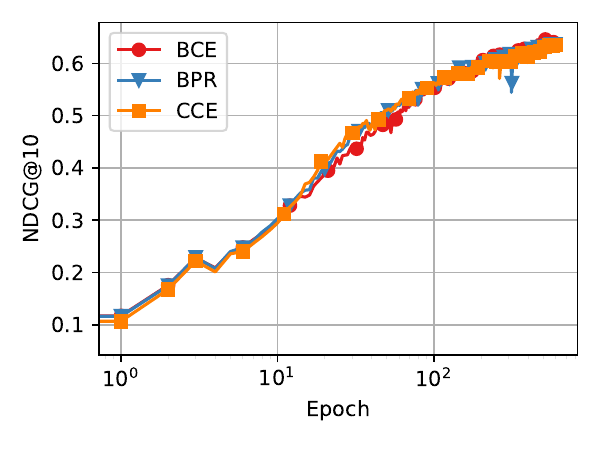}
        \caption{GRU4Rec}
    \end{subfigure}
    \begin{subfigure}{0.33\textwidth}
        \centering
        \includegraphics[width=\textwidth]{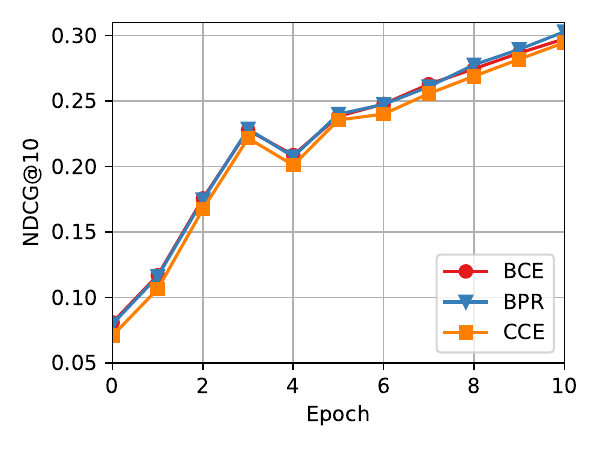}
        \caption{GRU4Rec - Start}
    \end{subfigure}
    \begin{subfigure}{0.33\textwidth}
        \centering
        \includegraphics[width=\textwidth]{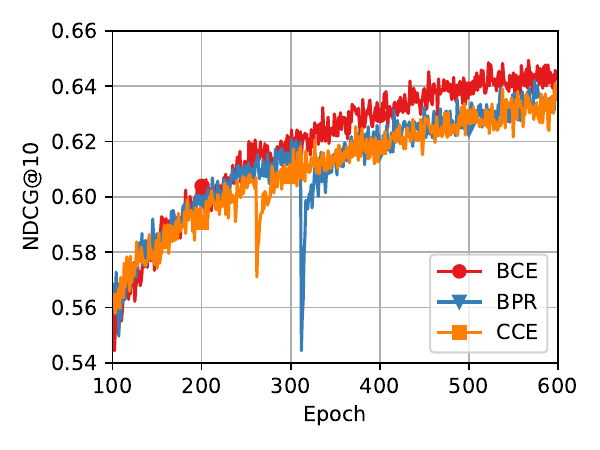}
        \caption{GRU4Rec - End}
    \end{subfigure}
    \begin{subfigure}{0.33\textwidth}
        \centering
        \includegraphics[width=\textwidth]{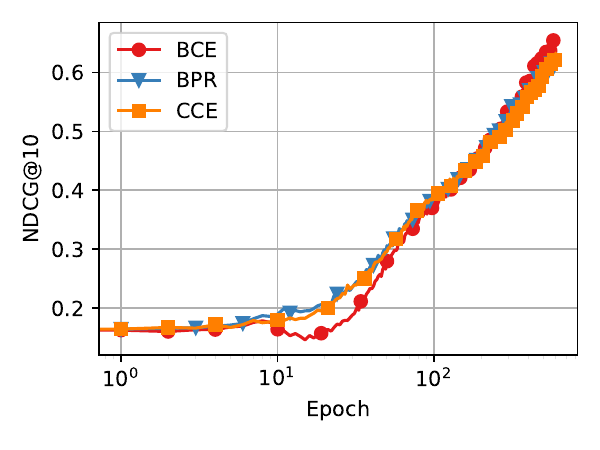}
        \caption{SASRec}
    \end{subfigure}
    \begin{subfigure}{0.33\textwidth}
        \centering
        \includegraphics[width=\textwidth]{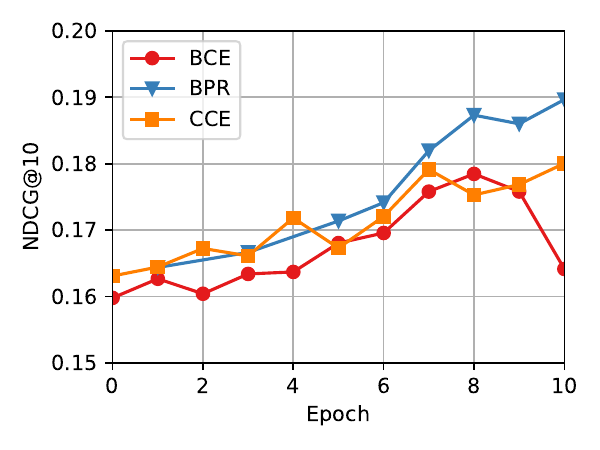}
        \caption{SASRec - Start}
    \end{subfigure}
    \begin{subfigure}{0.33\textwidth}
        \centering
        \includegraphics[width=\textwidth]{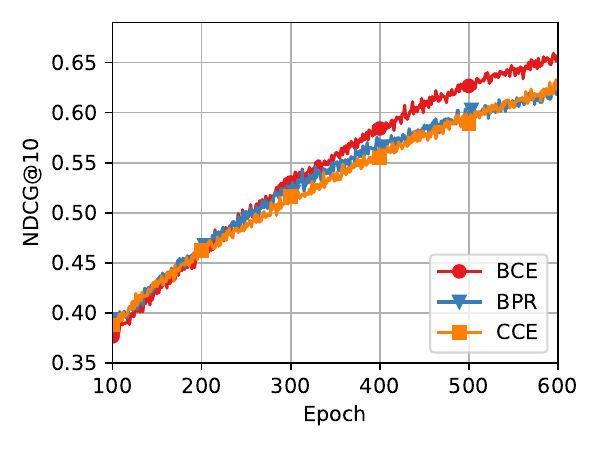}
        \caption{SASRec - End}
    \end{subfigure}
    \caption{SASRec and GRU4Rec NDCG@10 during training changing loss, using 1 negative item on Foursquare-NYC dataset.}
\end{figure}

    \begin{figure}[!ht]
        \begin{subfigure}{0.33\textwidth}
            \centering
            \includegraphics[width=\textwidth]{img/epochs_losses_NDCG10_100_GRU_SAS_ml-1m_0.pdf}
            \caption{GRU4Rec}
        \end{subfigure}
        \begin{subfigure}{0.33\textwidth}
            \centering
            \includegraphics[width=\textwidth]{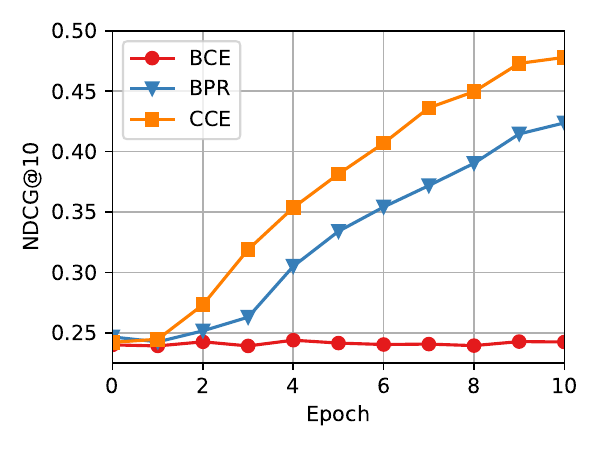}
            \caption{GRU4Rec - Start}
        \end{subfigure}
        \begin{subfigure}{0.33\textwidth}
            \centering
            \includegraphics[width=\textwidth]{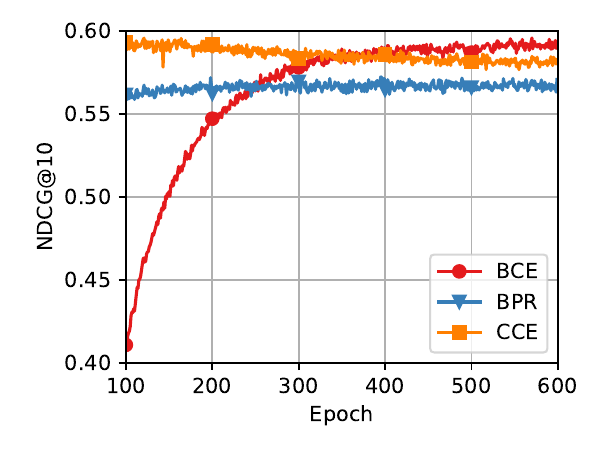}
            \caption{GRU4Rec - End}
        \end{subfigure}
        \begin{subfigure}{0.33\textwidth}
            \centering
            \includegraphics[width=\textwidth]{img/epochs_losses_NDCG10_100_GRU_SAS_ml-1m_1.pdf}
            \caption{SASRec}
        \end{subfigure}
        \begin{subfigure}{0.33\textwidth}
            \centering
            \includegraphics[width=\textwidth]{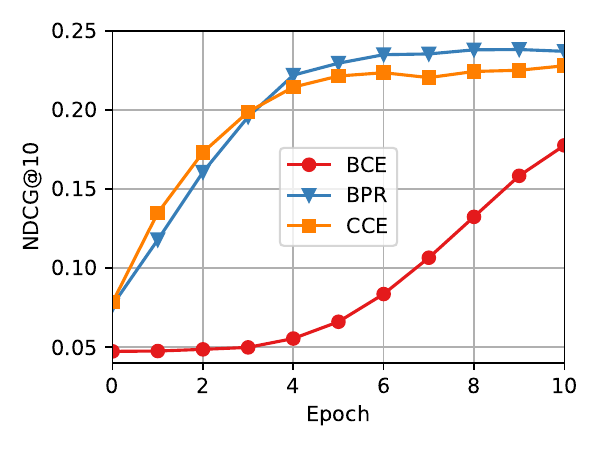}
            \caption{SASRec - Start}
        \end{subfigure}
        \begin{subfigure}{0.33\textwidth}
            \centering
            \includegraphics[width=\textwidth]{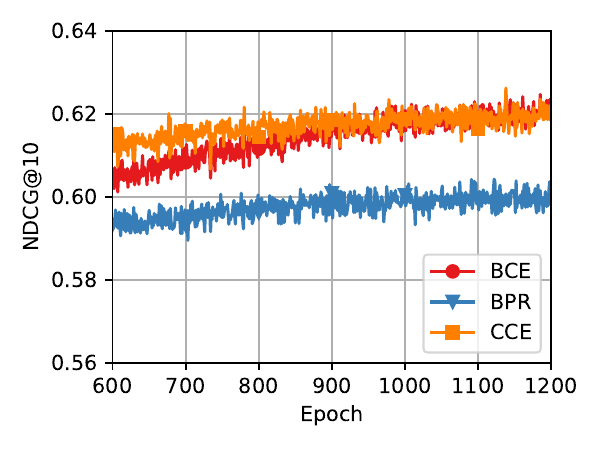}
            \caption{SASRec - End}
        \end{subfigure}
        \caption{SASRec and GRU4Rec NDCG@10 during training changing loss, using 100 negative items on ML-1M dataset.}
    \end{figure}
\begin{figure}[!ht]
    \begin{subfigure}{0.33\textwidth}
        \centering
        \includegraphics[width=\textwidth]{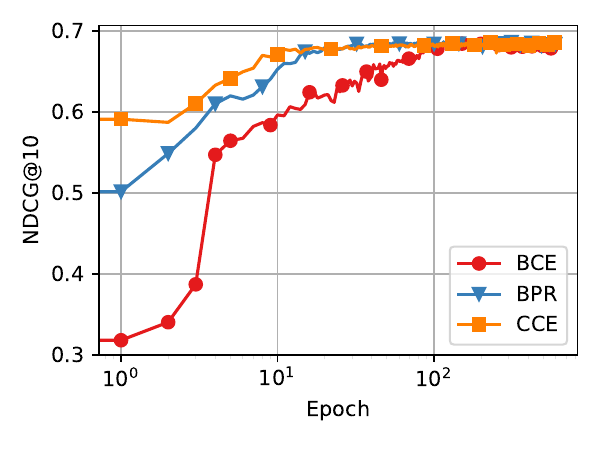}
        \caption{GRU4Rec}
    \end{subfigure}
    \begin{subfigure}{0.33\textwidth}
        \centering
        \includegraphics[width=\textwidth]{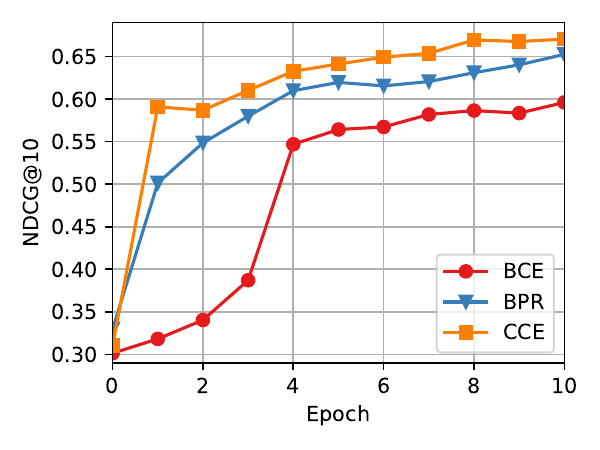}
        \caption{GRU4Rec - Start}
    \end{subfigure}
    \begin{subfigure}{0.33\textwidth}
        \centering
        \includegraphics[width=\textwidth]{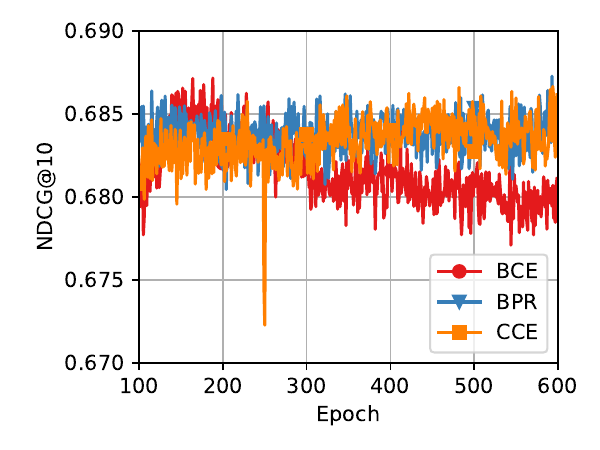}
        \caption{GRU4Rec - End}
    \end{subfigure}
    \begin{subfigure}{0.33\textwidth}
        \centering
        \includegraphics[width=\textwidth]{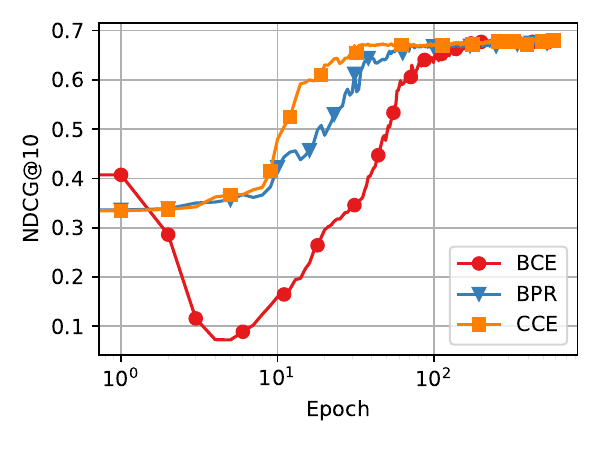}
        \caption{SASRec}
    \end{subfigure}
    \begin{subfigure}{0.33\textwidth}
        \centering
        \includegraphics[width=\textwidth]{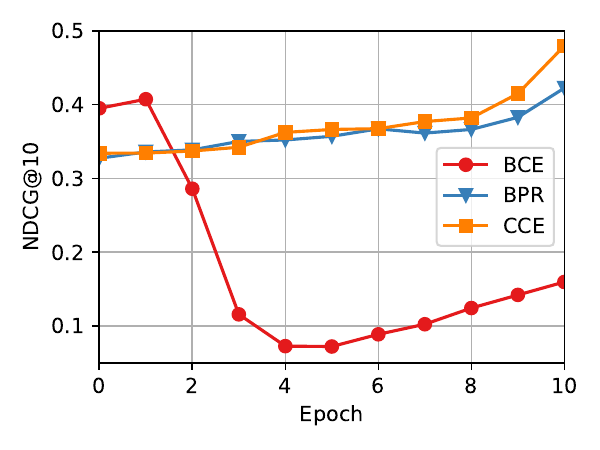}
        \caption{SASRec - Start}
    \end{subfigure}
    \begin{subfigure}{0.33\textwidth}
        \centering
        \includegraphics[width=\textwidth]{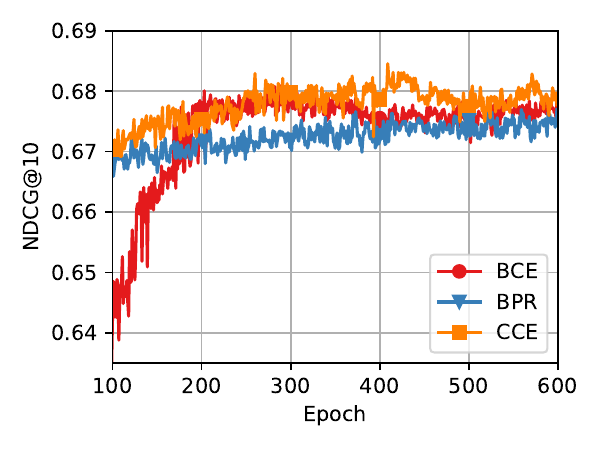}
        \caption{SASRec - End}
    \end{subfigure}
    \caption{SASRec and GRU4Rec NDCG@10 during training changing loss, using 100 negative items on Amazon Beauty dataset.}
\end{figure}

\begin{figure}[!ht]
    \begin{subfigure}{0.33\textwidth}
        \centering
        \includegraphics[width=\textwidth]{img/epochs_losses_NDCG10_100_GRU_SAS_foursquare-nyc_0.pdf}
        \caption{GRU4Rec}
    \end{subfigure}
    \begin{subfigure}{0.33\textwidth}
        \centering
        \includegraphics[width=\textwidth]{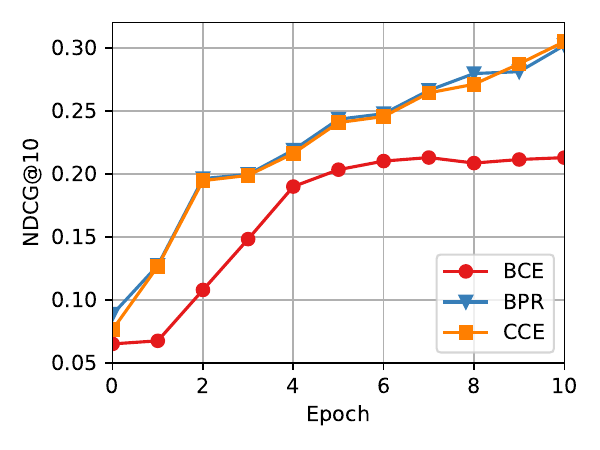}
        \caption{GRU4Rec - Start}
    \end{subfigure}
    \begin{subfigure}{0.33\textwidth}
        \centering
        \includegraphics[width=\textwidth]{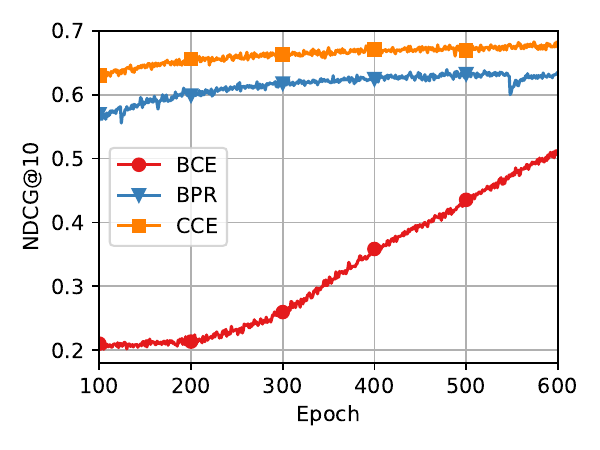}
        \caption{GRU4Rec - End}
    \end{subfigure}
    \begin{subfigure}{0.33\textwidth}
        \centering
        \includegraphics[width=\textwidth]{img/epochs_losses_NDCG10_100_GRU_SAS_foursquare-nyc_1.pdf}
        \caption{SASRec}
    \end{subfigure}
    \begin{subfigure}{0.33\textwidth}
        \centering
        \includegraphics[width=\textwidth]{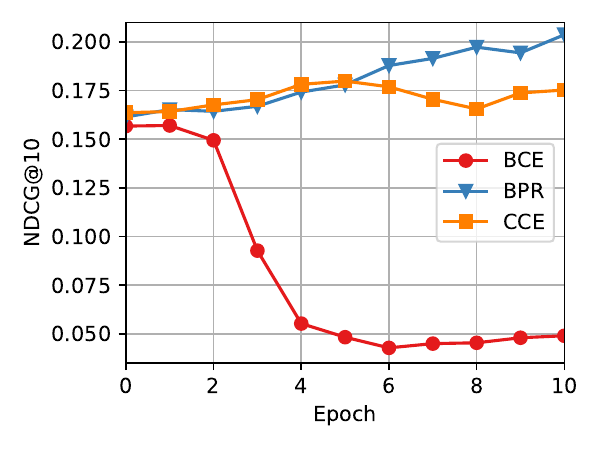}
        \caption{SASRec - Start}
    \end{subfigure}
    \begin{subfigure}{0.33\textwidth}
        \centering
        \includegraphics[width=\textwidth]{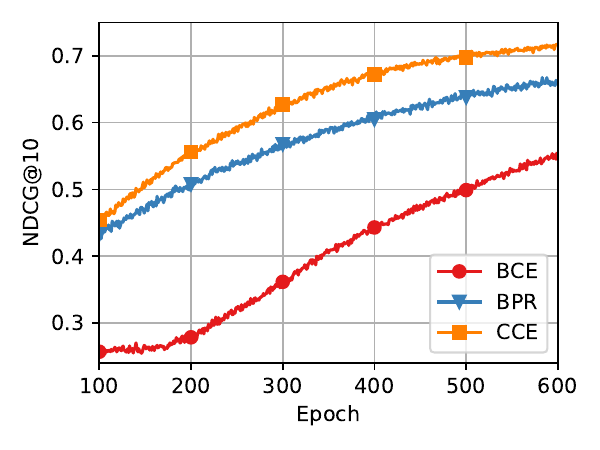}
        \caption{SASRec - End}
    \end{subfigure}
    \caption{SASRec and GRU4Rec NDCG@10 during training changing loss, using 100 negative items on Foursquare dataset.}
\end{figure}

    \begin{figure}[!ht]
        \begin{subfigure}{0.33\textwidth}
            \centering
            \includegraphics[width=\textwidth]{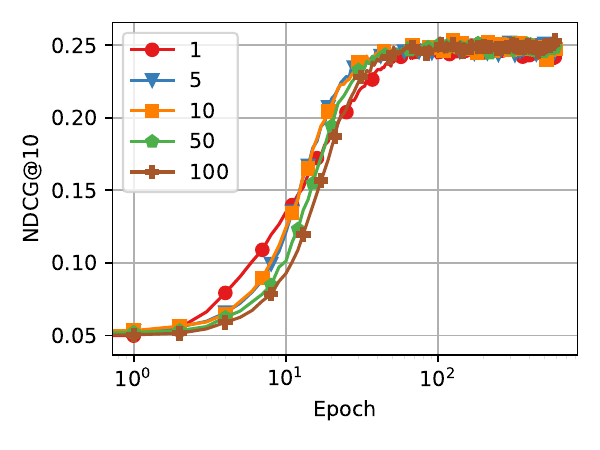}
            \caption{NCF}
        \end{subfigure}
        \begin{subfigure}{0.33\textwidth}
            \centering
            \includegraphics[width=\textwidth]{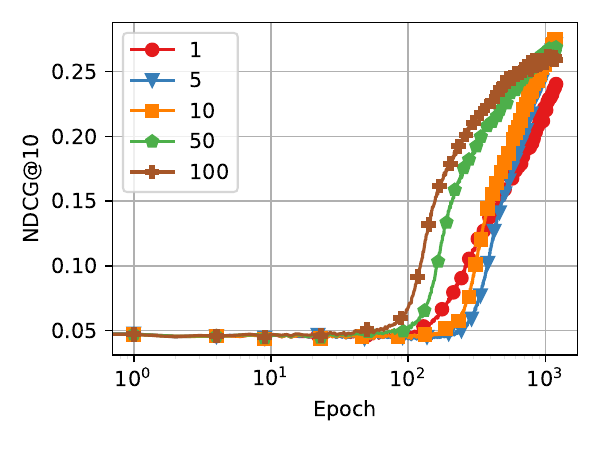}
            \caption{LightGCN}
        \end{subfigure}
    \caption{NCF and LightGCN NDCG@10 during training changing number of negative items, using BCE loss on ML-1M dataset.}
\end{figure}

\begin{figure}[!ht]
    \begin{subfigure}{0.33\textwidth}
        \centering
        \includegraphics[width=\textwidth]{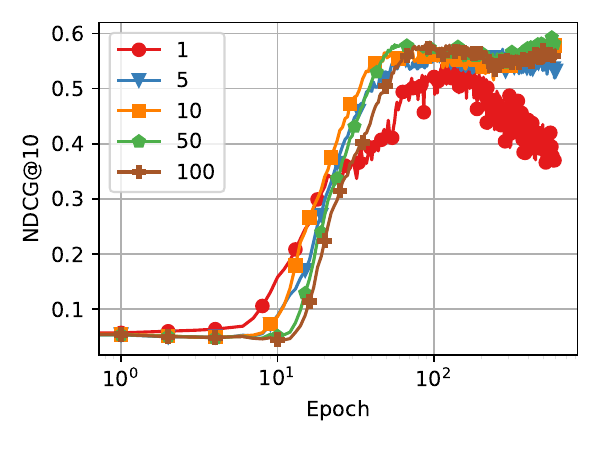}
        \caption{NCF}
    \end{subfigure}
    \begin{subfigure}{0.33\textwidth}
        \centering
        \includegraphics[width=\textwidth]{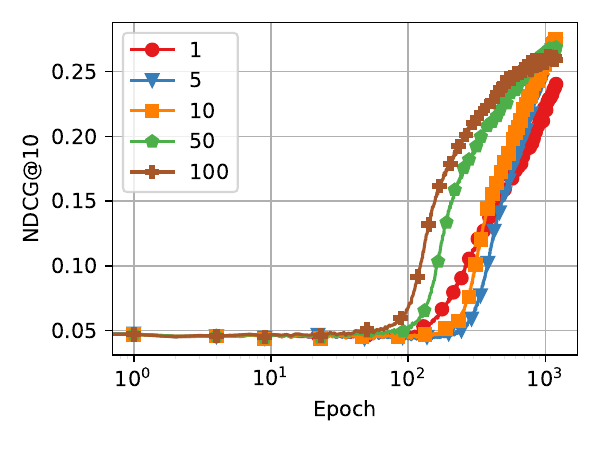}
        \caption{LightGCN}
    \end{subfigure}
    \caption{NCF and LightGCN NDCG@10 during training changing number of negative items, using BCE loss on Amazon Beauty dataset.}
\end{figure}

\begin{figure}[!ht]
    \begin{subfigure}{0.33\textwidth}
        \centering
        \includegraphics[width=\textwidth]{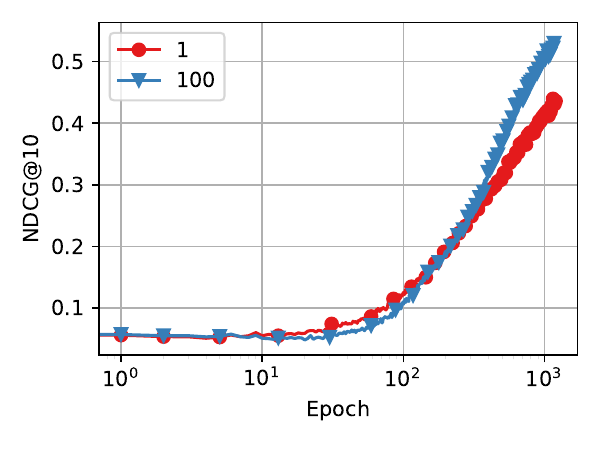}
        \caption{NCF}
    \end{subfigure}
    \begin{subfigure}{0.33\textwidth}
        \centering
        \includegraphics[width=\textwidth]{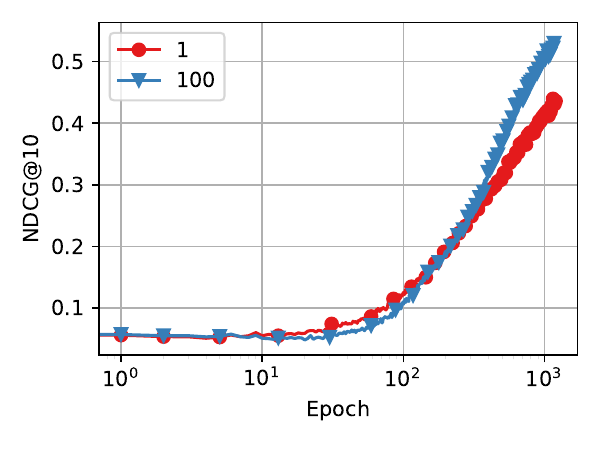}
        \caption{LightGCN}
    \end{subfigure}
    \caption{NCF and LightGCN NDCG@10 during training changing number of negative items, using BCE loss on Foursquare-NYC dataset.}
\end{figure}

    \begin{figure}[!ht]
        \begin{subfigure}{0.33\textwidth}
            \centering
            \includegraphics[width=\textwidth]{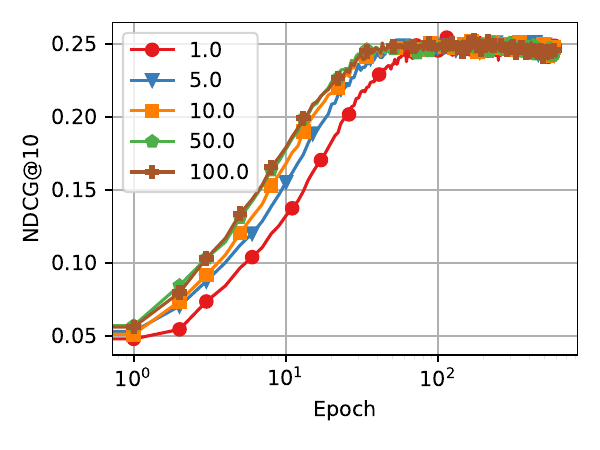}
            \caption{NCF}
        \end{subfigure}
        \begin{subfigure}{0.33\textwidth}
            \centering
            \includegraphics[width=\textwidth]{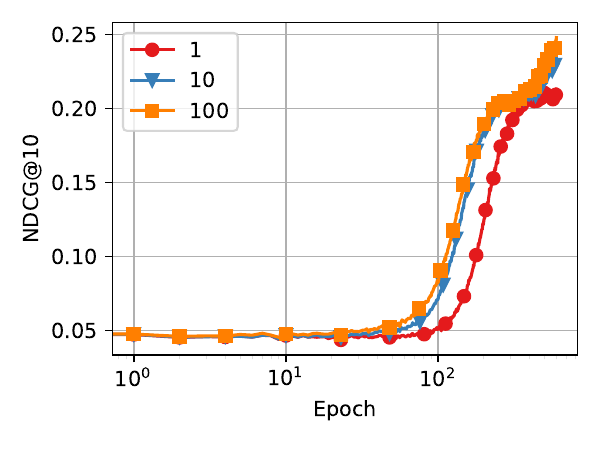}
            \caption{LightGCN}
        \end{subfigure}
        \caption{NCF and LightGCN NDCG@10 during training changing number of negative items, using BPR loss on ML-1M dataset.}
    \end{figure}

\begin{figure}[!ht]
    \begin{subfigure}{0.33\textwidth}
        \centering
        \includegraphics[width=\textwidth]{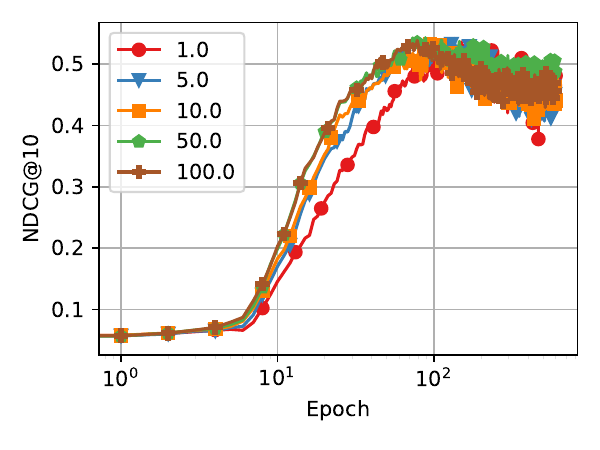}
        \caption{NCF}
    \end{subfigure}
    \begin{subfigure}{0.33\textwidth}
        \centering
        \includegraphics[width=\textwidth]{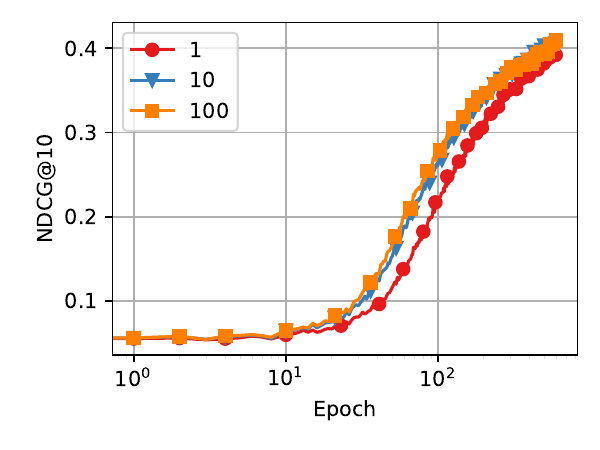}
        \caption{LightGCN}
    \end{subfigure}
    \caption{NCF and LightGCN NDCG@10 during training changing number of negative items, using BPR loss on Amazon Beauty dataset.}
\end{figure}

\begin{figure}[!ht]
    \begin{subfigure}{0.33\textwidth}
        \centering
        \includegraphics[width=\textwidth]{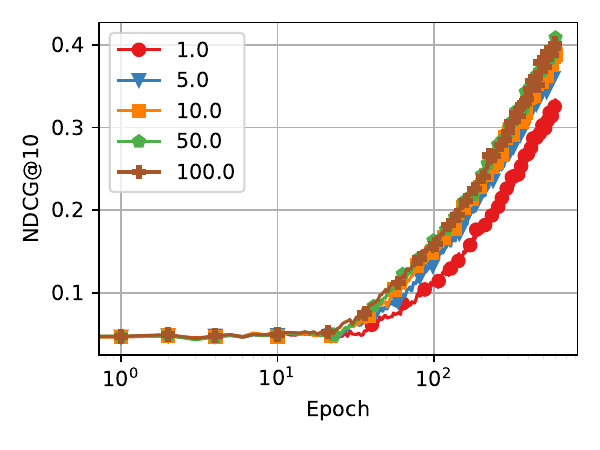}
        \caption{NCF}
    \end{subfigure}
    \begin{subfigure}{0.33\textwidth}
        \centering
        \includegraphics[width=\textwidth]{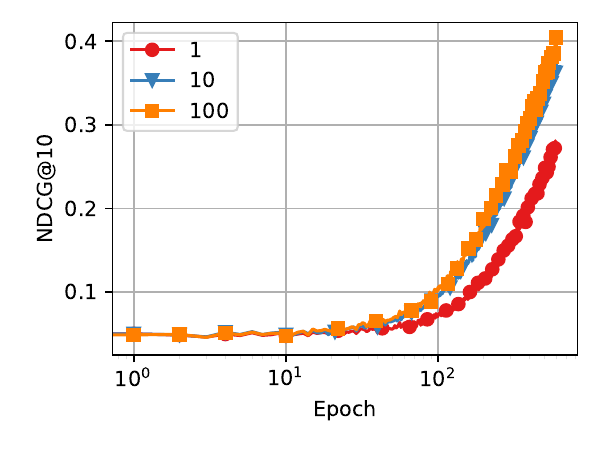}
        \caption{LightGCN}
    \end{subfigure}
    \caption{NCF and LightGCN NDCG@10 during training changing number of negative items, using BPR loss on Foursquare-NYC dataset.}
\end{figure}

    \begin{figure}[!ht]
        \begin{subfigure}{0.33\textwidth}
            \centering
            \includegraphics[width=\textwidth]{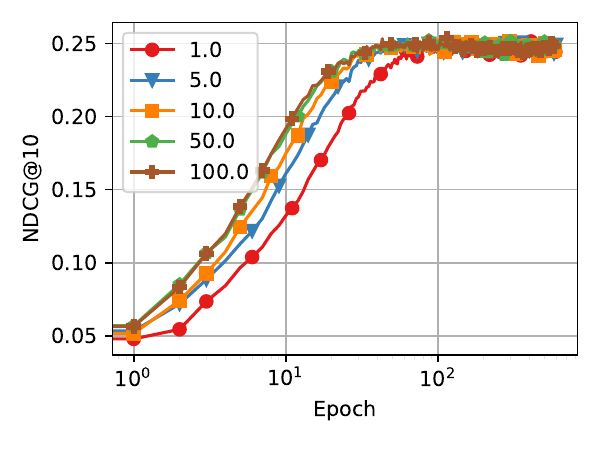}
            \caption{NCF}
        \end{subfigure}
        \begin{subfigure}{0.33\textwidth}
            \centering
            \includegraphics[width=\textwidth]{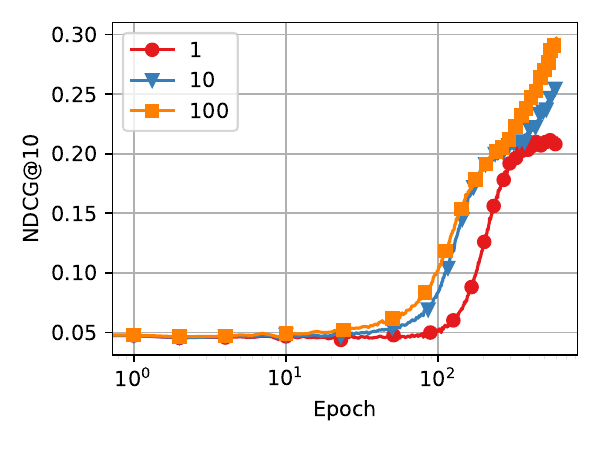}
            \caption{LightGCN}
        \end{subfigure}
        \caption{NCF and LightGCN NDCG@10 during training changing number of negative items, using CCE loss on ML-1M dataset.}
    \end{figure}
    
\begin{figure}[!ht]
    \begin{subfigure}{0.33\textwidth}
        \centering
        \includegraphics[width=\textwidth]{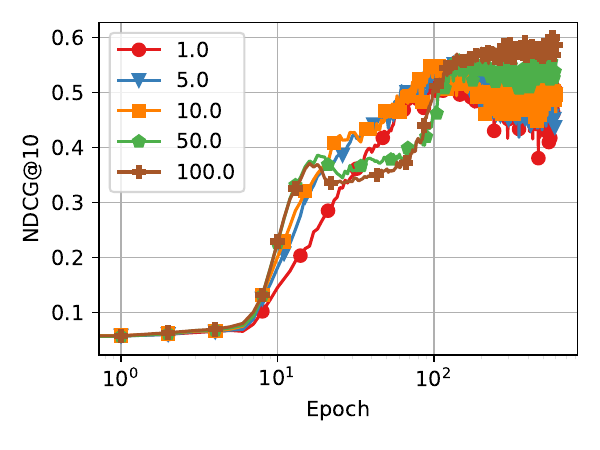}
        \caption{NCF}
    \end{subfigure}
    \begin{subfigure}{0.33\textwidth}
        \centering
        \includegraphics[width=\textwidth]{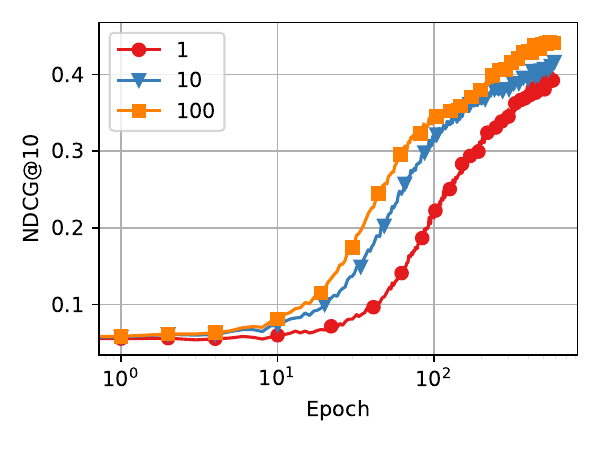}
        \caption{LightGCN}
    \end{subfigure}
    \caption{NCF and LightGCN NDCG@10 during training changing number of negative items, using CCE loss on Amazon Beauty dataset.}
\end{figure}

\begin{figure}[!ht]
    \begin{subfigure}{0.33\textwidth}
        \centering
        \includegraphics[width=\textwidth]{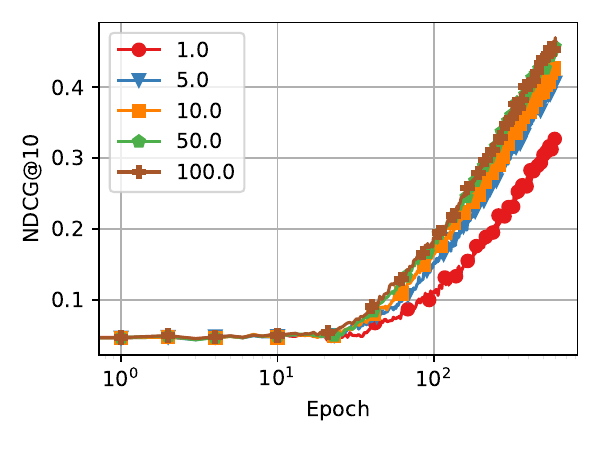}
        \caption{NCF}
    \end{subfigure}
    \begin{subfigure}{0.33\textwidth}
        \centering
        \includegraphics[width=\textwidth]{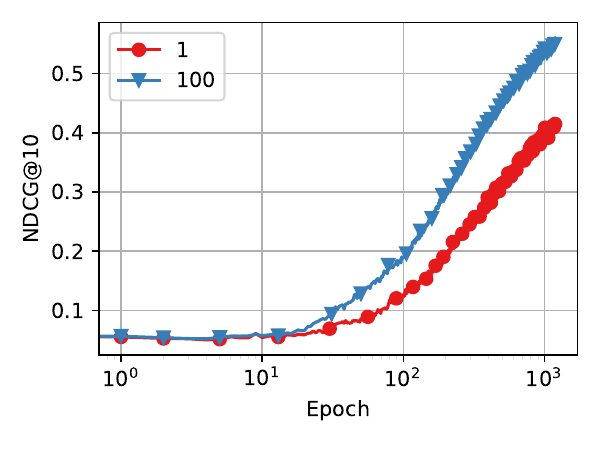}
        \caption{LightGCN}
    \end{subfigure}
    \caption{NCF and LightGCN NDCG@10 during training changing number of negative items, using CCE loss on Foursquare-NYC dataset.}
\end{figure}

    \begin{figure}[!ht]
        \begin{subfigure}{0.33\textwidth}
            \centering
            \includegraphics[width=\textwidth]{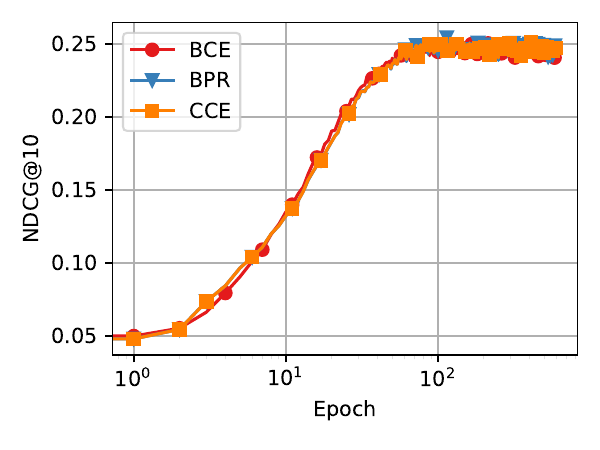}
            \caption{NCF}
        \end{subfigure}
        \begin{subfigure}{0.33\textwidth}
            \centering
            \includegraphics[width=\textwidth]{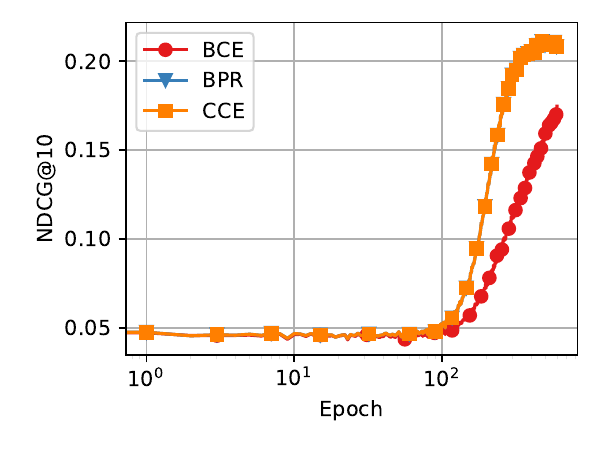}
            \caption{LightGCN}
        \end{subfigure}
        \caption{NCF and LightGCN NDCG@10 during training changing loss, using 1 negative item on ML-1M dataset.}
    \end{figure}
\begin{figure}[!ht]
    \begin{subfigure}{0.33\textwidth}
        \centering
        \includegraphics[width=\textwidth]{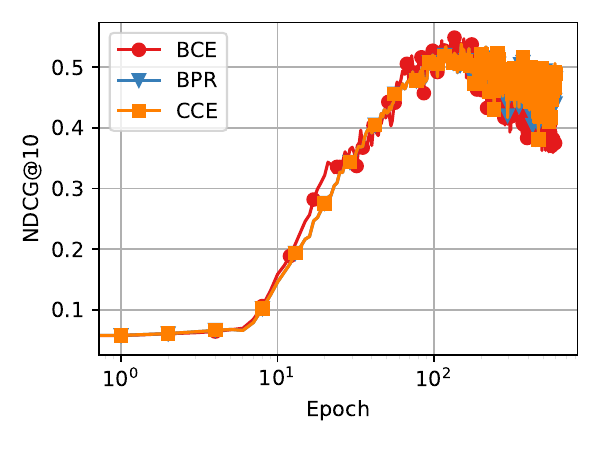}
        \caption{NCF}
    \end{subfigure}
    \begin{subfigure}{0.33\textwidth}
        \centering
        \includegraphics[width=\textwidth]{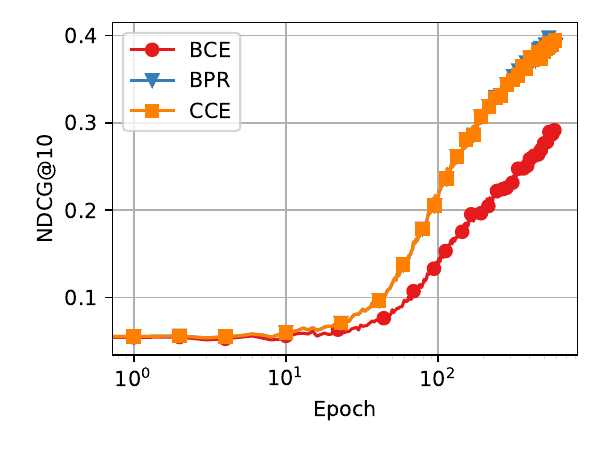}
        \caption{LightGCN}
    \end{subfigure}
    \caption{NCF and LightGCN NDCG@10 during training changing loss, using 1 negative item on Amazon Beauty dataset.}
\end{figure}

\begin{figure}[!ht]
    \begin{subfigure}{0.33\textwidth}
        \centering
        \includegraphics[width=\textwidth]{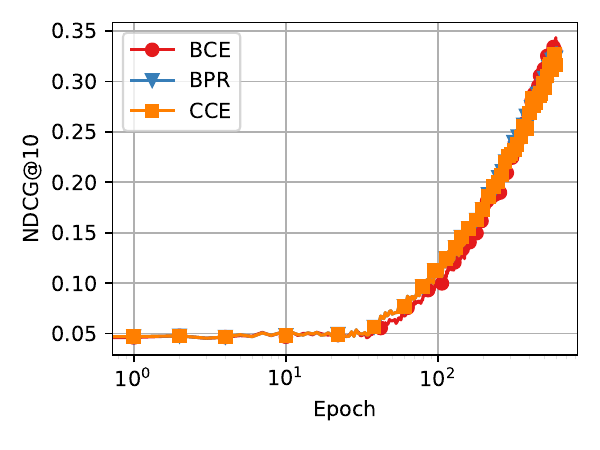}
        \caption{NCF}
    \end{subfigure}
    \begin{subfigure}{0.33\textwidth}
        \centering
        \includegraphics[width=\textwidth]{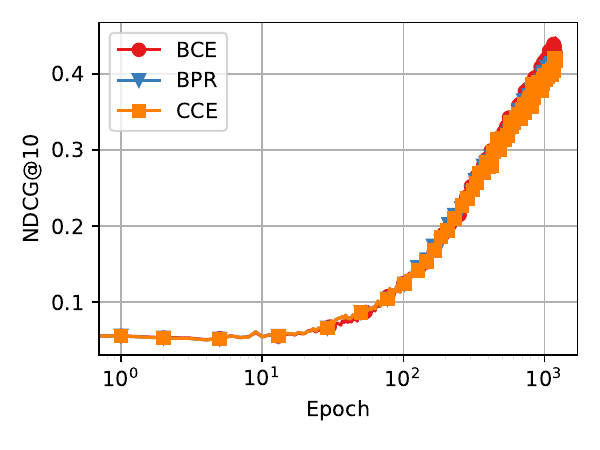}
        \caption{LightGCN}
    \end{subfigure}
    \caption{NCF and LightGCN NDCG@10 during training changing loss, using 1 negative item on Foursquare-NYC dataset.}
\end{figure}

    \begin{figure}[!ht]
        \begin{subfigure}{0.33\textwidth}
            \centering
            \includegraphics[width=\textwidth]{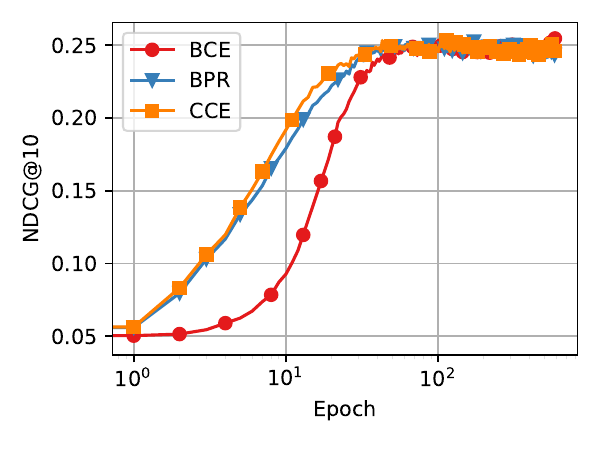}
            \caption{NCF}
        \end{subfigure}
        \begin{subfigure}{0.33\textwidth}
            \centering
            \includegraphics[width=\textwidth]{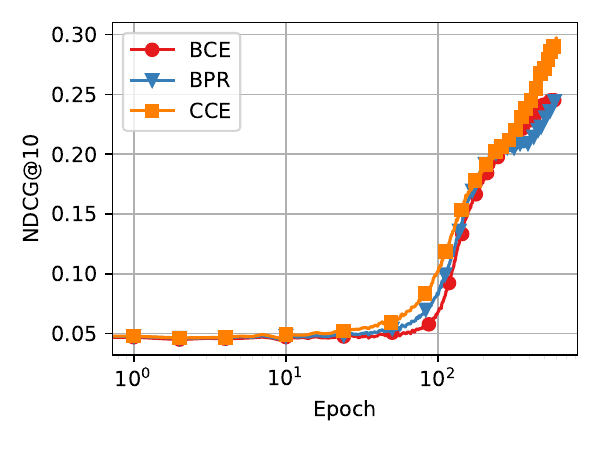}
            \caption{LightGCN}
        \end{subfigure}
        \caption{NCF and LightGCN NDCG@10 during training changing loss, using 100 negative items on ML-1M dataset.}
    \end{figure}
\begin{figure}[!ht]
    \begin{subfigure}{0.33\textwidth}
        \centering
        \includegraphics[width=\textwidth]{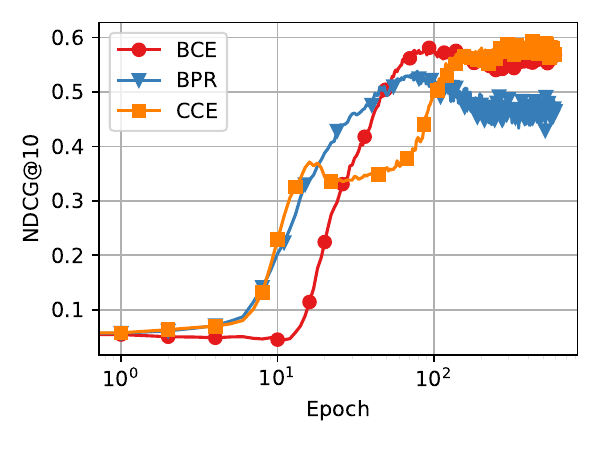}
        \caption{NCF}
    \end{subfigure}
    \begin{subfigure}{0.33\textwidth}
        \centering
        \includegraphics[width=\textwidth]{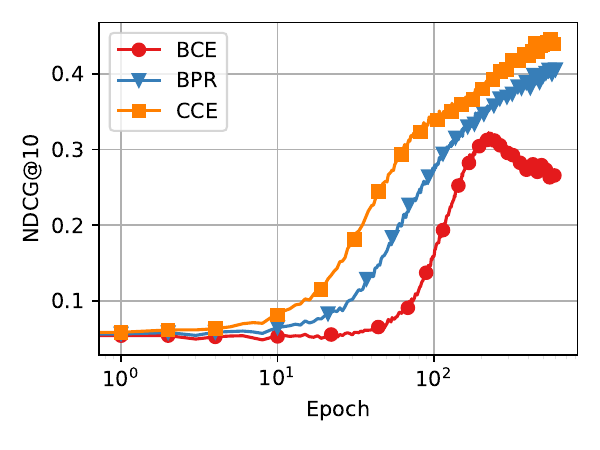}
        \caption{LightGCN}
    \end{subfigure}
    \caption{NCF and LightGCN NDCG@10 during training changing loss, using 100 negative items on Amazon Beauty dataset.}
\end{figure}

\begin{figure}[!ht]
    \begin{subfigure}{0.33\textwidth}
        \centering
        \includegraphics[width=\textwidth]{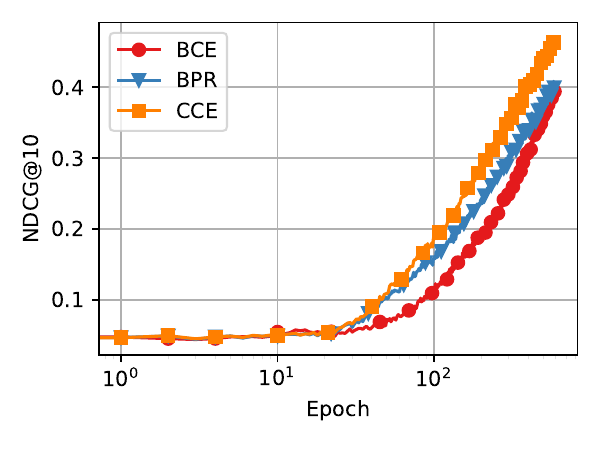}
        \caption{NCF}
    \end{subfigure}
    \begin{subfigure}{0.33\textwidth}
        \centering
        \includegraphics[width=\textwidth]{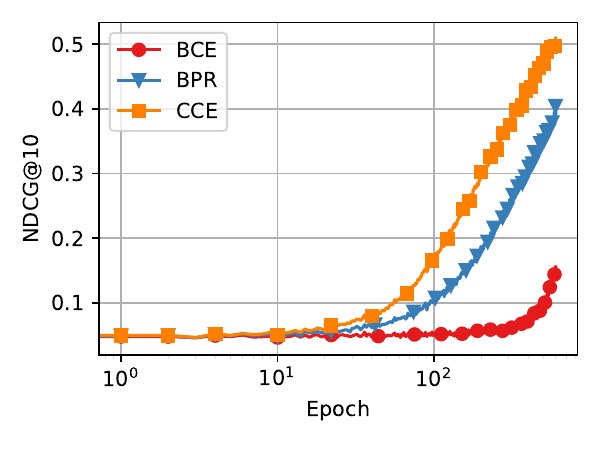}
        \caption{LightGCN}
    \end{subfigure}
    \caption{NCF and LightGCN NDCG@10 during training changing loss, using 100 negative items on Foursquare dataset.}
\end{figure}

\vfill

\clearpage

\end{document}